\pdfoutput=1

\documentclass[11pt,twoside,a4paper,cmspaper,final,collab]{cms-tdr}
\def\svnVersion{775a6a7-D}\def\svnDate{2019/09/24}\def\cmsCernNoTag{CERN-EP-2019-073}\def\cmsCernDate{\today}\def\cmsMessage{"Published in Physical Review D as \href{http://dx.doi.org/10.1103/PhysRevD.100.072002}{\doi{10.1103/PhysRevD.100.072002}.}"}

\begin{document}\cmsNoteHeader{TOP-18-006}

\hyphenation{had-ron-i-za-tion}
\hyphenation{cal-or-i-me-ter}
\hyphenation{de-vices}
\RCS$HeadURL$
\RCS$Id$
\newlength\cmsFigWidth
\ifthenelse{\boolean{cms@external}}{\setlength\cmsFigWidth{0.99\linewidth}}{\setlength\cmsFigWidth{0.5\linewidth}}
\ifthenelse{\boolean{cms@external}}{\providecommand{\cmsLeft}{left\xspace}}{\providecommand{\cmsLeft}{upper\xspace}}
\ifthenelse{\boolean{cms@external}}{\providecommand{\cmsRight}{right\xspace}}{\providecommand{\cmsRight}{lower\xspace}}
\ifthenelse{\boolean{cms@external}}{\providecommand{\NA}{\ensuremath{\cdots}\xspace}}{\providecommand{\NA}{\text{---}\xspace}}

\ifthenelse{\boolean{cms@external}}{\providecommand{\adash}{An ellipsis\xspace}}{\providecommand{\adash}{A dash\xspace}}
\ifthenelse{\boolean{cms@external}}{\providecommand{\chisqtext}{${\chi^2}$ (dof = 19)}}{\providecommand{\chisqtext}{${\chi^2}$}}

\ifthenelse{\boolean{cms@external}}{\providecommand{\one}{\mbox{1\hspace{-2.25pt}\textrm{l}}}}{\providecommand{\one}{\mathbb{1}}}
\ifthenelse{\boolean{cms@external}}{\providecommand{\CLnp}{\ensuremath{\text{C.L}}\xspace}}{\providecommand{\CLnp}{\ensuremath{\text{CL}}\xspace}}

\newlength\coordinateFigWidth
\ifthenelse{\boolean{cms@external}}{\setlength\coordinateFigWidth{0.485\linewidth}}{\setlength\coordinateFigWidth{0.4\linewidth}}
\newlength\coordinateFigSpace
\ifthenelse{\boolean{cms@external}}{\setlength\coordinateFigSpace{0.00\linewidth}}{\setlength\coordinateFigSpace{0.08\linewidth}}
\newlength\unfoldedFigWidth
\ifthenelse{\boolean{cms@external}}{\setlength\unfoldedFigWidth{0.92\linewidth}}{\setlength\unfoldedFigWidth{0.475\linewidth}}

\newlength\fSMWidth
\ifthenelse{\boolean{cms@external}}{\setlength\fSMWidth{1.35\linewidth}}{\setlength\fSMWidth{0.99\linewidth}}

\ifthenelse{\boolean{cms@external}}{\providecommand{\cmsTable}[1]{#1}}{\providecommand{\cmsTable}[1]{\resizebox{\textwidth}{!}{#1}}}
\newlength\cmsTabSkip\setlength{\cmsTabSkip}{1ex}

\newcommand{\mt}{\ensuremath{m_{\cPqt}}}
\newcommand{\Wjets}{{\PW}+jets\xspace}
\newcommand{\Zjets}{{\cPZ}+jets\xspace}
\newcommand{\Zjet}{{\cPZ}+jet\xspace}
\newcommand{\photonjet}{photon+jet\xspace}
\newcommand{\ee}{$\Pep\Pem$\xspace}
\newcommand{\mumu}{$\PGmp\PGmm$\xspace}
\newcommand{\emu}{$\Pe^{\pm}\PGm^{\mp}$\xspace}
\newcommand{\PythiaOnly}{\PYTHIA} 
\newcommand{\Pythia}{{\PYTHIA}{8}}
\newcommand{\Powheg}{\POWHEG} 
\newcommand{\Powhegvtwo}{{\POWHEG}{v2}}
\newcommand{\MadSpin}{{\textsc{MadSpin}}} 
\newcommand{\Rivet}{{\textsc{Rivet}}} 
\newcommand{\MGaMCatNLO}{\textsc{MG5}\_a\MCATNLO} 
\newcommand{\llbar}{\ensuremath{\ell\overline{\ell}}}
\newcommand{\mll}{\ensuremath{m_{\llbar}}\xspace}
\newcommand{\otg}{\ensuremath{O_\text{tG}}}
\newcommand{\ctg}{\ensuremath{C_\text{tG}}}
\newcommand{\ctgl}{\ensuremath{C_\text{tG}/\Lambda^{2}}}
\newcommand{\ctgi}{\ensuremath{C^{I}_\text{tG}}}
\newcommand{\ctgil}{\ensuremath{C^{I}_\text{tG}/\Lambda^{2}}}

\newcommand{\lumivalue}{35.9\fbinv} 
\newcommand{\xsectheo}{\ensuremath{ 831.8\,^{+19.8}_{-29.2}\,\text{(scale)} \pm 35.1\,(\text{PDF}+\alpS)\unit{pb}}}

\newcommand{\ctgfitCIold}{$-0.10 < \ctgl\ < 0.22\TeV^{-2}$}
\newcommand{\ctgfitCI}{$-0.24 < \ctgl\ < 0.07\TeV^{-2}$}  
\newcommand{\ctgifitCI}{$-0.33 < \ctgil\ < 0.20\TeV^{-2}$}

\newcommand{\Trule}{\rule{0pt}{2.3ex}}
\newcommand{\sTrule}{\rule{0pt}{1.95ex}}

\newcommand{\cospk}{\ensuremath{\cos\theta_{1}^{k}}}
\newcommand{\cosmk}{\ensuremath{\cos\theta_{2}^{k}}}
\newcommand{\cospr}{\ensuremath{\cos\theta_{1}^{r}}}
\newcommand{\cosmr}{\ensuremath{\cos\theta_{2}^{r}}}
\newcommand{\cospn}{\ensuremath{\cos\theta_{1}^{n}}}
\newcommand{\cosmn}{\ensuremath{\cos\theta_{2}^{n}}}
\newcommand{\cospks}{\ensuremath{\cos\theta_{1}^{k*}}}
\newcommand{\cosmks}{\ensuremath{\cos\theta_{2}^{k*}}}
\newcommand{\cosprs}{\ensuremath{\cos\theta_{1}^{r*}}}
\newcommand{\cosmrs}{\ensuremath{\cos\theta_{2}^{r*}}}
\newcommand{\coskk}{\ensuremath{\cospk\cosmk}}
\newcommand{\cosrr}{\ensuremath{\cospr\cosmr}}
\newcommand{\cosnn}{\ensuremath{\cospn\cosmn}}
\newcommand{\cosPrk}{\ensuremath{\cospr\cosmk+\cospk\cosmr}}
\newcommand{\cosMrk}{\ensuremath{\cospr\cosmk-\cospk\cosmr}}
\newcommand{\cosPnr}{\ensuremath{\cospn\cosmr+\cospr\cosmn}}
\newcommand{\cosMnr}{\ensuremath{\cospn\cosmr-\cospr\cosmn}}
\newcommand{\cosPnk}{\ensuremath{\cospn\cosmk+\cospk\cosmn}}
\newcommand{\cosMnk}{\ensuremath{\cospn\cosmk-\cospk\cosmn}}
\newcommand{\cosphi}{\ensuremath{\cos\varphi}}
\newcommand{\cosphilab}{\ensuremath{\cosphi_{\mathrm{lab}}}}
\newcommand{\dphi}{\ensuremath{\abs{\Delta\phi_{\ell\ell}}}}

\newcommand{\bpk}{\ensuremath{B_{1}^{k}}}
\newcommand{\bmk}{\ensuremath{B_{2}^{k}}}
\newcommand{\bpr}{\ensuremath{B_{1}^{r}}}
\newcommand{\bmr}{\ensuremath{B_{2}^{r}}}
\newcommand{\bpn}{\ensuremath{B_{1}^{n}}}
\newcommand{\bmn}{\ensuremath{B_{2}^{n}}}
\newcommand{\bpks}{\ensuremath{B_{1}^{k*}}}
\newcommand{\bmks}{\ensuremath{B_{2}^{k*}}}
\newcommand{\bprs}{\ensuremath{B_{1}^{r*}}}
\newcommand{\bmrs}{\ensuremath{B_{2}^{r*}}}
\newcommand{\ckk}{\ensuremath{C_{kk}}}
\newcommand{\crr}{\ensuremath{C_{rr}}}
\newcommand{\cnn}{\ensuremath{C_{nn}}}
\newcommand{\cPrk}{\ensuremath{C_{rk}+C_{kr}}}
\newcommand{\cMrk}{\ensuremath{C_{rk}-C_{kr}}}
\newcommand{\cPnr}{\ensuremath{C_{nr}+C_{rn}}}
\newcommand{\cMnr}{\ensuremath{C_{nr}-C_{rn}}}
\newcommand{\cPnk}{\ensuremath{C_{nk}+C_{kn}}}
\newcommand{\cMnk}{\ensuremath{C_{nk}-C_{kn}}}
\newcommand{\Aphilab}{\ensuremath{A_{\cosphi}^{\text{lab}}}}
\newcommand{\Adphi}{\ensuremath{A_{\dphi}}}

\newcommand{\mut}{\ensuremath{\hat{\mu}_{\cPqt}}}
\newcommand{\dt}{\ensuremath{\hat{d}_{\cPqt}}}

\newcommand{\cmm}{\ensuremath{\hat{c}_{-\,-}}}
\newcommand{\cmp}{\ensuremath{\hat{c}_{-\,+}}}
\newcommand{\cVV}{\ensuremath{\hat{c}_{\text{VV}}}}
\newcommand{\cVA}{\ensuremath{\hat{c}_{\text{VA}}}}
\newcommand{\cAV}{\ensuremath{\hat{c}_{\text{AV}}}}
\newcommand{\cAA}{\ensuremath{\hat{c}_{\text{AA}}}}
\newcommand{\cone}{\ensuremath{\hat{c}_{1}}}
\newcommand{\ctwo}{\ensuremath{\hat{c}_{2}}}
\newcommand{\cthree}{\ensuremath{\hat{c}_{3}}}

\newcolumntype{X}[1]{D{,}{\,\pm\,}{#1}}
\newcolumntype{Y}[1]{D{,}{\,.\,}{#1}}
\newcolumntype{Z}[1]{D{,}{,\;}{#1}}

\cmsNoteHeader{TOP-18-006}
\title{Measurement of the top quark polarization and \texorpdfstring{\ttbar}{ttbar} spin correlations using dilepton final states in proton-proton collisions at \texorpdfstring{$\sqrt{s}=13\TeV$}{sqrt(s) = 13 TeV}}

\date{\today}

\abstract{
 Measurements of the top quark polarization and top quark pair
 (\ttbar) spin correlations are presented using events
 containing two oppositely charged leptons (\ee, \emu, or \mumu)
 produced in proton-proton collisions at a
 center-of-mass energy of $13\TeV$. 
 The data were recorded by the CMS experiment at the LHC in 2016
 and correspond to an integrated luminosity of
 \lumivalue. A set of parton-level normalized
 differential cross sections, sensitive to each of the independent
 coefficients of the spin-dependent parts of the \ttbar\
 production density matrix, is measured for the first time at
 $13\TeV$. The measured distributions and extracted
 coefficients are compared with standard model predictions from
 simulations at next-to-leading-order (NLO) accuracy in
 quantum chromodynamics (QCD), and from NLO QCD calculations including
 electroweak corrections. All measurements are found to be
 consistent with the expectations of the standard model.
 The normalized differential cross sections are used
 in fits to constrain the anomalous chromomagnetic
 and chromoelectric dipole moments of the top quark to \ctgfitCI\ and \ctgifitCI, 
 respectively, at 95\% confidence level. 
}

\hypersetup{%
pdfauthor={CMS Collaboration},%
pdftitle={Measurement of the top quark polarization and ttbar spin correlations using dilepton final states in proton-proton collisions at sqrt(s) = 13 TeV},%
pdfsubject={CMS},%
pdfkeywords={CMS, physics, top, spin, polarization, spin correlation, asymmetry}}

\maketitle

\section{Introduction}
\label{sec:intro}

The top quark is the heaviest known fundamental particle and has a lifetime on the order of $10^{-25}\unit{s}$~\cite{PhysRevD.98.030001}.
This is shorter than the quantum chromodynamic (QCD) hadronization time scale $1/\Lambda_{\text{QCD}}\approx 10^{-24}\unit{s}$, and much shorter than the spin decorrelation time scale
$\mt/\Lambda^{2}_{\text{QCD}}\approx 10^{-21}\unit{s}$~\cite{MahlonParke2010} (where $\mt$ is the top quark mass).
Thus, not only does the top quark decay before hadronization occurs, but also its spin information is preserved in the angular distribution of its decay products.

At the CERN LHC, top quarks are produced mostly in pairs via
gluon fusion ($\cPg\cPg \to \ttbar$).
The quarks
are unpolarized at leading order (LO), owing
to the parity-conserving nature (longitudinal polarization) and
approximate time invariance (transverse polarization) of QCD interactions. 
In the standard model (SM), a small longitudinal polarization arises
from electroweak (EW) corrections, while a small transverse polarization
comes from absorptive terms at one loop (both
${<}1\%$~\cite{Bernreuther:2013aga,Bernreuther:2015yna}). 
The spins of the top quarks and antiquarks are strongly
correlated, and the configuration of spins depends on
the invariant mass of the \ttbar pair ($m_{\ttbar}$), with like
(unlike) helicity pairs dominating at low (high) $m_{\ttbar}$.

This paper presents a measurement of all the independent
coefficients of the top quark spin-dependent parts of the \ttbar
production density matrix, as described in
Ref.~\cite{Bernreuther:2015yna}, 
using events (labeled dileptonic) in which the decay of the \ttbar\ pair leads to two oppositely charged leptons (\ee, \emu, or \mumu) in the final state.
The analysis uses a data sample of proton-proton ($\Pp\Pp$) collision events
collected by the CMS experiment at a center-of-mass (CM) energy of $13\TeV$ in 2016, corresponding to an integrated luminosity of
\lumivalue~\cite{bib:CMS-PAS-LUM-17-001}.  
Similar measurements have been made by the ATLAS Collaboration at $\sqrt{s}=8\TeV$~\cite{bib:ATLASspindensity8TeV}. 
Differential \ttbar\ cross sections corresponding to a subset of the
coefficients, and other observables sensitive to the top quark
polarization and \ttbar spin correlations, have been measured by the
ATLAS and CMS Collaborations at $\sqrt{s}=7$,
8, and 13\TeV~\cite{bib:Chatrchyan:2013wua,bib:PhysRevD.90.112016,bib:PhysRevD.93.012002,bib:Khachatryan:2016xws,Aaboud:2019hwz}. 

In this analysis, each coefficient is extracted from a measured normalized differential \ttbar\ cross section,
using the same event selection and reconstruction as described in Ref.~\cite{Sirunyan:2018ucr}.
The distributions are corrected to the parton level and extrapolated to the full phase space,
using a refined unfolding procedure with no regularization bias.
In addition to full statistical and systematic covariance matrices for each measured distribution, matrices are provided for the combined set of all measured bins, allowing constraints to be placed 
using several measured distributions simultaneously.

The absence of direct signals of beyond-the-SM (BSM) particles in the LHC
data analyzed so far suggests that BSM phenomena might only be directly
observed at an energy scale larger than that probed at the LHC.  
However, BSM physics could still indirectly manifest itself in new
vertices and modified couplings. Such effects can be
accommodated by adding higher-dimensional operators to the SM
Lagrangian in an effective field theory (EFT) approach.
The coefficients measured in this analysis are sensitive to
all but one of the operators of mass dimension six relevant for hadronic \ttbar production~\cite{Bernreuther:2015yna}.
We set limits on contributions from these operators using simultaneous fits to the measured normalized differential cross sections,
including constraints on the chromomagnetic and chromoelectric dipole moments
of the top quark.

\section{Formalism and observables}
\label{sec:observables}

The square of the matrix element for \ttbar production and decay to two leptons (with appropriate color and spin summation implied)~\cite{bib:1212.4888} can be written as
\begin{linenomath}
\begin{align}
\abs{\mathcal{M}(\cPq\cPaq/\cPg\cPg \to \ttbar \to \ell^+ \cPgn \cPqb \; \ell^- \cPagn \cPaqb )}^2 \propto \rho R \overline{\rho}.
 \label{eq:ME}
\end{align}
\end{linenomath}
Here, $\ell$ refers to an electron or muon, $R$ is the spin density matrix related to on-shell \ttbar\ production, and $\rho$ and $\overline{\rho}$ are the decay spin density matrices for the top quark and antiquark, respectively. The narrow width of the top quark compared to its mass allows factorization of the production and decay processes.

The aim of this analysis is to study the properties of the $R$ matrix, 
which is purely a function of the partonic initial state and production kinematic variables, and is therefore sensitive to BSM phenomena in \ttbar production~\cite{bib:1212.4888}.
While the analysis is also sensitive to BSM effects in \ttbar decays, 
these effects are heavily constrained~\cite{Fabbrichesi:2014wva,Cao:2015doa}, 
and therefore have a minimal effect on the measured distributions~\cite{bib:1212.4888,Bernreuther:2015yna}. 

The production spin density matrix $R$ can be decomposed in the \cPqt\ and \cPaqt\ spin spaces using a Pauli matrix basis:
\begin{linenomath}
\begin{align}
R \propto \tilde{A} \one \otimes  \one + \tilde{B}_i^+ \sigma^i \otimes \one + \tilde{B}_i^- \one \otimes \sigma^i + \tilde{C}_{ij} \sigma^i \otimes \sigma^j,
\label{eq:prodspindensity}
\end{align}
\end{linenomath}
where
$\one$ is the $2{\times}2$ unit matrix, $\sigma^i$ are the Pauli matrices, and
the first (second) matrix in each tensor product refers to the top quark (antiquark) spin space.
The function $\tilde{A}$ determines the total \ttbar\ production cross section and the top quark kinematic distributions,
$\mathbf{\tilde{B}^\pm}$ are three-dimensional vectors of functions that characterize the degree of top quark or antiquark polarization in each direction, 
and $\tilde{C}$ is a $3{\times}3$ matrix of functions that characterize the correlation between the top quark and antiquark spins.

\begin{figure}[!htpb]
\centering
\includegraphics[width=\coordinateFigWidth]{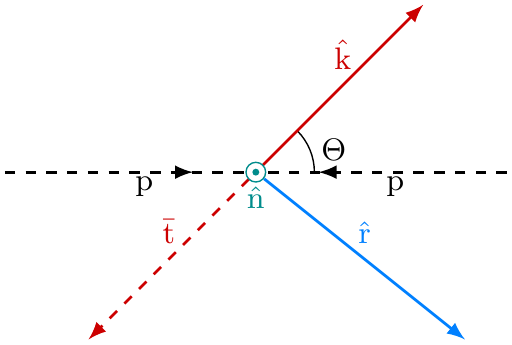}
\hspace{\coordinateFigSpace}
\includegraphics[width=\coordinateFigWidth]{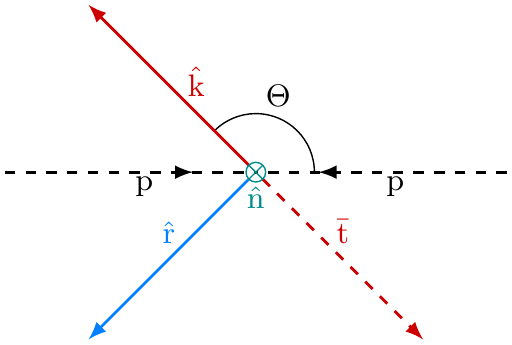}
\caption{\label{fig:coordinate}\protect
Coordinate system used for the spin measurements, illustrated in the scattering plane for $\Theta<\pi/2$ (left) and $\Theta>\pi/2$ (right),
where the signs of $\hat{r}$ and $\hat{n}$ are flipped at $\Theta=\pi/2$ as shown in Eq.~(\ref{eq:basis}).
The $\hat{k}$ axis is defined by the top quark direction, measured in the \ttbar CM frame.
For the basis used to define the coefficient functions in Eq.~(\ref{eq:prodspindensityexpanded}), the incoming particles $\text{p}$ represent the incoming partons, while for the basis used to measure the coefficients in Eqs.~(\ref{eq:theodist1})--(\ref{eq:theodist3}) they represent the incoming protons.
}
\end{figure}

We choose an orthonormal basis to decompose the top quark spin, where these functions have definite properties with respect to discrete symmetries~\cite{Degrande2011,Bernreuther:2015yna}.
This basis is illustrated in Fig.~\ref{fig:coordinate}.
Working in the \ttbar CM frame, we use the helicity axis $\hat{k}$ defined by the top quark direction
and the direction $\hat{p}$ of the incoming parton to define the
direction perpendicular to the scattering plane $\hat{n}=(\hat{p}\times\hat{k})/\sin\Theta$, where $\Theta$ is the top quark scattering angle.
The direction in the scattering plane mutually perpendicular to $\hat{k}$ and the transverse axis $\hat{n}$ is given by ${\hat{r}=(\hat{p} - \hat{k} \cos\Theta)/\sin\Theta}$.

We expand $\tilde{B}_i^\pm$ and $ \tilde{C}_{ij} $ in terms of the orthonormal basis $\{\hat{k},\hat{r},\hat{n}\}$:
\begin{linenomath}
\ifthenelse{\boolean{cms@external}}
{
\begin{align}
 & \tilde{B}_i^\pm =  b_k^\pm \hat{k}_i + b_r^\pm \hat{r}_i + b_n^\pm \hat{n}_i, \nonumber \\
 \label{eq:prodspindensityexpanded}
 & \tilde{C}_{ij}  =  c_{kk} \hat{k}_i\hat{k}_j + c_{rr} \hat{r}_i\hat{r}_j + c_{nn} \hat{n}_i\hat{n}_j \\ 
 &+  c_{rk} (\hat{r}_i\hat{k}_j + \hat{k}_i\hat{r}_j)  +  c_{nr} (\hat{n}_i\hat{r}_j + \hat{r}_i\hat{n}_j) +  c_{kn} (\hat{k}_i\hat{n}_j + \hat{n}_i\hat{k}_j) \nonumber \\
 &+  c_{n} (\hat{r}_i\hat{k}_j - \hat{k}_i\hat{r}_j)+  c_{k} (\hat{n}_i\hat{r}_j - \hat{r}_i\hat{n}_j) +  c_{r} (\hat{k}_i\hat{n}_j - \hat{n}_i\hat{k}_j). \nonumber
\end{align}
}
{
\begin{equation}
\begin{aligned}
 \tilde{B}_i^\pm &=  b_k^\pm \hat{k}_i + b_r^\pm \hat{r}_i + b_n^\pm \hat{n}_i, \\
 \label{eq:prodspindensityexpanded}
 \tilde{C}_{ij}  &=  c_{kk} \hat{k}_i\hat{k}_j + c_{rr} \hat{r}_i\hat{r}_j + c_{nn} \hat{n}_i\hat{n}_j \\ 
 &+  c_{rk} (\hat{r}_i\hat{k}_j + \hat{k}_i\hat{r}_j)  +  c_{nr} (\hat{n}_i\hat{r}_j + \hat{r}_i\hat{n}_j) +  c_{kn} (\hat{k}_i\hat{n}_j + \hat{n}_i\hat{k}_j) \\
 &+  c_{n} (\hat{r}_i\hat{k}_j - \hat{k}_i\hat{r}_j)+  c_{k} (\hat{n}_i\hat{r}_j - \hat{r}_i\hat{n}_j) +  c_{r} (\hat{k}_i\hat{n}_j - \hat{n}_i\hat{k}_j).
\end{aligned}
\end{equation}
}
\end{linenomath}
The coefficient functions $b_{i}^\pm$, $c_{ij}$, and $c_{i}$ are functions of the partonic CM energy squared $s$ and $\cos\Theta$.
They can each be classified with respect to P, CP, T, and Bose--Einstein symmetry,
and their P and CP symmetry properties are summarized in Table~\ref{tab:obscoef}.
The approximate CP invariance of the SM requires the $\tilde{C}$ matrix to be symmetric (i.e., the CP-odd coefficient functions vanish: $c_{k}=c_{r}=c_{n}=0$) and the top quark and antiquark to have the same polarization coefficient functions (i.e., $b_i^+=b_i^-$). 
The P invariance of QCD forces the P-odd coefficient functions to vanish in the absence of EW interactions, so large values are allowed only for the P- and CP-even spin correlations $c_{kk}$, $c_{rr}$, $c_{nn}$, and $c_{rk}$ (and the transverse polarization coefficient functions $b_n^\pm$, but these are zero at tree level in QCD by T invariance).
Any deviation from these expectations would be a sign of BSM phenomena.

\begin{table*}[!htb]
\centering
\topcaption{\label{tab:obscoef} Observables and their corresponding measured coefficients, production spin density matrix coefficient functions,
and P and CP symmetry properties.
For the laboratory-frame asymmetries
shown in the last two rows,
there is no direct correspondence with the coefficient functions.
}
\cmsTable{
\begin{scotch}{l l l Z{7.8}}
{Observable} & {Measured coefficient} & {Coefficient function}  & \multicolumn{1}{c}{{Symmetries}} \\
\hline
\Trule
$\cospk$ & $\bpk$ & $b^+_k$ & \text{P-odd}, \text{CP-even} \\[0.2ex]
$\cosmk$ & $\bmk$ & $b^-_k$ & \text{P-odd}, \text{CP-even} \\[0.2ex]
$\cospr$ & $\bpr$ & $b^+_r$ & \text{P-odd}, \text{CP-even} \\[0.2ex]
$\cosmr$ & $\bmr$ & $b^-_r$ & \text{P-odd}, \text{CP-even} \\[0.2ex]
$\cospn$ & $\bpn$ & $b^+_n$ & \text{P-even}, \text{CP-even} \\[0.2ex]
$\cosmn$ & $\bmn$ & $b^-_n$ & \text{P-even}, \text{CP-even} \\[\cmsTabSkip]
$\cospks$ & $\bpks$ & $b^+_k$ & \text{P-odd}, \text{CP-even} \\[0.2ex]
$\cosmks$ & $\bmks$ & $b^-_k$ & \text{P-odd}, \text{CP-even} \\[0.2ex]
$\cosprs$ & $\bprs$ & $b^+_r$ & \text{P-odd}, \text{CP-even} \\[0.2ex]
$\cosmrs$ & $\bmrs$ & $b^-_r$ & \text{P-odd}, \text{CP-even} \\ [\cmsTabSkip]
$\coskk$ & $\ckk$ & $c_{kk}$ & \text{P-even}, \text{CP-even} \\[0.2ex]
$\cosrr$ & $\crr$ & $c_{rr}$ & \text{P-even}, \text{CP-even} \\[0.2ex]
$\cosnn$ & $\cnn$ & $c_{nn}$ & \text{P-even}, \text{CP-even} \\ [\cmsTabSkip]
$\cosPrk$ & $\cPrk$ & $c_{rk}$ & \text{P-even}, \text{CP-even} \\[0.2ex]
$\cosMrk$ & $\cMrk$ & $c_{n}$ & \text{P-even}, \text{CP-odd} \\[0.2ex]
$\cosPnr$ & $\cPnr$ & $c_{nr}$ & \text{P-odd}, \text{CP-even} \\[0.2ex]
$\cosMnr$ & $\cMnr$ & $c_{k}$ & \text{P-odd}, \text{CP-odd} \\[0.2ex]
$\cosPnk$ & $\cPnk$ & $c_{kn}$ & \text{P-odd}, \text{CP-even} \\[0.2ex]
$\cosMnk$ & $\cMnk$ & $-c_{r}$ & \text{P-odd}, \text{CP-odd} \\ [\cmsTabSkip]
$\cosphi$ & $D$ & $-(c_{kk}+c_{rr}+c_{nn})/3$ & \text{P-even}, \text{CP-even} \\ [\cmsTabSkip]
$\cosphilab$ & $\Aphilab$ & \NA & \multicolumn{1}{c}{\NA} \\[0.2ex]
$\dphi$ & $\Adphi$ & \NA & \multicolumn{1}{c}{\NA} \\ [0.3ex]
\end{scotch}
}
\end{table*}

The Bose--Einstein symmetry of the $\cPg\cPg$ initial state requires a redefinition of the $\hat{r}$ and $\hat{n}$ axes (which are odd under Bose--Einstein symmetry) to allow nonzero values of the relevant coefficient functions~\cite{Bernreuther:2015yna}:
\begin{linenomath}
\begin{align}
\{\hat{k},\hat{r},\hat{n}\} \to \{\hat{k}, \,\sign(\cos\Theta) \hat{r} , \,\sign(\cos\Theta) \hat{n}\},
 \label{eq:basis}
\end{align}
\end{linenomath}
i.e., we have used the sign of the cosine of the top quark scattering angle, which is odd under Bose--Einstein symmetry, to define a ``forward'' direction for each event.

The top quark spin cannot be measured directly, but the angular distribution of the decay products of a top quark is correlated with its spin axis~\cite{MahlonParke2010}:
\begin{linenomath}
\begin{align}
\frac{1}{\Gamma}\frac{\rd \Gamma}{\rd \cos\chi_a} = \frac{1}{2} \left( 1 + \kappa_a \cos\chi_a \right),
 \label{eq:topdecay}
\end{align}
\end{linenomath}
where $\Gamma$ is the top quark decay width, $\chi_a$ is the angle between the direction of decay product $a$ and the top quark spin axis in the top quark rest frame, and $\kappa_a$ is the spin analyzing power.
The charged lepton has maximal spin analyzing power, $\kappa_{\ell^+} \approx 1$~\cite{Brandenburg2002235}.
For top antiquark decay, the sign is reversed: $\kappa_{\ell^-}=-\kappa_{\ell^+}$.

Each of the 15 coefficient functions from Eq.~(\ref{eq:prodspindensityexpanded}) (six $b_i^\pm$ and nine $c_{ij/i'}$) is probed by a normalized differential cross section at the parton level,
using the charged lepton directions measured in the rest frames of their parent top quark and antiquark as proxies for the top quark and antiquark spins.
Since the measurements are made in $\Pp\Pp$ collisions, the basis is adjusted from that of Eq.~(\ref{eq:basis}) by defining
$\hat{p}=(0,0,1)$, the direction of the proton beam in the positive $z$ direction in the laboratory frame, in the derivation of $\hat{r}$ and $\hat{n}$~\cite{Bernreuther:2015yna}.
This basis is illustrated in Fig.~\ref{fig:coordinate}.

The four-fold angular distribution for the two leptons 
follows from Eqs.~(\ref{eq:ME}) and~(\ref{eq:prodspindensity}), and is given by
\begin{linenomath}
\begin{align}
\frac{1}{\sigma}\frac{\rd^4 \sigma}{\rd \Omega_1 \, \rd \Omega_2} = \frac{1}{(4\pi)^2} \left( 1 + \mathbf{B_1} \cdot \hat{\ell}_1 + \mathbf{B_2} \cdot \hat{\ell}_2 - \hat{\ell}_1 \cdot C \cdot \hat{\ell}_2  \right),
\label{eq:fulldist}
\end{align}
\end{linenomath}
where $\sigma$ is the \ttbar\ production cross section, $\Omega_{1,2}$ are the solid angles of the leptons in their parent top quark and antiquark rest frames, and $\hat{\ell}_{1,2}$ are the corresponding unit vectors. 
The negative sign in front of the matrix $C$ is chosen to define same-helicity top quarks as having positive spin correlation.
The elements of the vectors $\mathbf{B_{1,2}}$ and the matrix $C$ are the following coefficients [in whose definitions the factors of $\kappa_{\ell^+}$ and $\kappa_{\ell^-}$ from Eq.~(\ref{eq:topdecay}) are absorbed]:
\begin{itemize}
\item  $B_{1}^{i}$ and $B_{2}^{i}$, the top quark and antiquark polarization coefficients with respect to each reference axis $i$  (sensitive to $b^+_i$ and $b^-_i$).

\item  $C_{ii}$, the ``diagonal'' spin correlation coefficient for each reference axis $i$ (sensitive to $c_{ii}$).

\item  $C_{ij}$, the ``cross'' spin correlation coefficients for each pair of axes $i \ne j$, whose sums and differences $C_{ij} \pm C_{ji}$ are sensitive to $c_{ij}$ and $c_{i'}$.
\end{itemize}

These measurable coefficients are closely related to the production spin density matrix coefficient functions from Eq.~(\ref{eq:prodspindensityexpanded}), but are not identical, owing to the different basis used for the spin measurement.
We do not measure the coefficients differentially or attempt to separate the contributions from different initial states.
The association between the measured coefficients and the coefficient functions is given in Table~\ref{tab:obscoef}.

For each of the 15 coefficients that make up $\mathbf{B_{1,2}}$ and $C$, 
a change of variables can be made to obtain a single-differential cross section that depends only on that coefficient.
After integrating out the azimuthal angles, Eq.~(\ref{eq:fulldist}) reduces to
\begin{linenomath}
\ifthenelse{\boolean{cms@external}}
{
\begin{multline}
\frac{1}{\sigma}\frac{\rd^2 \sigma}{\rd \cos\theta^i_1 \, \rd \cos\theta^j_2} =  \frac{1}{4} \Big( 1 + B_{1}^{i} \cos\theta^i_1 + B_{2}^{j} \cos\theta^j_2 \\
- C_{ij} \cos\theta^i_1\cos\theta^j_2  \Big),
\end{multline}
}
{
\begin{align}
\frac{1}{\sigma}\frac{\rd^2 \sigma}{\rd \cos\theta^i_1 \, \rd \cos\theta^j_2} =  \frac{1}{4} \left( 1 + B_{1}^{i} \cos\theta^i_1 + B_{2}^{j} \cos\theta^j_2 - C_{ij} \cos\theta^i_1\cos\theta^j_2  \right),
\end{align}
}
\end{linenomath}
where $\theta^i_1$ ($\theta^j_2$) is the angle of the positively (negatively) charged lepton, measured with respect to axis $i$ ($j$) in the rest frame of its parent top quark (antiquark).
By changing variables (if necessary) and integrating out one of the angles, we can derive single-differential cross sections with respect to $\cos\theta_1^i$, $\cos\theta_2^i$, and $\cos\theta_1^i\cos\theta_2^j$:
\begin{linenomath}
\begin{align}
\label{eq:theodist1}
\frac{1}{\sigma}\frac{\rd \sigma}{\rd \cos\theta^i_1} &= \frac{1}{2} \left( 1 + B_{1}^{i} \cos\theta^i_1  \right), \nonumber \\
\frac{1}{\sigma}\frac{\rd \sigma}{\rd \cos\theta^i_2} &= \frac{1}{2} \left( 1 + B_{2}^{i} \cos\theta^i_2  \right), \\
\ifthenelse{\boolean{cms@external}}
{
\frac{1}{\sigma}\frac{\rd \sigma}{\rd x} &= \frac{1}{2} \left( 1 - C_{ij} \, x \right) \ln \left(\frac{1}{\abs{ x } }\right), \nonumber \\
x&=\cos\theta^i_1\cos\theta^j_2. \nonumber
}
{
\frac{1}{\sigma}\frac{\rd \sigma}{\rd \cos\theta^i_1\cos\theta^j_2} &= \frac{1}{2} \left( 1 - C_{ij} \cos\theta^i_1\cos\theta^j_2 \right) \ln \left(\frac{1}{\abs{ \cos\theta^i_1\cos\theta^j_2 } }\right). \nonumber
}
\end{align}
\end{linenomath}
Similarly, starting from Eq.~(\ref{eq:fulldist}), we obtain (for $i \ne j$)
\begin{linenomath}
\begin{equation}
\begin{aligned}
\label{eq:theodist2}
\frac{1}{\sigma}\frac{\rd \sigma}{\rd x_\pm} &= \frac{1}{2} \left( 1 - \frac{C_{ij} \pm C_{ji}}{2} \, x_\pm \right)  \cos ^{-1}\abs{ x_\pm }, \\
x_\pm&=\cos\theta_1^i\cos\theta_2^j\pm \cos\theta_1^j\cos\theta_2^i.
\end{aligned}
\end{equation}
\end{linenomath}
Thus, in order to determine the 15 coefficients, the \ttbar\ production cross section is measured as a function of each of the following 15 observables at the parton level:
\begin{itemize}

\item  The three $\cos\theta_1^i$ terms and three $\cos\theta_2^i$ terms to measure $B_{1}^{i}$ and $B_{2}^{i}$,  the top quark and antiquark polarization coefficients with respect to each reference axis $i$.

\item  The three $\cos\theta_1^i\cos\theta_2^i$ terms to measure $C_{ii}$, the diagonal spin correlation coefficient for each axis $i$.

\item  The six sum and difference terms $\cos\theta_1^i\cos\theta_2^j\pm \cos\theta_1^j\cos\theta_2^i$ to measure $C_{ij} \pm C_{ji}$, the sums and differences of the cross spin correlation coefficients for each pair of axes $i \ne j$.
\end{itemize}
We do not measure the separate cross spin correlation $\cos\theta_1^i\cos\theta_2^j$ distributions, because it is the sums and differences that are sensitive to the $c_{ij}$ and $c_{i}$ coefficients of Eq.~(\ref{eq:prodspindensityexpanded}) (see Table~\ref{tab:obscoef}).

In addition, we measure four further $\cos\theta_{1,2}^i$ distributions based on modified axes $\hat{k}^*$ and $\hat{r}^*$, equal to $\pm \hat{k}$ or $\pm \hat{r}$, depending on the sign of $\abs{y_\cPqt}-\abs{y_{\cPaqt}}$, the difference of the moduli of the top quark and antiquark rapidities in the laboratory frame.
The use of the modified axes probes the coefficient functions in different areas of phase space, providing sensitivity to different combinations of four-quark operators~\cite{Bernreuther:2015yna}.

The spin correlation coefficient $D$ is related to the diagonal $C$ coefficients as $D=-\mathrm{Tr}[C]/3 = -(\ckk + \crr + \cnn)/3$.
We make a direct measurement of the $D$ coefficient using the distribution of the dot product of the two lepton directions measured in their parent top quark and antiquark rest frames~\cite{Bernreuther:2015yna}, $\hat{\ell}_1 \cdot \hat{\ell}_2=\cosphi$:
\begin{linenomath}
\begin{align}
\label{eq:theodist3}
\frac{1}{\sigma}\frac{\rd\sigma}{\rd\cosphi} = \frac{1}{2}(1-D\cosphi).
\end{align}
\end{linenomath}
We also measure two related laboratory-frame distributions, using the following observables:
\begin{itemize}
\item $\cosphilab=\hat{\ell}^{\mathrm{lab}}_1 \cdot \hat{\ell}^{\mathrm{lab}}_2$, defined by analogy to $\cosphi$, but using the lepton directions measured in the laboratory frame, which have excellent experimental resolution.
\item $\dphi$, the absolute value of the difference in azimuthal angle $\phi$ between the two leptons in the laboratory frame.
\end{itemize}
The association between the 22 measured observables and the coefficients and coefficient functions is given in Table~\ref{tab:obscoef}.
Except for the laboratory-frame distributions, the distribution shapes are completely determined by the coefficients, following the functional forms of Eqs.~(\ref{eq:theodist1})--(\ref{eq:theodist3}).
The laboratory-frame observables (given in the last two rows of Table~\ref{tab:obscoef}) do not directly relate to any of the coefficients, and we instead quantify the shapes of their distributions by calculating the asymmetry ($A$) in the number of events ($N$) about the center of the distribution:
\begin{linenomath}
\begin{equation}
\begin{aligned}
\label{eq:asym}
\Aphilab &= \frac{N (  \cosphilab > 0 )-N ( \cosphilab < 0 )}{N ( \cosphilab > 0 ) +N ( \cosphilab < 0 )}, \\
\Adphi &= \frac{N (  \dphi > \pi/2 )-N ( \dphi < \pi/2 )}{N ( \dphi > \pi/2 ) +N ( \dphi < \pi/2 )}.
\end{aligned}
\end{equation}
\end{linenomath}

\section{The CMS detector}
\label{sec:cms}
The central feature of the CMS apparatus is a superconducting solenoid of 6\unit{m} internal diameter, providing a magnetic field of 3.8\unit{T}. Within the solenoid volume are a silicon pixel and strip tracker, a lead tungstate crystal electromagnetic calorimeter (ECAL), and a brass and scintillator hadron calorimeter, each composed of a barrel and two endcap sections. Forward calorimeters extend the pseudorapidity ($\eta$) coverage provided by the barrel and endcap detectors. Muons are measured in gas-ionization detectors embedded in the steel flux-return yoke outside the solenoid. Events of interest are selected using a two-tiered trigger system~\cite{Khachatryan:2016bia}. The first level, composed of custom hardware processors, uses information from the calorimeters and muon detectors to select events at a rate of around 100\unit{kHz} within a time interval of less than 4\mus. The second level, known as the high-level trigger (HLT), consists of a farm of processors running a version of the full event reconstruction software optimized for fast processing, and reduces the event rate to around 1\unit{kHz} before data storage.
    A more detailed description of the CMS detector, together with a definition of the coordinate system used and the relevant kinematic variables, can be found in Ref.~\cite{bib:Chatrchyan:2008zzk}.

\section{Event simulation}
\label{sec:simulation}
 
Simulated \ttbar\ events with a top quark mass of $\mt=172.5\GeV$
are produced
at next-to-leading order (NLO) in QCD at the matrix element (ME) level by the \Powheg\ v.\,2~\cite{Frixione:2007nw,bib:powheg0,bib:powheg,bib:powheg2} generator (\Powhegvtwo).
The $h_\mathrm{damp}$ parameter of \Powhegvtwo, which regulates the damping of real emissions in the NLO calculation when matching to the parton shower (PS), is set to 272.72\GeV~\cite{bib:CMS:2016kle}.
The PS and hadronization are performed by \PythiaOnly\ 8.219~\cite{Sjostrand:2007gs}  (referred to as \Pythia\ in the following) with the CUETP8M2T4 tune~\cite{bib:CMS:2016kle,bib:CUETP8tune,Skands:2014pea}.

  In order to assess the level of variation when using an alternative ME and matching procedure,
    an alternative \ttbar sample is generated using the \MGvATNLO~2.3.3~\cite{Alwall:2014hca} generator including up to two extra partons at the ME level with NLO precision. The decays of the top quarks are modeled using \MadSpin~\cite{bib:madspin}, 
and events are matched to \Pythia\ for PS and hadronization using the FxFx jet merging prescription~\cite{Frederix:2012ps}.
This sample is referred to as \MGaMCatNLO\ + \Pythia\ [FxFx].

Signal \ttbar\ events are defined as those with two charged leptons (\ee, \emu, or \mumu), originating from \PW~boson decays and not from $\tau$ lepton decays. All other \ttbar\ events are regarded as a background.
The largest background contributions originate from \ttbar events with leptonically decaying $\tau$ leptons, single top quarks produced in association with a \PW~boson ($\cPqt\PW$), and, in events with same-flavor leptons, \cPZ/$\gamma^{*}$ bosons produced with additional jets (\Zjets). 
Additional significant backgrounds include \PW~boson production with additional jets (\Wjets), diboson ($\PW\PW$, $\PW\cPZ$, and $\cPZ\cPZ$) events, and the production of a \ttbar\ pair in association with a \PW or a \cPZ~boson ($\ttbar+\PW/\cPZ$). Other sources of background are negligible in comparison to the uncertainties in the main backgrounds, and are not included in this analysis.

The \Wjets process is simulated at LO precision using \MGvATNLO\ with up to four additional partons at the ME level and matched to \Pythia\ using the MLM jet merging prescription~\cite{Alwall:2007fs}. The \Zjets process is simulated at NLO precision using \MGvATNLO\ with up to two additional partons at the ME level and matched to \Pythia\ using the FxFx prescription. The $\ttbar+\PW/\cPZ$ processes are simulated with \MGvATNLO\ with NLO precision at the ME level and matched to \Pythia. In the case of $\ttbar+\PW$, one extra parton is simulated at the ME level and the calculation is matched to \Pythia\ using the FxFx prescription. Single top quark production is simulated with \Powheg\ v.\,1~\cite{bib:powheg1,bib:powheg3} with the $h_\mathrm{damp}$ parameter set to 172.5\GeV and using the CUETP8M2T4 tune in \Pythia. Diboson events are simulated at LO with \Pythia.
The NNPDF3.0\_lo\_as\_0130 and NNPDF3.0\_nlo\_as\_0118~\cite{bib:NNPDF,Ball2015} parton distribution function (PDF) sets are used for the LO and NLO simulations, respectively.

The cross sections used to normalize the simulated predictions are calculated at the highest orders of perturbative QCD currently available: next-to-next-to-leading order (NNLO) for \Wjets and \Zjets~\cite{Li:2012wna}; approximate NNLO for single top quark production in the $\cPqt\PW$ channel~\cite{bib:twchan}; and NLO for diboson~\cite{bib:mcfm:diboson} and $\ttbar+\PW/\cPZ$~\cite{bib:Maltoni:2015ena}. The \ttbar\ simulation is normalized to a cross section of \xsectheo\ (where $\alpS$ is the strong coupling constant), calculated with the \textsc{Top++}2.0 program~\cite{Czakon:2011xx} at NNLO, including resummation of next-to-next-to-leading-logarithmic soft-gluon terms and assuming $\mt = 172.5\GeV$.

Additional $\Pp\Pp$ interactions within the same or nearby bunch crossings (pileup) are simulated for all samples,
using a pileup multiplicity distribution that reflects the distribution of reconstructed vertices 
in data.
The interactions of particles with the CMS detector are simulated using \GEANTfour (v.\,9.4)~\cite{bib:geant}.

\section{Event selection}
\label{sec:selection}

The event selection~\cite{Sirunyan:2018ucr} targets the dileptonic decay $\ttbar \to \cPqb \ell^{+} \cPgn \: \cPaqb \ell^{-} \cPagn$.
    To maximize the trigger efficiency, both single-lepton and dilepton trigger paths are used.
    For the single-electron (single-muon) trigger, a transverse momentum threshold of $\pt=27$ (24)\GeV is applied. The same-flavor dilepton triggers require either an electron pair with $\pt > 23$ (12)\GeV for the leading (trailing) electron, or a muon pair with $\pt > 17$ (8)\GeV for the leading (trailing) muon, where leading (trailing) refers to the electron or muon with the highest (second-highest) \pt in the event. The different-flavor dilepton triggers require either an electron with $\pt > 12\GeV$ and a muon with  $\pt > 23\GeV$, or an electron with $\pt > 23\GeV$ and a muon with $\pt > 8 \GeV$.

The events selected by the HLT are reconstructed offline using a particle-flow algorithm~\cite{bib:Sirunyan:2017ulk},
which aims at reconstructing each individual particle in an event using an optimized combination of information from the various elements of the CMS detector.
Electron candidates are reconstructed from a combination of the track momentum at the main interaction vertex and the corresponding clusters in the ECAL with a Gaussian sum filter algorithm~\cite{1748-0221-10-06-P06005}.
Electron candidates with ECAL clusters in the region between the barrel and endcap ($1.44<\abs{\eta_{\mathrm{cluster}}}<1.57$) have a reduced reconstruction efficiency and are excluded. A relative isolation criterion $I_{\text{rel}} < 0.0588$~(0.0571) is applied for electron candidates in the barrel (endcap).
The $I_{\text{rel}}$ is defined as the \pt sum of all neutral hadron, charged hadron, and photon candidates within a distance of 0.3 from the electron candidate in $\eta$--$\phi$~space, divided by the \pt of the electron candidate, with
a correction to suppress the residual effect of pileup.
Additional electron identification requirements are applied to reject misidentified electron candidates and candidates originating from photon conversions~\cite{bib:Sirunyan:2017ulk,1748-0221-10-06-P06005}. Muon candidates are reconstructed using the track information from the tracker and the muon system~\cite{Sirunyan:2018fpa}.
A relative isolation requirement of $I_{\text{rel}} < 0.15$ within a distance of 0.4 in $\eta$--$\phi$~space from the muon candidate is applied.
In addition, muon identification requirements are used to reject misidentified muon candidates and candidates originating from decay-in-flight processes~\cite{Sirunyan:2018fpa}.
Both electron and muon candidates are required to have $\pt > 25$~(20)\GeV for the leading (trailing) candidate and $\abs{\eta}<2.4$.

Jets are reconstructed by clustering the particle-flow candidates using the anti-\kt clustering algorithm with a distance parameter of 0.4~\cite{bib:antikt,bib:Cacciari:2011ma}. The jet momentum is determined as the vector sum of all particle momenta in the jet, and is found from simulation to be within 5--10\% of the true momentum over the whole \pt spectrum and detector acceptance. Pileup can contribute additional tracks and calorimetric energy depositions to the jet momentum. To mitigate this effect, tracks identified to originate from pileup vertices are discarded, and an offset correction is applied to correct for remaining contributions from neutral particles from pileup~\cite{bib:Sirunyan:2017ulk}. Jet energy corrections are derived from simulation to bring the measured response of jets to that of particle-level jets on average. In situ measurements of the momentum imbalance in dijet, \photonjet, \Zjet, and multijet events are used to account for any residual differences in the jet energy scale (JES) between data and simulation.
Additional selection criteria are applied to remove badly reconstructed jets~\cite{Chatrchyan:2009hy,bib:Sirunyan:2017ulk}.
Jets are selected if they have $\pt > 30\GeV$ and $\abs{\eta} < 2.4$. Jets are rejected if the distance in $\eta$--$\phi$~space between the jet and the closest lepton is less than 0.4. Jets originating from the hadronization of \cPqb~quarks (\cPqb~jets) are identified (\cPqb~tagged)
by combining information related to secondary decay vertices reconstructed within the jets and track-based lifetime information in an algorithm (CSVv2) that provides a \cPqb~jet identification efficiency of 79--87\% and a probability to
misidentify light- and charm-flavor jets as \cPqb~jets of approximately 10 and 40\%, respectively~\cite{Sirunyan:2017ezt}.

The missing transverse momentum vector \ptvecmiss is defined as the projection on the plane perpendicular to the beam axis of the negative vector sum of the momenta of all reconstructed particles in an event. Its magnitude is referred to as \ptmiss.

The selected events are required to have
exactly two isolated electrons or muons of opposite electric charge and at least two jets. At least one of the jets is required to be \cPqb~tagged. Events with a lepton-pair invariant mass $\mll <20\GeV$ are removed in order to suppress contributions from heavy-flavor resonance decays and low-mass Drell--Yan processes. In the \ee and \mumu channels, backgrounds from \Zjets processes are further suppressed by requiring $\ptmiss > 40\GeV$ and vetoing events with $76<\mll<106\GeV$. The remaining background yield from \Zjets events, which is large in the \ee and \mumu channels, is determined by applying a factor derived from simulation to the number of \Zjets events observed in data in a control region where \mll\ is close to the \cPZ~boson mass~\cite{bib:TOP-12-028_paper,bib:TOP-15-003_paper}. A correction to account for non-\Zjets backgrounds in the control region is derived from the \emu channel.
The simulated \Zjets yield is corrected by up to 5\% in each channel to match the determination from data.

The four-momenta of the top quark and antiquark in each event are estimated using a kinematic reconstruction algorithm~\cite{bib:TOP-12-028_paper,Sirunyan:2018ucr}. The algorithm considers all possible assignments of reconstructed jets and leptons to the \cPqb~quarks and leptons from top quark decay,
and solves for the unknown neutrino momenta using
the following assumptions and constraints: \ptmiss is assumed to originate solely from the two neutrinos; the invariant mass of each reconstructed \PW~boson ($m_{\PW}$) must equal $80.4\GeV$~\cite{PhysRevD.98.030001}; and the invariant mass of each reconstructed top quark must equal $172.5\GeV$.
Effects of detector resolution are accounted for by randomly smearing the measured energies and directions of the reconstructed jets and leptons according to their simulated resolutions. The assumed value of $m_{\PW}$ is varied according to a simulated Breit--Wigner distribution, with a width of $2.1\GeV$~\cite{PhysRevD.98.030001}. For a given application of the smearing, the solution of the equations for the neutrino momenta yielding the smallest reconstructed $m_{\ttbar}$ is chosen. For each solution, a weight is calculated based on the spectrum of the true invariant mass of the lepton and \cPqb~jet system from top quark decay at the particle level~\cite{Sirunyan:2018ucr}. The weights are summed over 100 applications of the smearing, and the top quark and antiquark four-momenta are calculated as a weighted average.
Considering only the combinations with the most \cPqb-tagged jets, the assignment of jets and leptons that yields the maximum sum of weights is chosen.
The efficiency of the kinematic reconstruction, defined as the fraction of the selected \ttbar events where a solution is found, is about 90\% in both data and simulation. Events with no real solution for the neutrino momenta are excluded from further analysis.
 
After applying the full event selection, 34\,890 events in the \ee channel, 70\,346 events in the \mumu channel, and 150\,410 events in the \emu channel are observed.
The difference in the \ee and \mumu channel yields is attributable to the lower efficiencies of the electron identification and isolation requirements. 
The differential cross section measurements are made using the combination of events from the three channels,
where
the fraction of signal events in the data sample, estimated from simulation, is 79\%.

\begin{figure*}[!htpb]
\centering
 \includegraphics[width=0.325\linewidth]{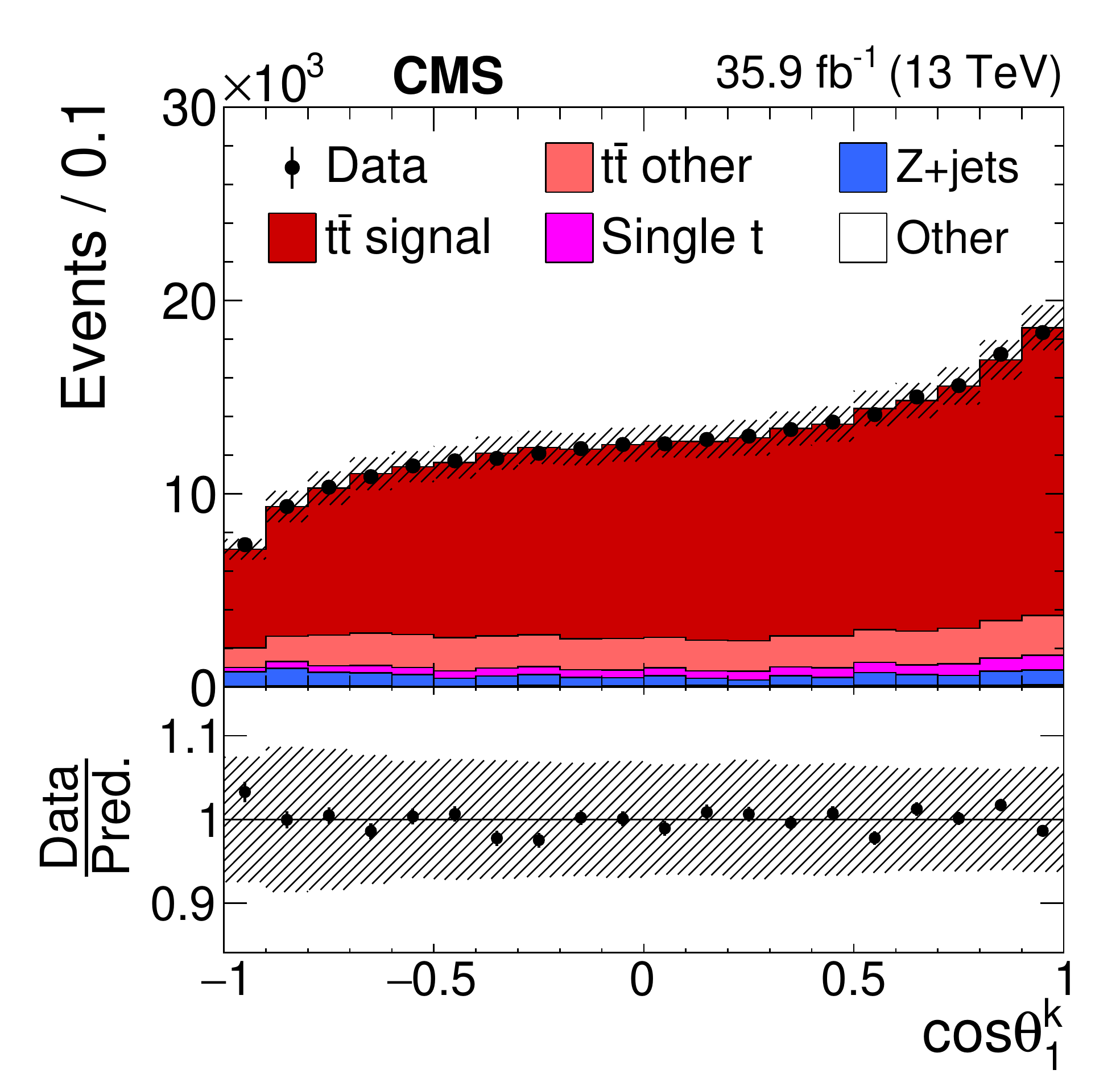}
	        \hfill
 \includegraphics[width=0.325\linewidth]{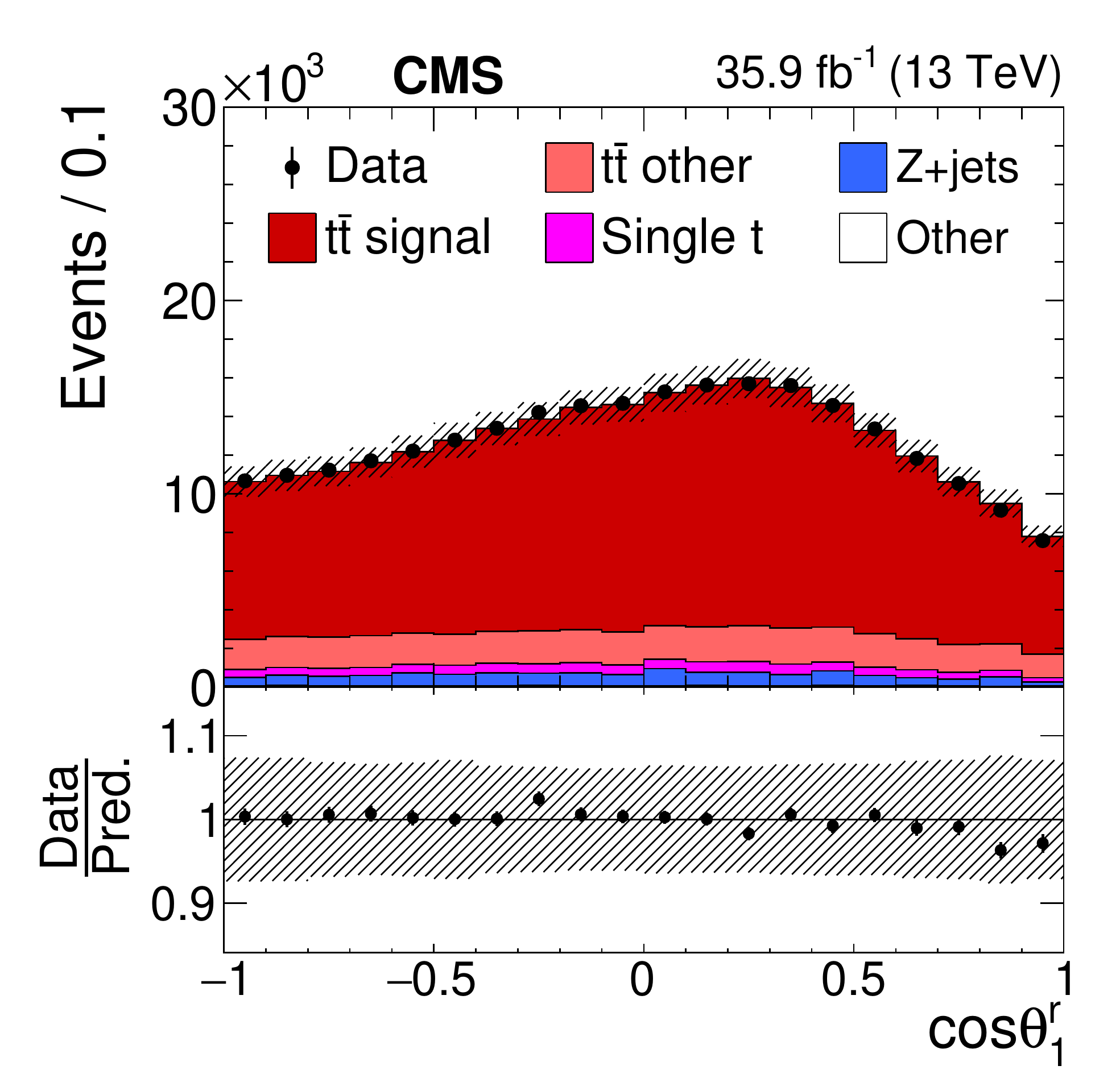} 
	        \hfill
\includegraphics[width=0.325\linewidth]{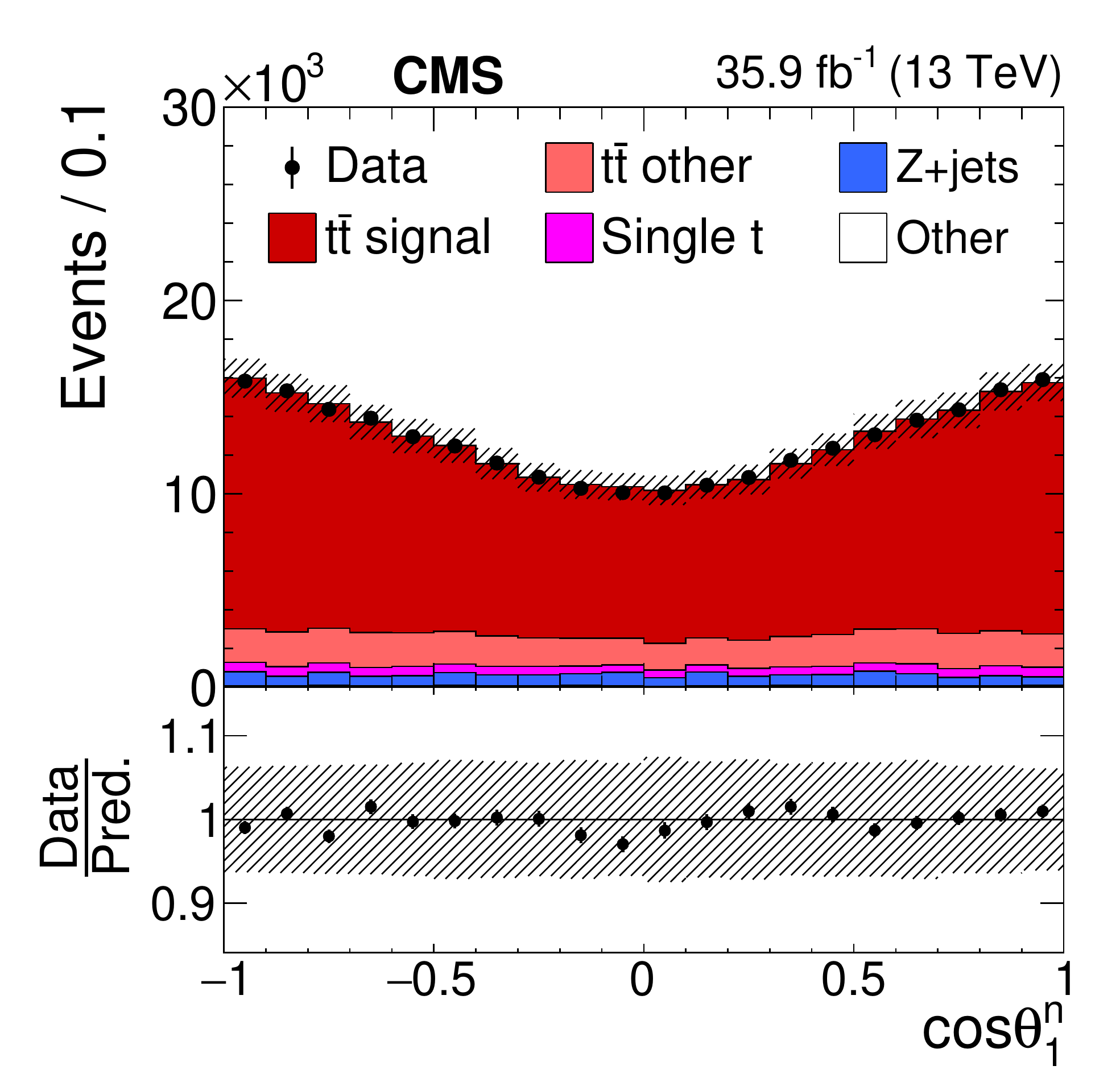} \\
 \includegraphics[width=0.325\linewidth]{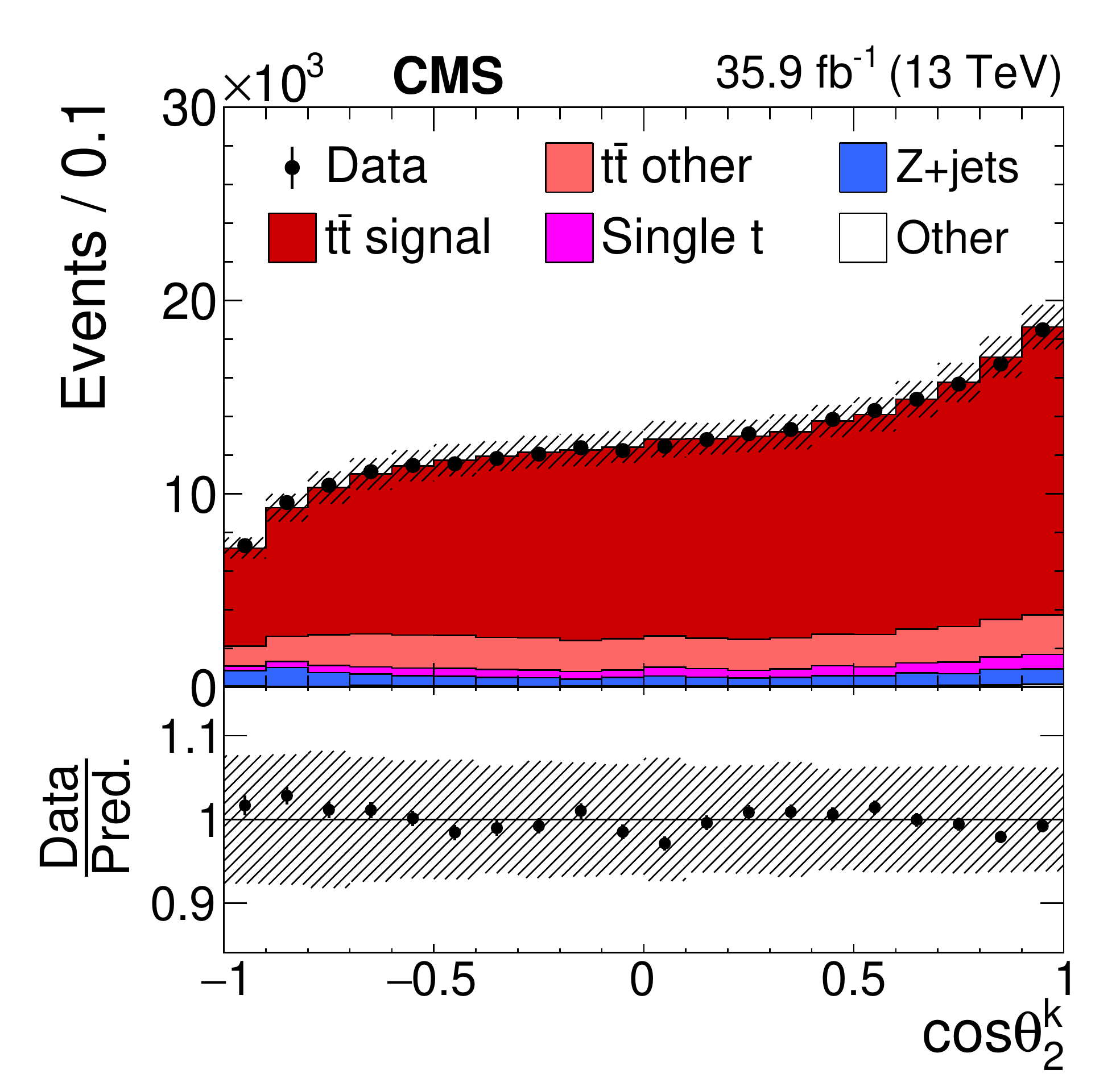}
	        \hfill
  \includegraphics[width=0.325\linewidth]{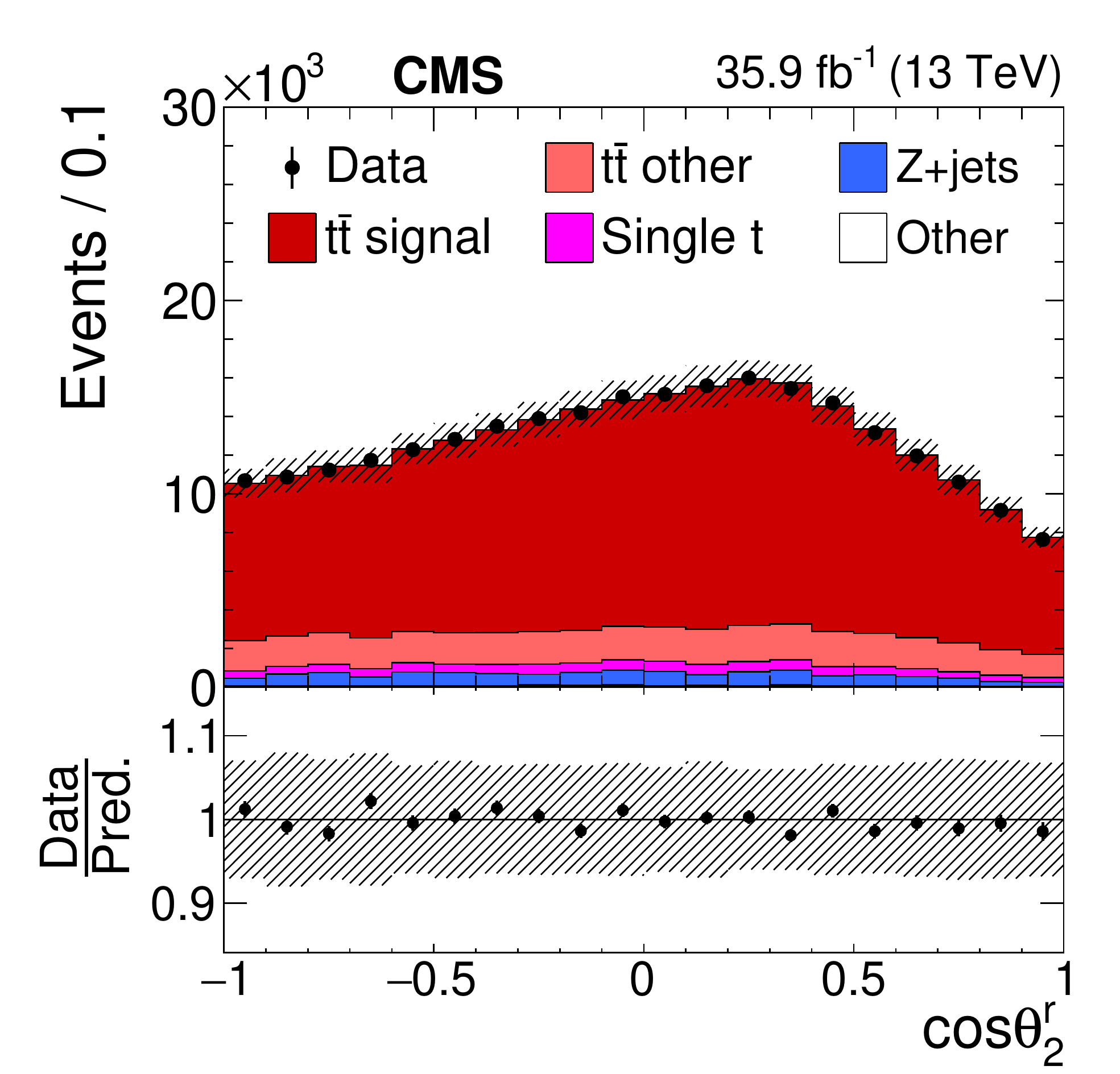}
	        \hfill
  \includegraphics[width=0.325\linewidth]{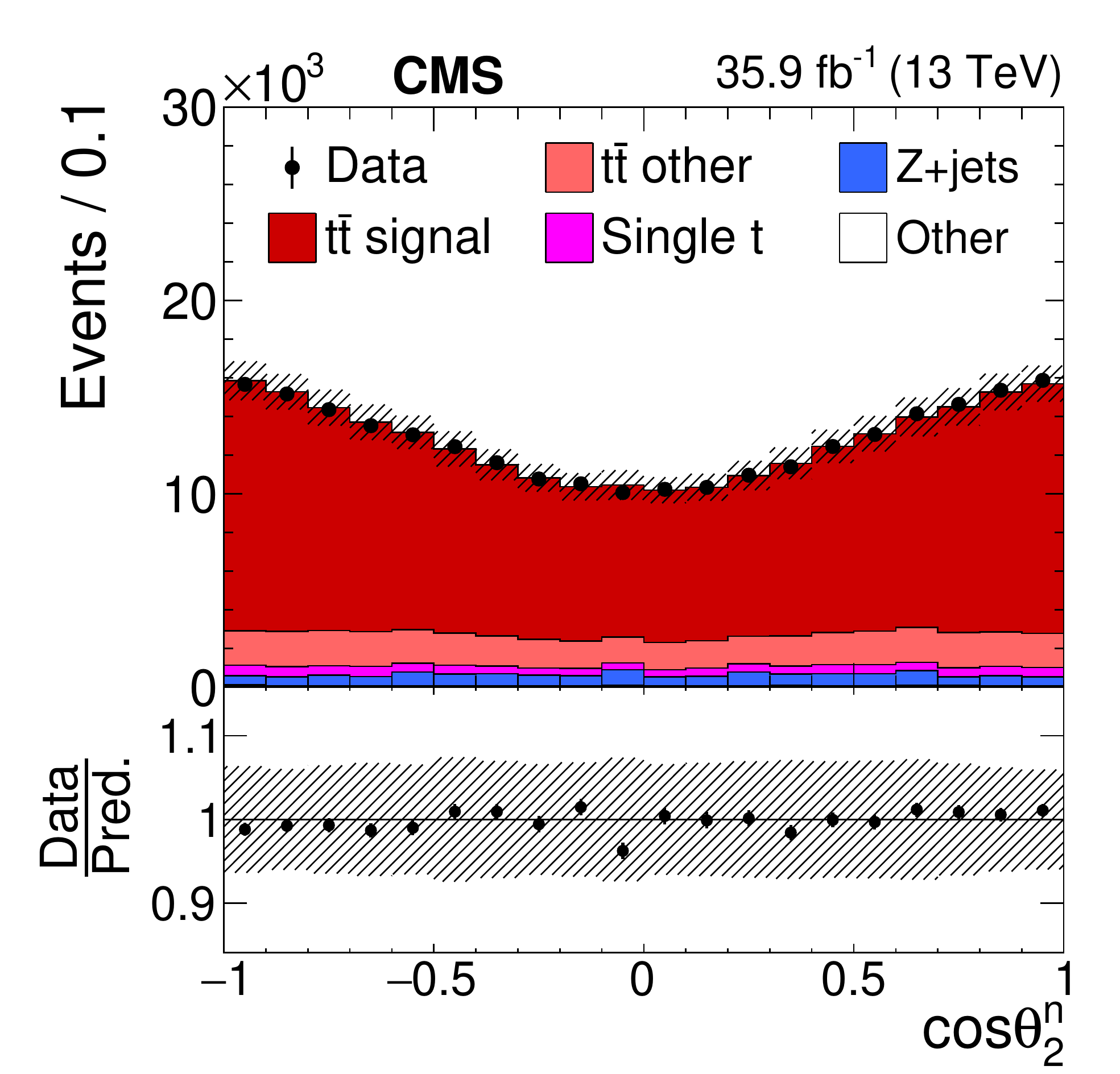} \\
 \includegraphics[width=0.325\linewidth]{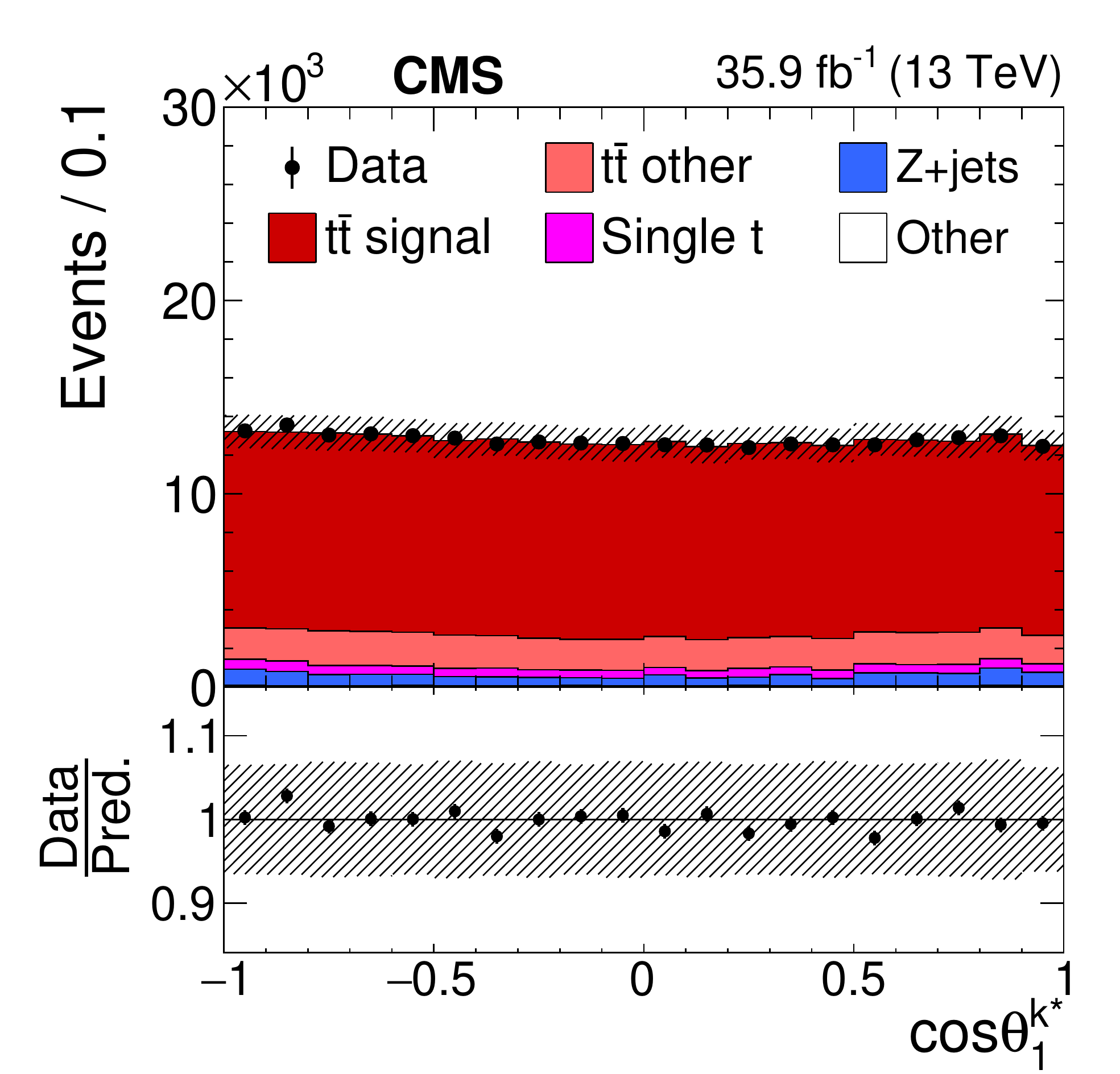}
			\hspace{0.005\linewidth}
 \includegraphics[width=0.325\linewidth]{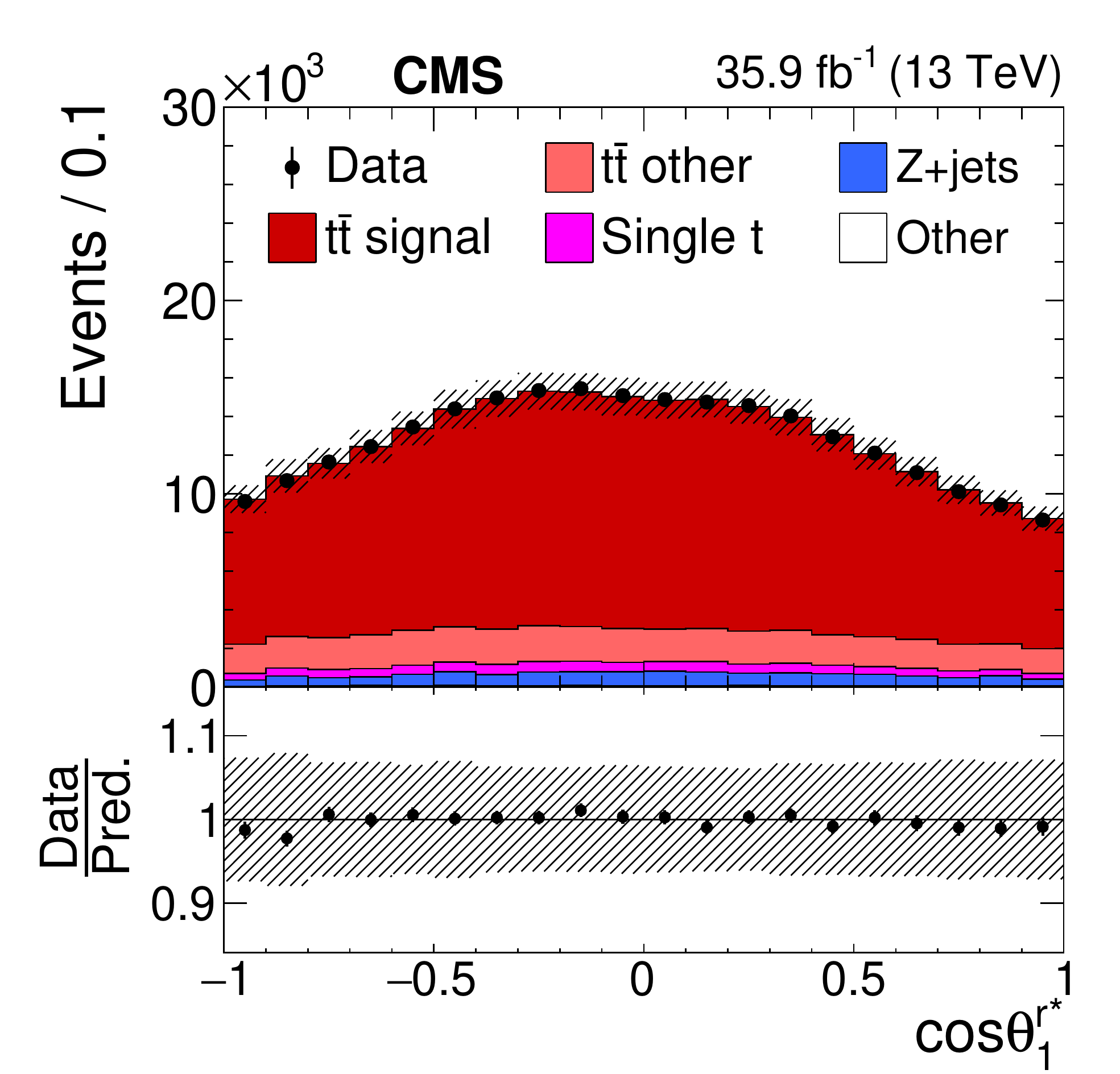} \\
 \includegraphics[width=0.325\linewidth]{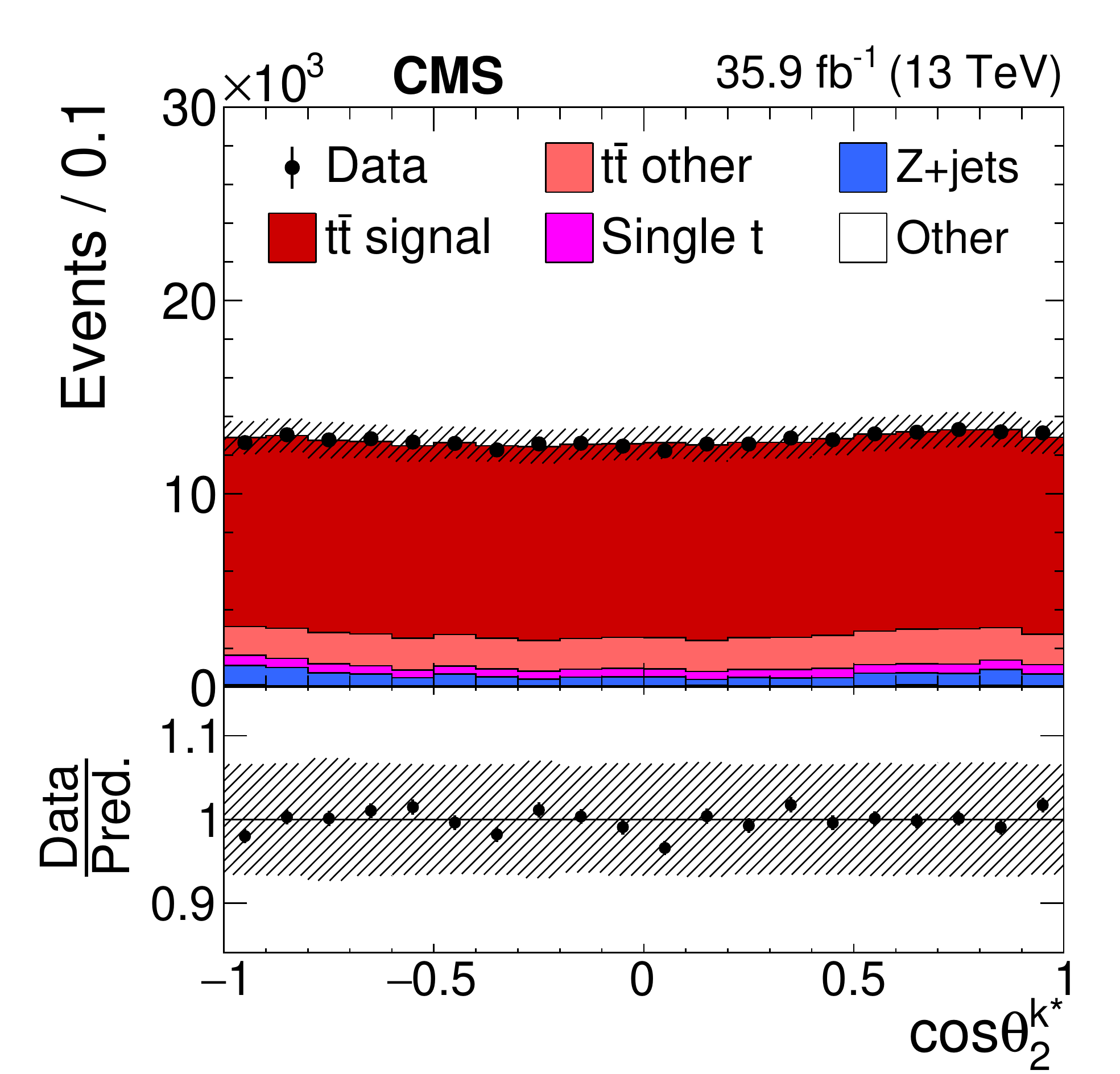}
			\hspace{0.005\linewidth}
  \includegraphics[width=0.325\linewidth]{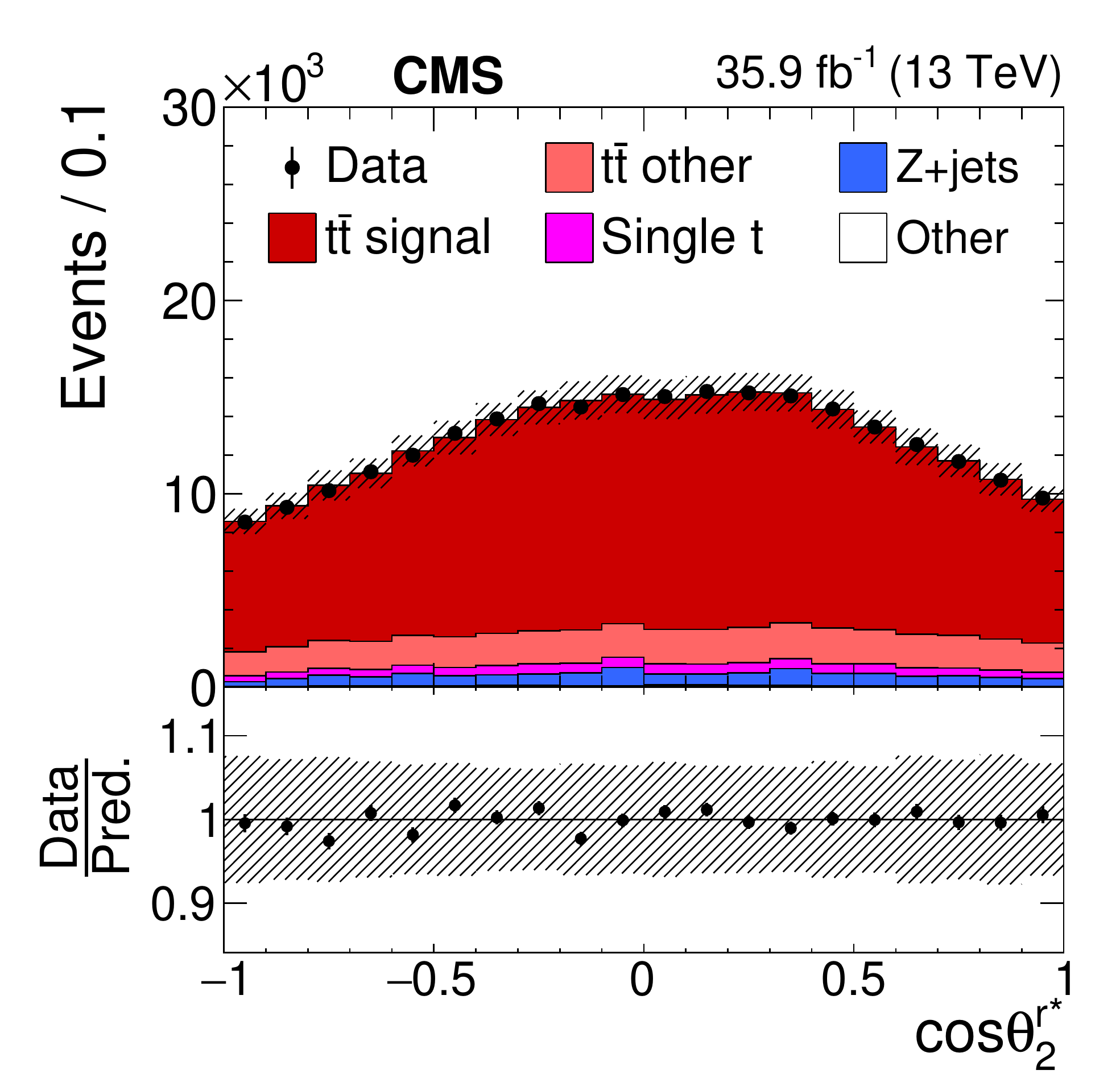} \\
\caption{\label{fig:B1B2Dist}\protect
Reconstructed distributions of $\cos\theta^i$ for top quarks (antiquarks) in the first and third (second and fourth) rows,
where $i$ refers to the reference axis with which the angle $\theta^i$ is measured.
From left to right, $i=\hat{k}$, $\hat{r}$, $\hat{n}$ (upper two rows), and $i=\hat{k}^*$,~$\hat{r}^*$ (lower two rows).
The data (points) are compared to the simulated predictions (histograms).
The vertical bars on the points represent statistical uncertainties, and the estimated systematic uncertainties in the simulated histograms are indicated by hatched bands.
The ratio of the data to the sum of the predicted signal and background is shown in the lower panels.
}
\end{figure*}

\begin{figure*}[!htpb]
\centering
 \includegraphics[width=0.325\linewidth]{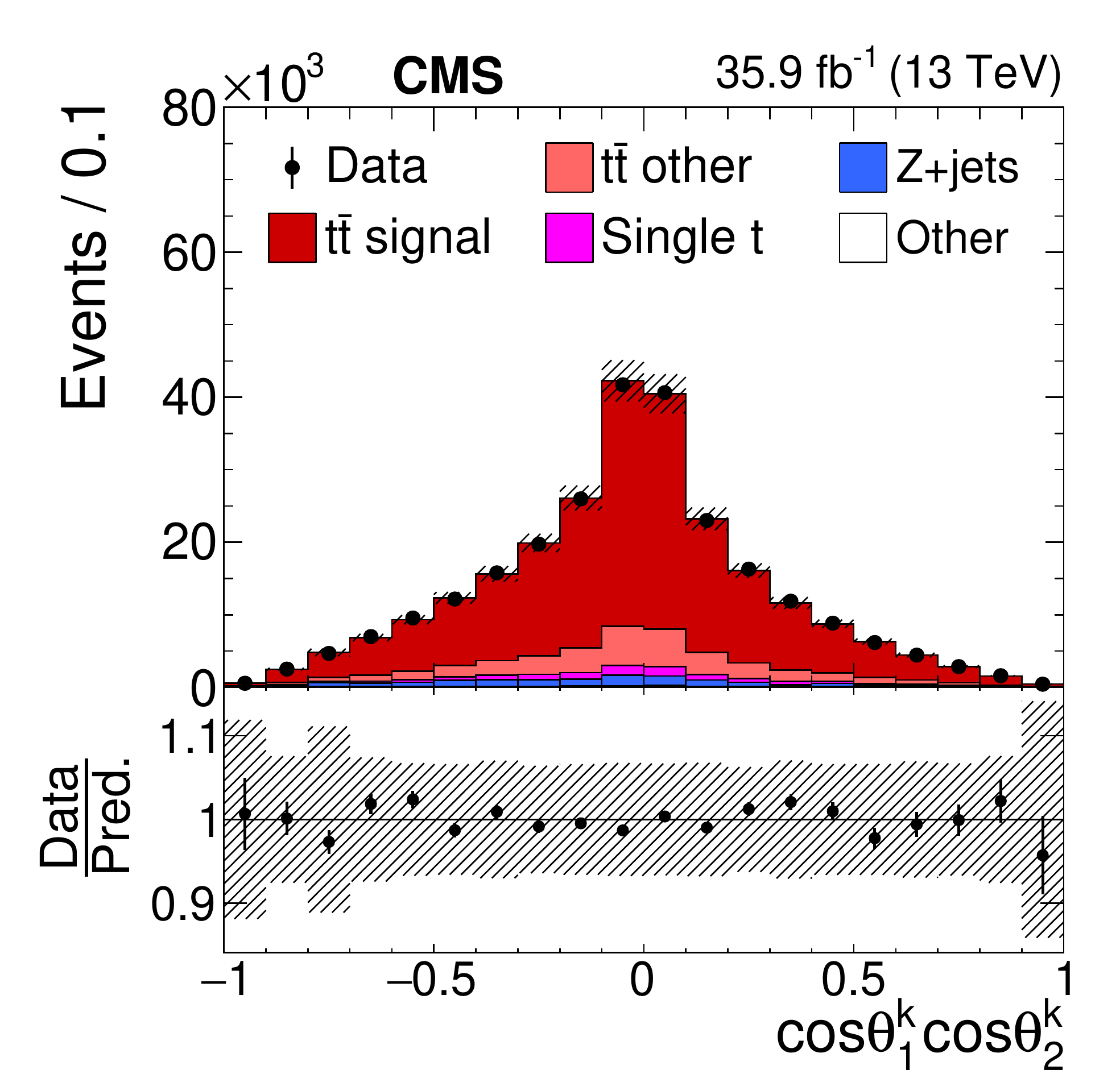}
 	        \hfill
 \includegraphics[width=0.325\linewidth]{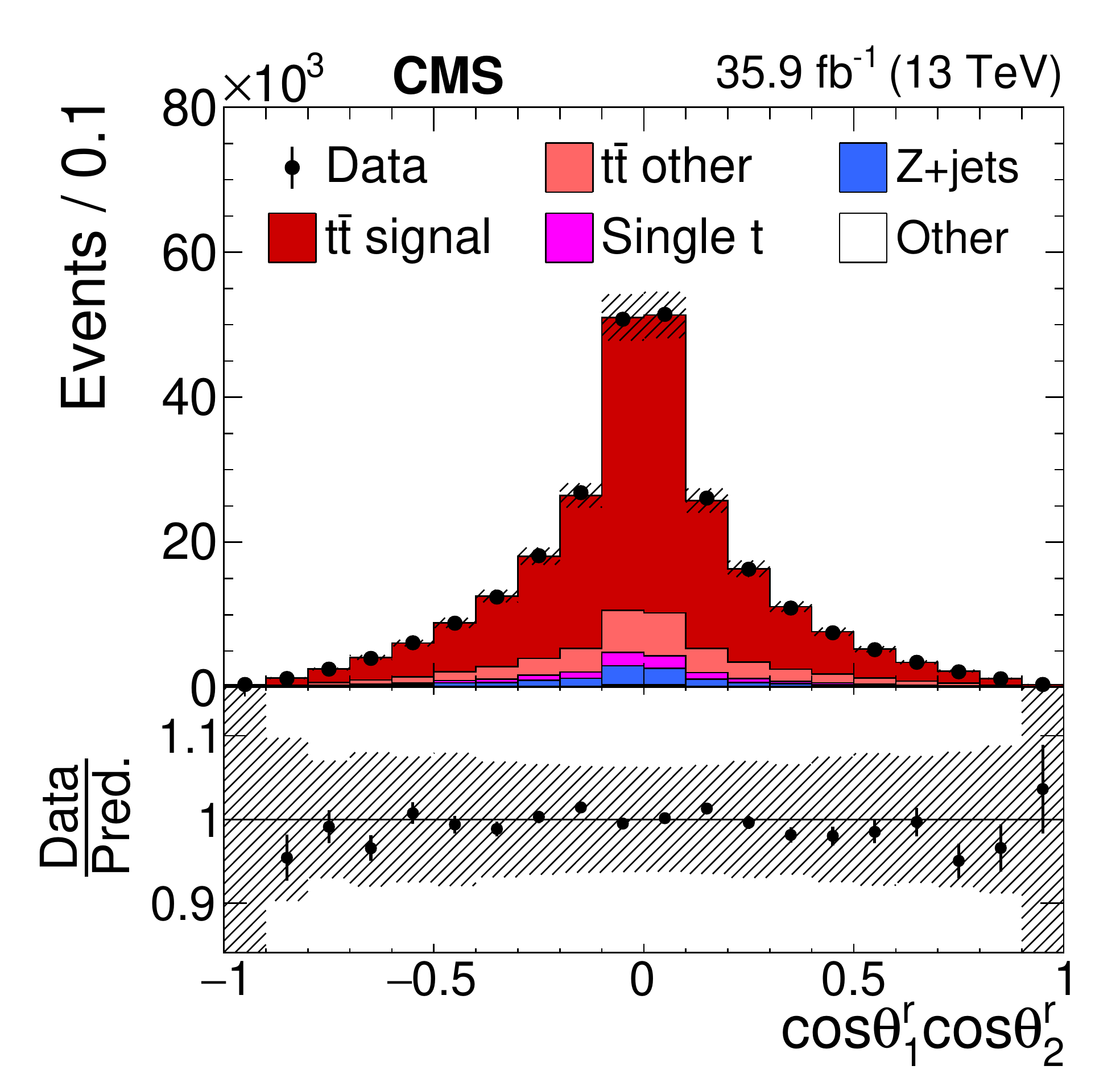}
 	        \hfill
 \includegraphics[width=0.325\linewidth]{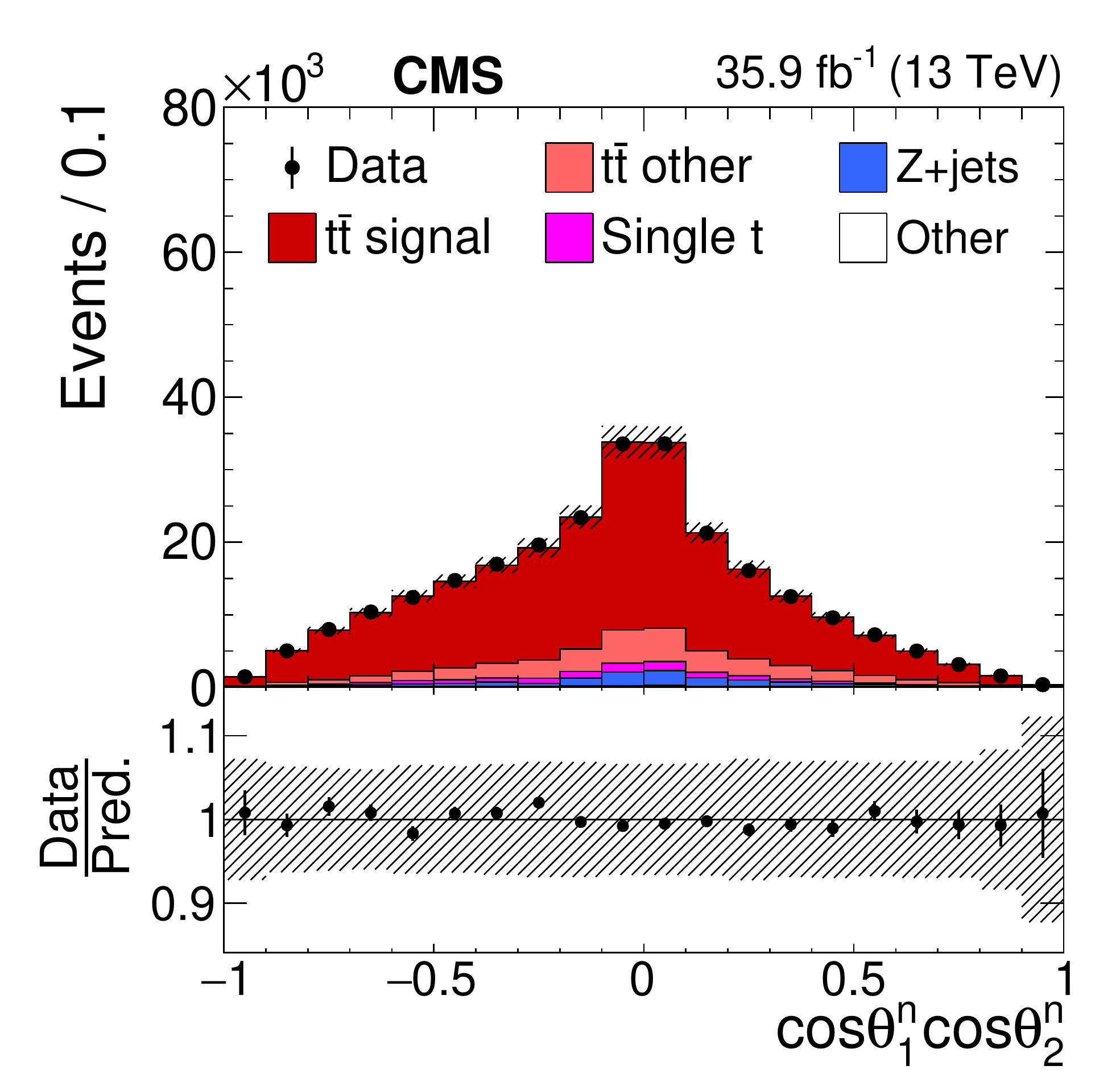} \\
 \includegraphics[width=0.325\linewidth]{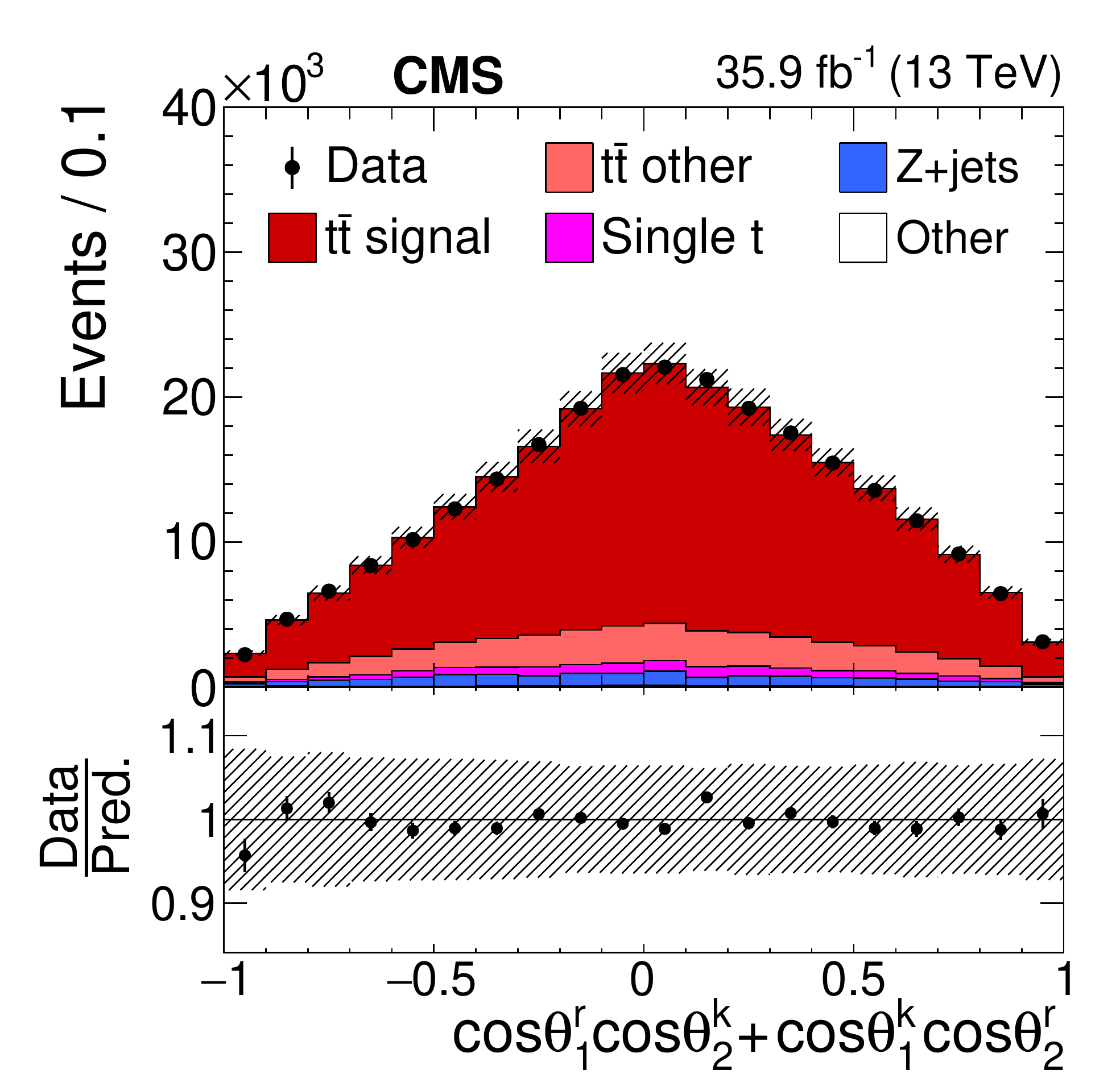}
 	        \hfill
 \includegraphics[width=0.325\linewidth]{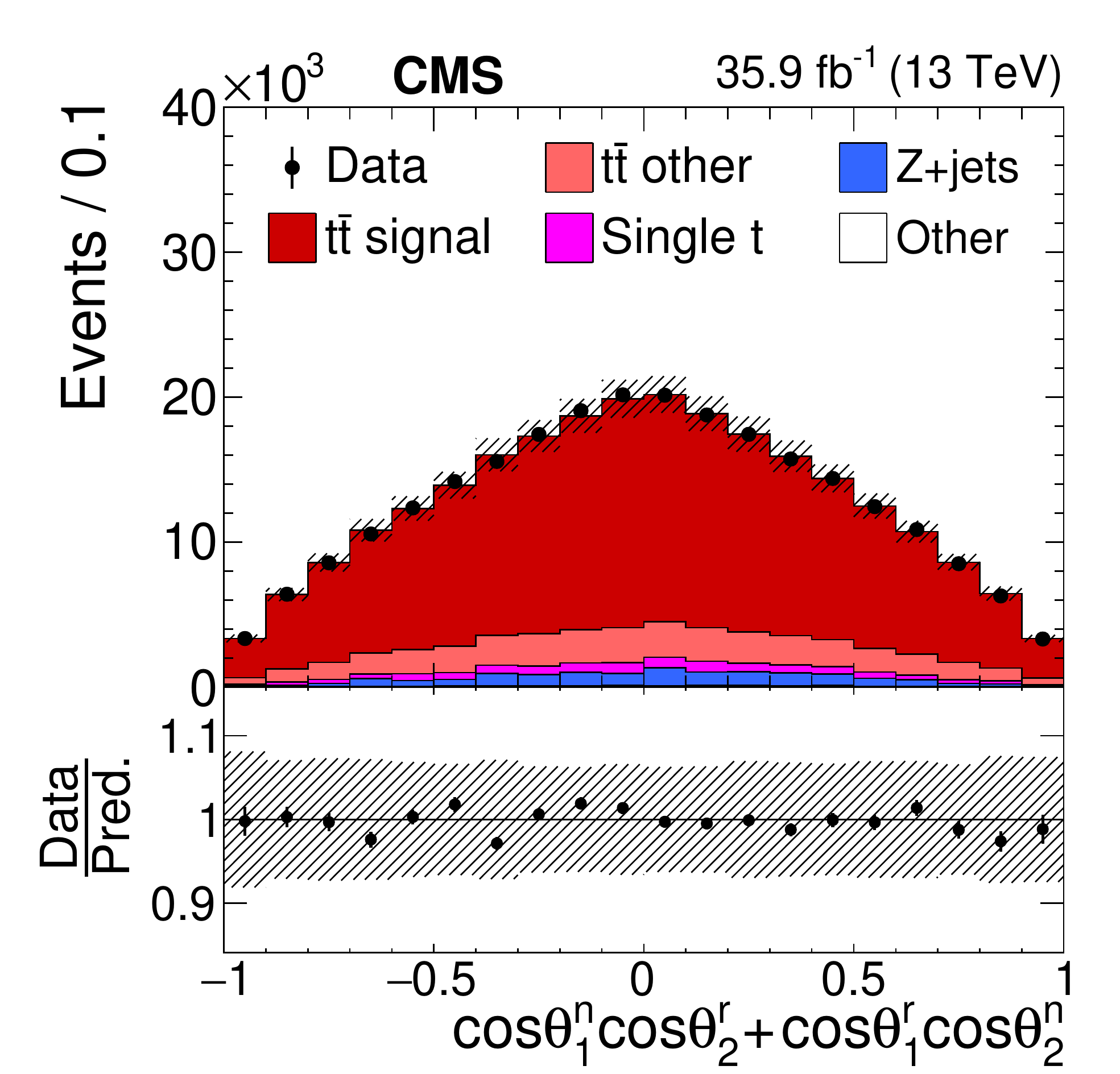}
 	        \hfill
 \includegraphics[width=0.325\linewidth]{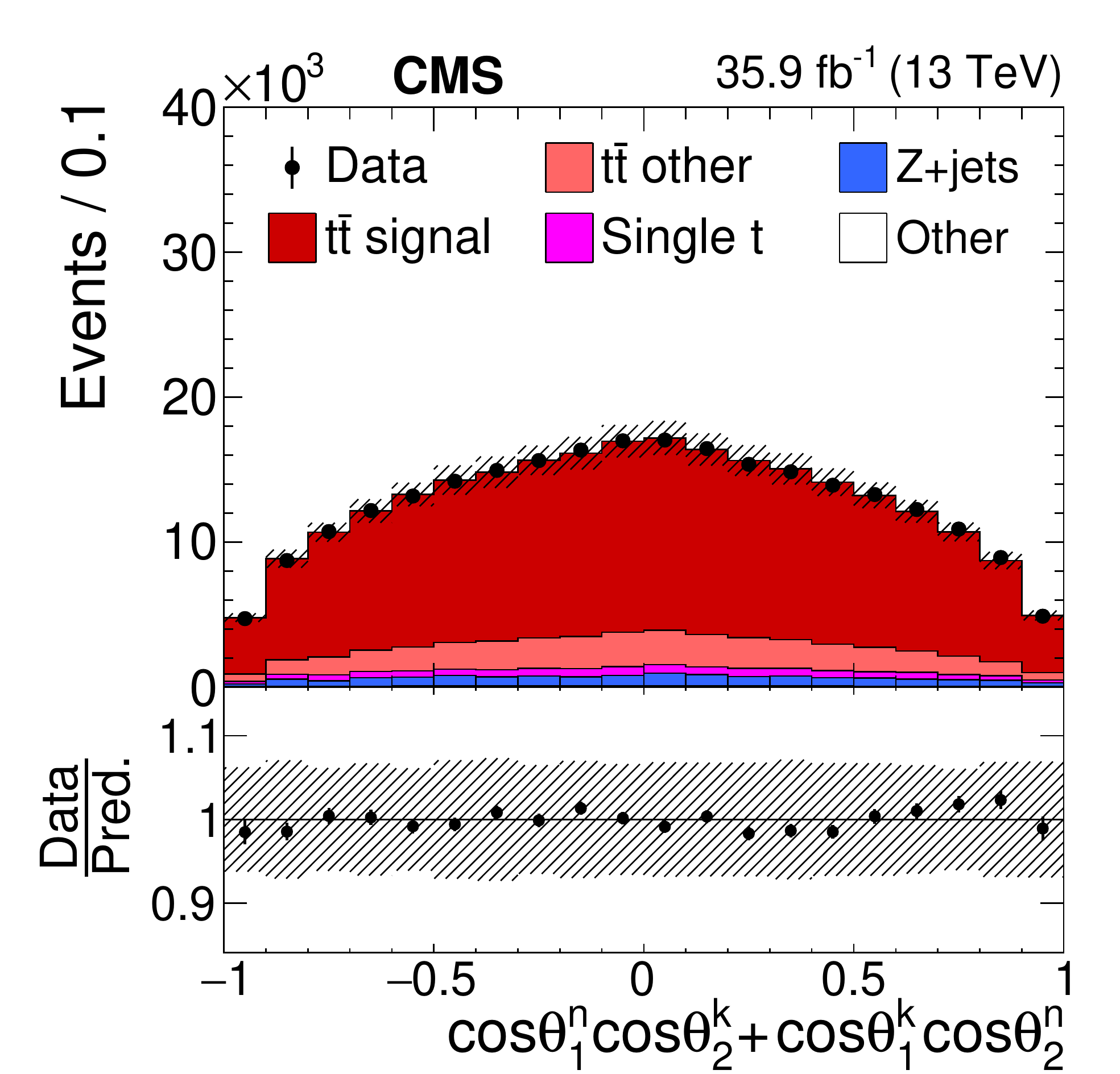} \\
 \includegraphics[width=0.325\linewidth]{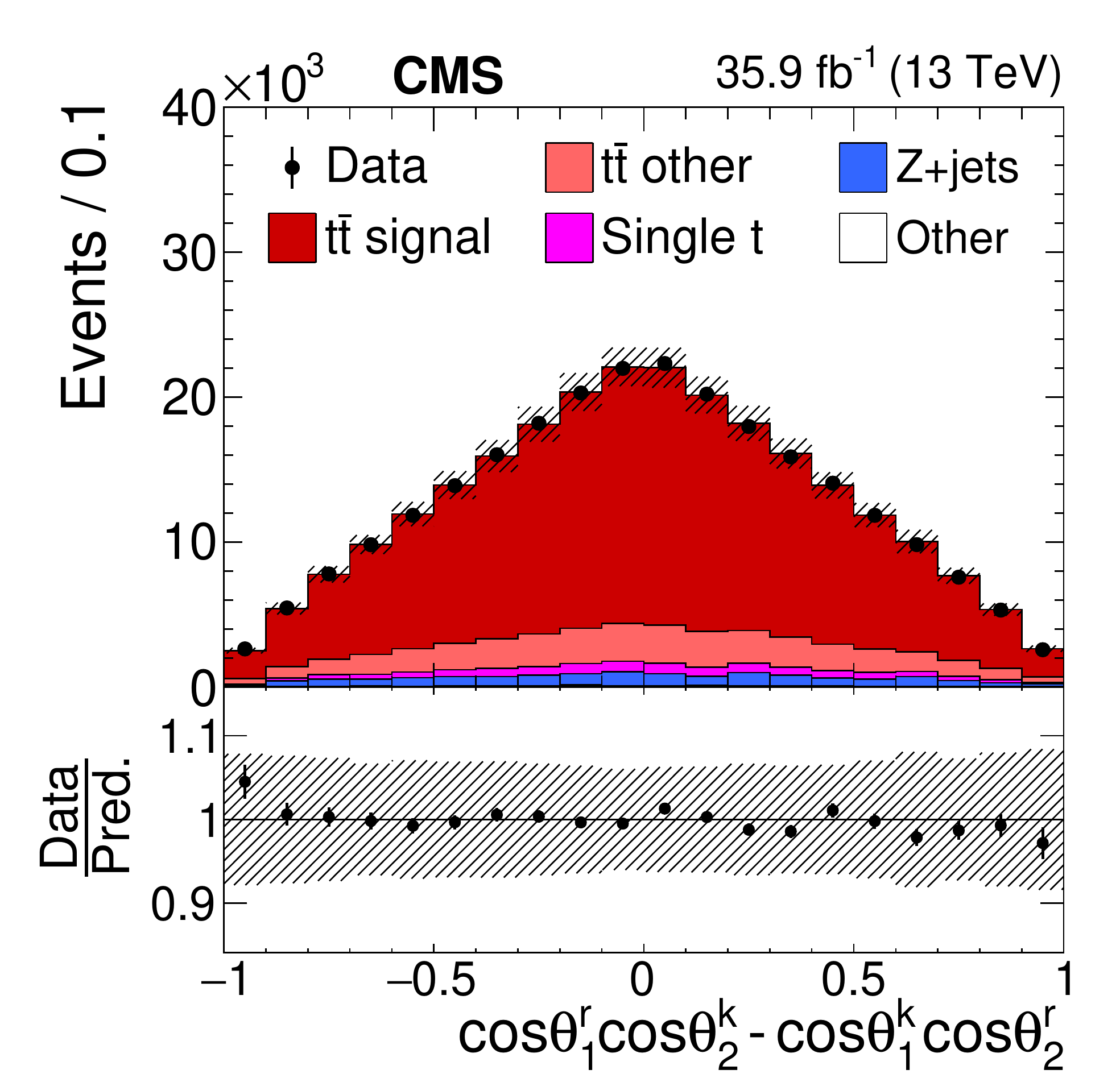}
 	        \hfill
 \includegraphics[width=0.325\linewidth]{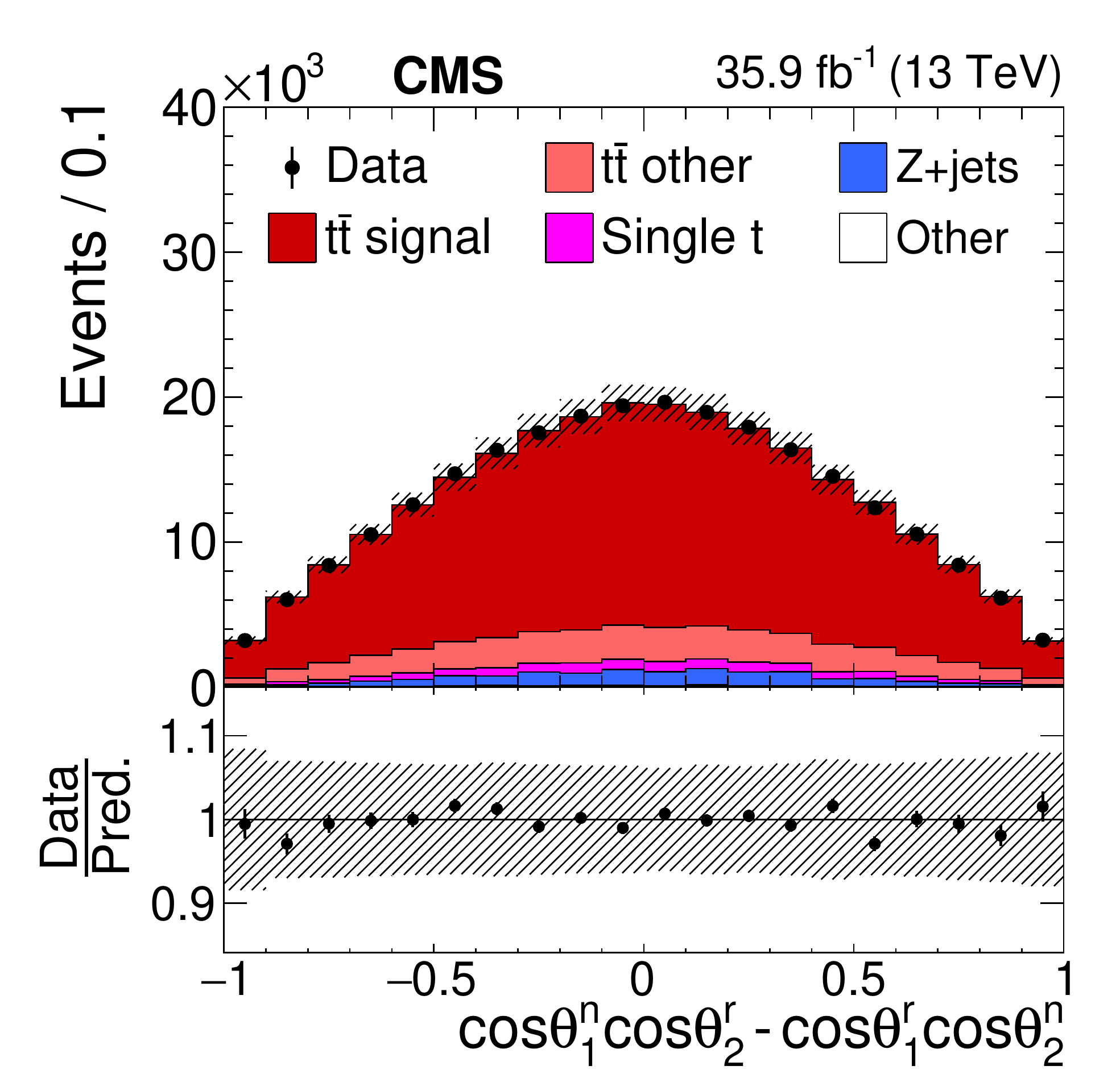}
 	        \hfill
 \includegraphics[width=0.325\linewidth]{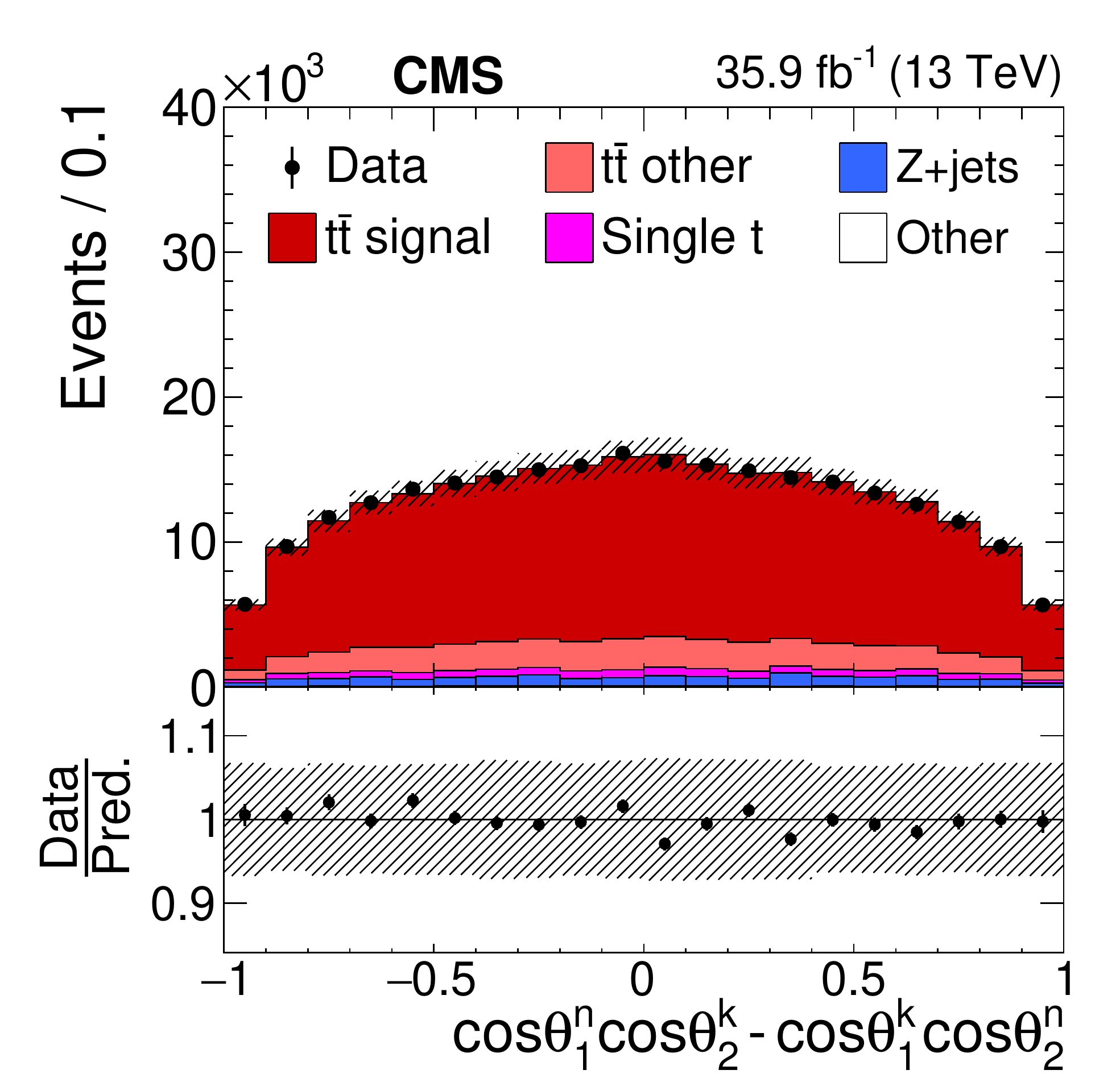} \\
 \includegraphics[width=0.325\linewidth]{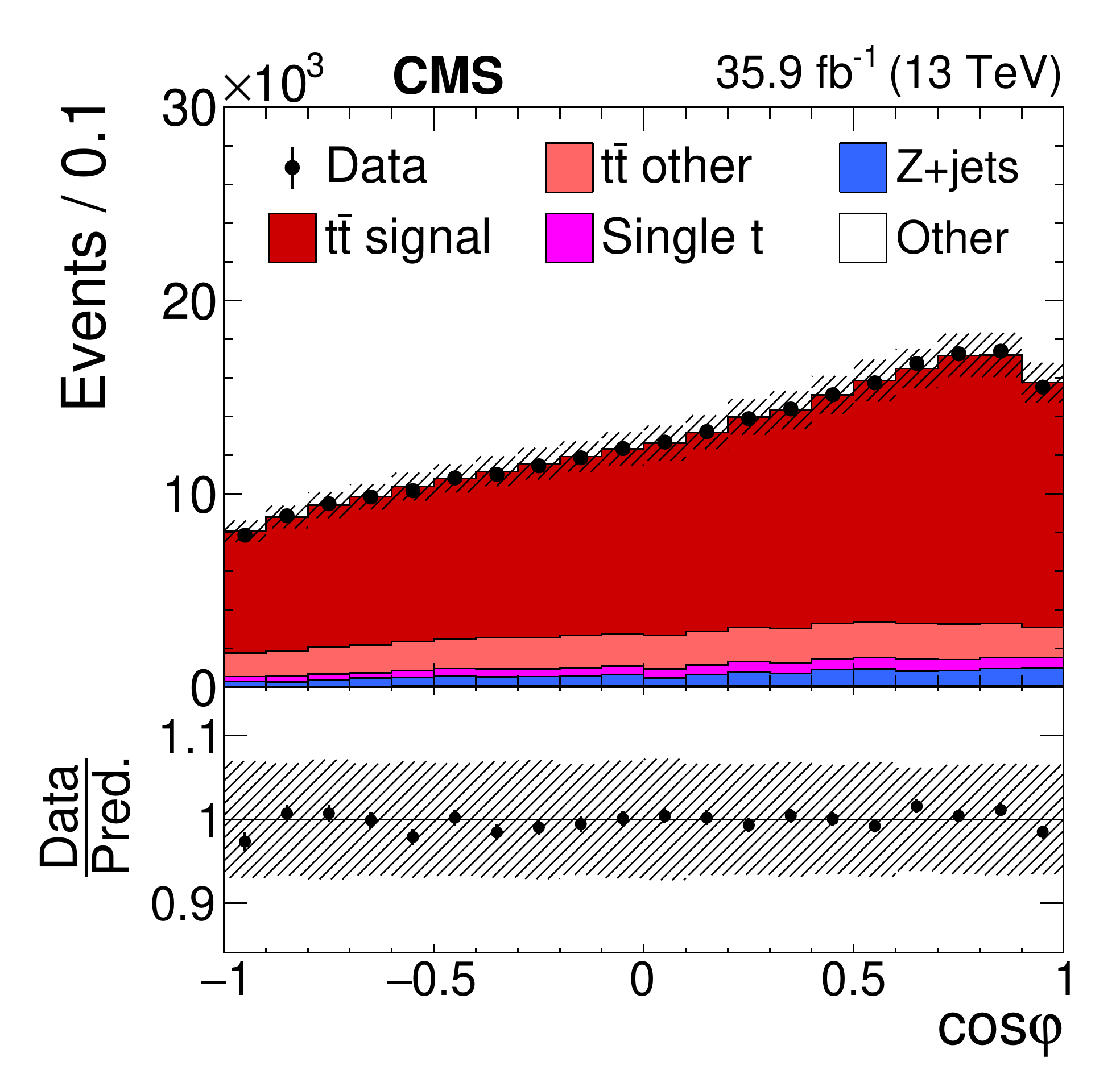}
 	        \hfill
 \includegraphics[width=0.325\linewidth]{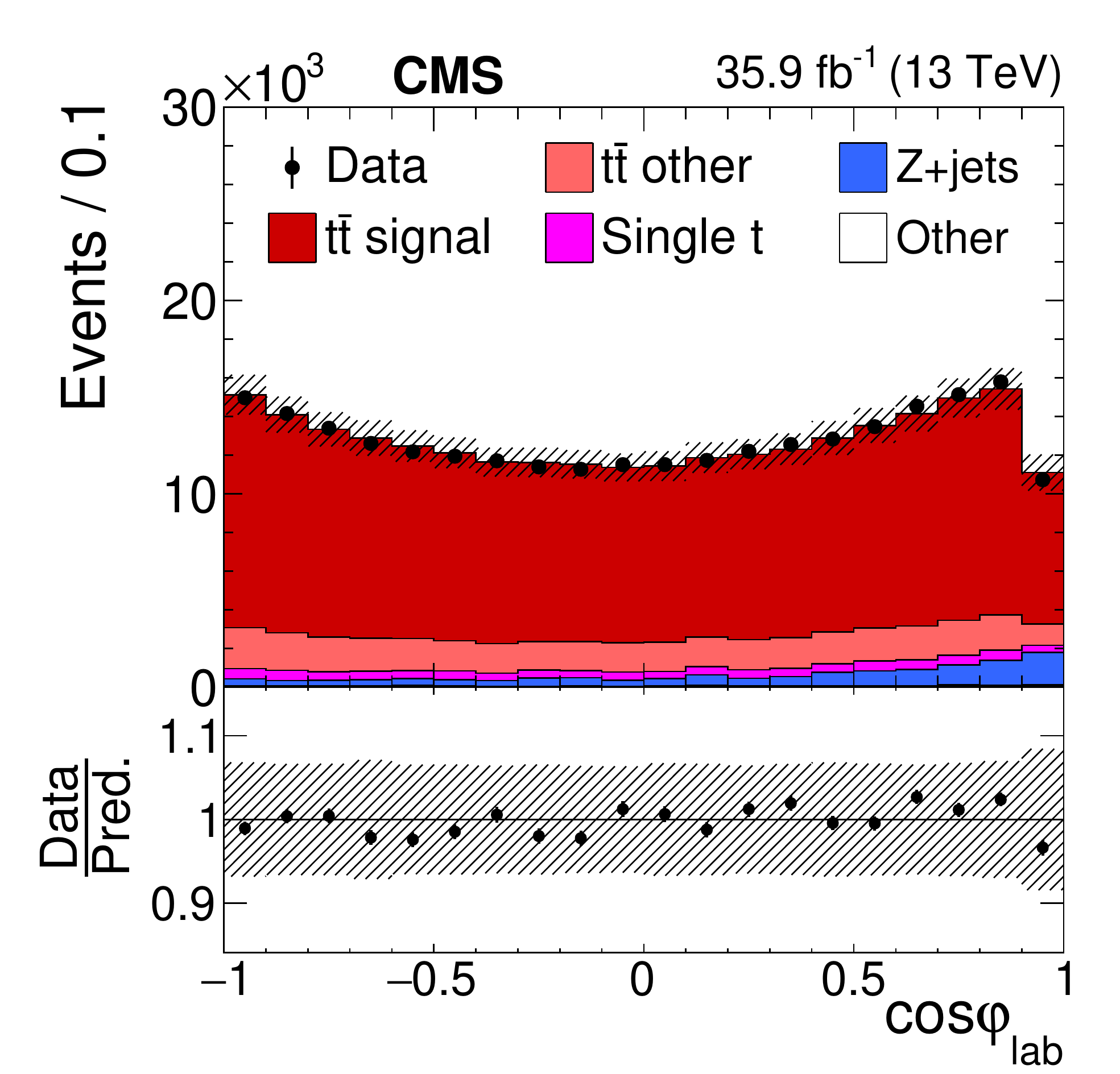}
 	        \hfill
 \includegraphics[width=0.325\linewidth]{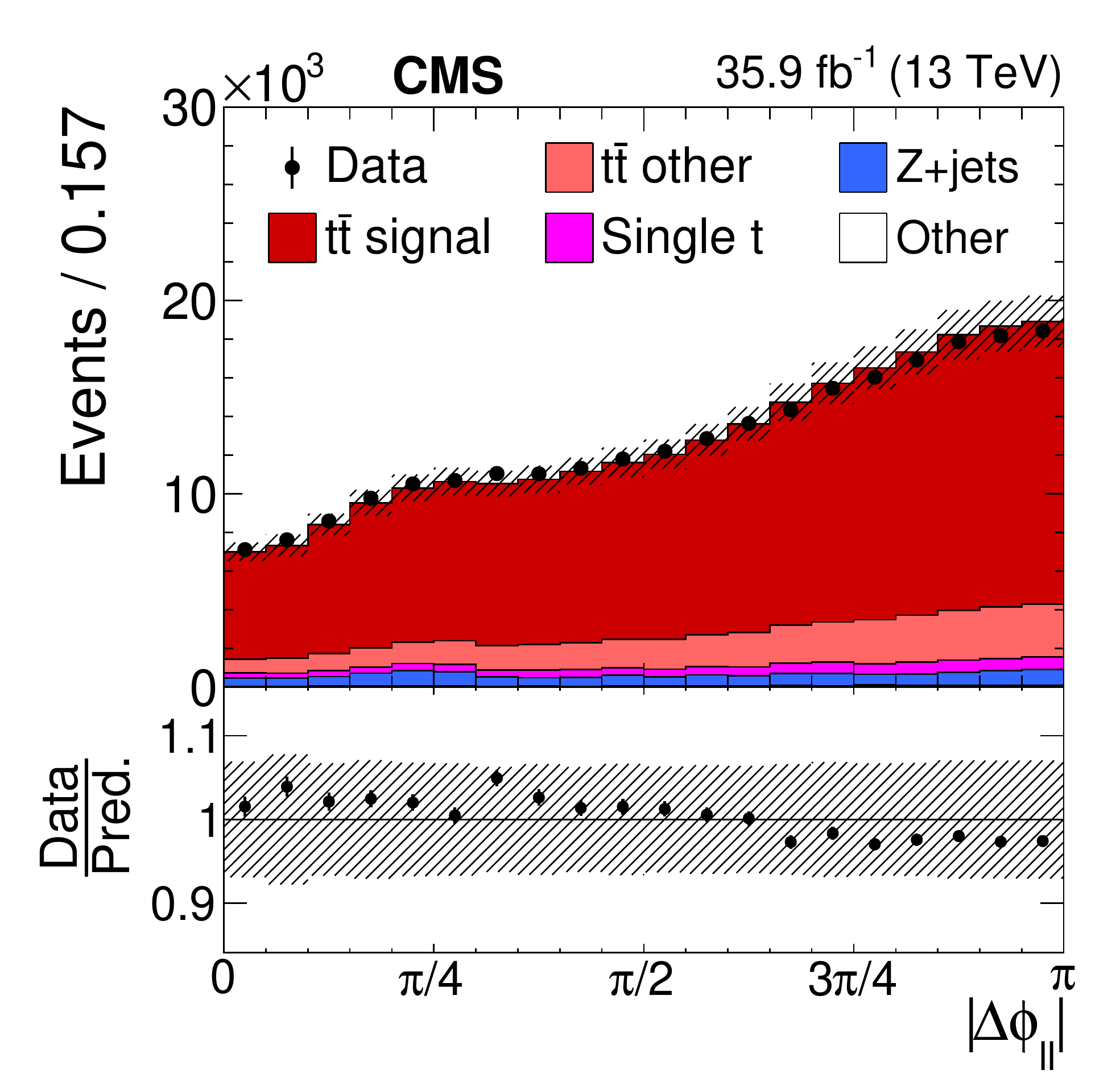} \\
\caption{\label{fig:CDist}\protect
Reconstructed angular distributions used in the measurement of the \ttbar spin correlation observables.
The data (points) are compared to the simulated predictions (histograms).
The vertical bars on the points represent statistical uncertainties, and the estimated systematic uncertainties in the simulated histograms are indicated by hatched bands.
The ratio of the data to the sum of the predicted signal and background is shown in the lower panels.
}
\end{figure*}

In Figs.~\ref{fig:B1B2Dist} and~\ref{fig:CDist}, distributions of all the reconstructed angular observables (defined in Section~\ref{sec:observables}) are shown.
There is reasonable agreement between the data and the sum of the expected signal and background contributions given the systematic uncertainties.
In addition to the systematic uncertainties discussed in Section~\ref{sec:errors}, two uncertainties that affect only the normalization of the measured differential cross section are considered:
the $2.5\%$ uncertainty in the integrated luminosity of the data sample~\cite{bib:CMS-PAS-LUM-17-001}
is applied to the normalization of all simulated predictions,
and a $1.5\%$ normalization uncertainty is applied to the \ttbar\ prediction to account for the uncertainty in the dileptonic branching fraction (BF)~\cite{PhysRevD.98.030001}.
The shapes of the reconstructed distributions differ substantially from the expected parton-level
functional forms of Eqs.~(\ref{eq:theodist1})--(\ref{eq:theodist3})
owing to the effects of bin migration, detector acceptance and efficiency, and background events, which are described in Section~\ref{sec:diffxsec}.

\section{Unfolding the differential cross sections}
\label{sec:diffxsec}

The effects of detector acceptance and efficiency sculpt the reconstructed distributions, and
the smearing introduced by the detector response, kinematic reconstruction algorithm, PS, and hadronization leads to the migration of events across bins.
In order to measure the differential cross sections at the parton level in the full phase space, 
these effects are
accounted for by using the TUnfold regularized unfolding method~\cite{bib:TUnfold}.
The response matrix used in the unfolding is calculated for each measured distribution using the default \ttbar\ simulation, where
the momenta of the parton-level top quarks are defined after QCD radiation has been simulated but before the top quark decays. 

\ifthenelse{\boolean{cms@external}}{}{\pagebreak}

To keep the bin-to-bin migrations small (to avoid strong bin-to-bin correlations in the unfolded distributions), the widths of the measurement bins are chosen according 
to the reconstruction resolution of the observable.
This is quantified both directly by comparing the generator level and detector level observables in simulation, and by measuring the purity and stability. Purity is defined as the fraction of events in a given bin at the detector level that originate from the same bin at the generator level, and stability is defined as the fraction of events in a given bin at the generator level that are reconstructed in the same bin at the detector level.
For all observables measured in the top quark rest frame, the use of six bins of uniform width is found to be well-matched to the reconstruction resolution. The purities and stabilities are typically $40\%$.
For the observables measured in the laboratory frame ($\cosphilab$ and $\dphi$), six uniform-width bins are also used. These observables have excellent experimental resolution, and the purities and stabilities are ${>}99\%$.

The presence of background events is accounted for prior to performing the unfolding.
After subtracting all other background components,
the background from dileptonic \ttbar\ events with leptonically decaying $\tau$ leptons  
is subtracted as a fraction of the total remaining events.
The fraction is evaluated per bin as the ratio of the background to the total dileptonic \ttbar\ events in simulation.
Thus, the shapes of the distributions for dileptonic \ttbar\ events are taken from data,
and any dependence on the total cross section used in the normalization of the simulated \ttbar\ sample is avoided.

In TUnfold, a procedure based on matrix inversion is used 
 to obtain an unfolded distribution from the measured distribution by applying a $\chi^2$ minimization technique.
 The potential large statistical fluctuations and strong anticorrelations between adjacent bins arising from the matrix inversion
are suppressed by introducing a term in the $\chi^2$ expression that smooths (regularizes) the shape of the unfolded distribution~\cite{bib:TUnfold}.
The regularization term penalizes the curvature of a vector constructed from the product of the difference between the unfolded and simulated bin values and
a factor calculated using the expected functional form [Eqs.~(\ref{eq:theodist1})--(\ref{eq:theodist3})] such that a deviation in the coefficient corresponds to a linear change in the vector.
Since linear changes are unconstrained by regularization of the curvature, and the functional forms at the parton level (which are unaffected by BSM phenomena in \ttbar\ production) depend only on the coefficient, this ensures that the regularization cannot introduce a bias in the unfolded distribution.
For the laboratory-frame distributions there are no such simple functional forms, and no factor is applied to the difference vector. However, this choice is of little consequence because the regularization is very weak owing to the low level of bin migration.

The use of wide bins for the response matrix loses 
information about its dependence inside each bin, meaning the unfolding can be biased if the physical process density differs from the simulation.
Since the curvature regularization is unbiased, we make use of narrower bins in 
the TUnfold $\chi^2$ minimization; a factor 4 narrower is found to be sufficient to reduce the bias from binning to a negligible level. We have thus replaced the biased implicit regularization from binning with an unbiased regularization of the curvature within each of the original bins.

The regularization level is determined for each distribution by minimizing the average global correlation coefficient ($\rho_{\mathrm{avg}}$)~\cite{bib:TUnfold},
where $\rho_{\mathrm{avg}}$ is determined after rebinning to the original six bins.

For each measured bin, we perform tests using pseudo-data to confirm a linear response of the method to variations in the coefficient, and confirm that the distribution of the difference between the nominal bin value and that measured in pseudo-data, normalized to the measured uncertainty, is consistent with having zero mean and unit width.

The data in the three channels are combined before unfolding in order to model correlations between channels in situ and reduce statistical uncertainties in poorly populated regions of the response matrix.
After unfolding, each distribution is normalized to unit area to measure the normalized differential cross section.

\section{Systematic uncertainties}
\label{sec:errors}

The systematic uncertainties arising from the detector performance and the modeling of the signal and background processes are evaluated from the
difference between the nominal measurement and that obtained by repeating the unfolding procedure using simulated events with the appropriate systematic variation.
Each source of systematic uncertainty is represented by a covariance matrix for the bins of the measured normalized differential cross sections.
The total systematic uncertainty is derived from the sum of these covariance matrices.
In this section, each of the applied variations is detailed and categorized into experimental and theoretical sources of uncertainty.

\subsection{Experimental sources of uncertainty}

Many of the experimental sources of uncertainty relate to the scale factors (SFs), defined as the ratio of the efficiencies in data and simulation, that are applied to the simulation
in order to accurately model the data.

The efficiencies of the triggers in data are measured as the fraction of events passing alternative triggers based on a \ptmiss requirement that also satisfy the criteria of the trigger of interest~\cite{TOP-17-001,Sirunyan:2018ucr}.
As the efficiency of the \ptmiss requirement is only weakly correlated with the dilepton trigger efficiencies, the bias introduced by the \ptmiss requirement is negligible. The efficiencies are close to unity in both data and simulation, as are the corresponding SFs.
To estimate the uncertainty from the modeling of the trigger efficiency, the SFs are varied within their uncertainties, both globally for all bins and depending on the $\eta$ of the leptons. The total trigger uncertainty is derived by taking the maximum deviation produced by the two variations in each  unfolded bin.

The SFs for the lepton identification and isolation efficiencies are determined with a tag-and-probe method using \Zjets event samples~\cite{bib:tp,bib:TOP-15-003_paper}. 
Measured in bins of $\eta$ and \pt, the SFs are generally within 10\% of unity for electrons, and consistent with unity for muons. The lepton identification and isolation uncertainty is estimated by varying the SFs within their uncertainties.
The efficiency of the kinematic reconstruction of the top quarks is found to be consistent between data and simulation within around 0.2\%. An associated uncertainty is derived by varying the corresponding SFs by ${\pm}0.2\%$. 

The uncertainty from the modeling of the number of pileup events is obtained by changing the inelastic $\Pp\Pp$ cross section assumed in simulation by ${\pm}4.6\%$, consistent with the cross section uncertainty presented in Ref.~\cite{Aaboud:2016mmw}.

The uncertainty arising from the imperfect modeling of the \cPqb~tagging efficiency is determined by varying the measured SFs 
within their uncertainties, both globally and depending on the \pt and $\eta$ of the \cPqb~jets.
The total uncertainty is derived by taking the maximum observed deviation in each unfolded bin. The \cPqb~tagging uncertainties for heavy-flavor (\cPqb and \cPqc) and light-flavor (\cPqu, \cPqd, \cPqs, and gluon) jets are calculated separately, and combined in quadrature to give the total \cPqb~tagging uncertainty.
To avoid double-counting of the uncertainty related to the \cPqb~tagging efficiency, when necessary, the SFs for \cPqb~tagging efficiency are recalculated in the evaluation of the remaining sources of experimental and theoretical uncertainty described in this section, using the procedure given in Ref.~\cite{Sirunyan:2018ucr}.

The uncertainty arising from the JES is determined by varying the individual sources of uncertainty in the JES in bins of the jet \pt and $\eta$, and taking the quadrature sum of the differences~\cite{Khachatryan:2016kdb}. The JES variations are propagated to the uncertainties in \ptmiss.
An additional uncertainty in the calculation of \ptmiss\ is estimated by varying the energies of reconstructed particles not clustered into jets (unclustered energy).
The uncertainty from the jet energy resolution (JER) is determined by the variation of the JER in simulation within its uncertainty in different $\eta$ regions~\cite{Khachatryan:2016kdb}.

\subsection{Theoretical sources of uncertainty}

The uncertainty arising from the missing higher-order terms in the simulation of the signal process at the ME level is assessed by varying the renormalization and factorization scales ($\mu_\mathrm{R}$ and $\mu_\mathrm{F}$) in the \Powhegvtwo\ simulation up and down by a factor of 2 with respect to their nominal values,
both individually and simultaneously (six variations in total).
 The nominal choice for the scales is $\mT^{\cPqt} = \sqrt{\smash[b]{m^2_{\cPqt} + p^2_{\mathrm{T,\cPqt}}}}$, where $p_{\mathrm{T},\cPqt}$ denotes the \pt of the top quark in the \ttbar\ rest frame. 
 In the PS simulation, the corresponding uncertainty is estimated by four additional variations: 
changing the scale of initial- and final-state radiation individually up and down by factors of 2 and $\sqrt{2}$, respectively, as suggested in Ref.~\cite{Skands:2014pea}.
The total scale uncertainty is taken as the maximum deviation from the nominal prediction from all ten variations. 

The uncertainty originating from the scheme used to match the ME-level calculation to the PS simulation is derived by varying the $h_\mathrm{damp}$ parameter in \Powhegvtwo\ by factors of 1.42 and 0.63, according to the results of tuning this parameter from Ref.~\cite{bib:CMS:2016kle}.

The default setup in \Pythia\ 
includes a multiple parton interaction (MPI) based model of color
 reconnection (CR) with early resonance decays switched off. To estimate the uncertainty from this choice of model, the analysis is repeated with three other CR models within \Pythia: the MPI-based scheme with early resonance decays switched on, a gluon-move scheme~\cite{Argyropoulos:2014zoa}, and a QCD-inspired scheme~\cite{Christiansen:2015yqa}. The total uncertainty from CR modeling is estimated by taking the maximum deviation from the nominal result.
The uncertainty related to modeling of the underlying event is estimated by varying the parameters used to derive the CUETP8M2T4 tune in the default setup. 

The uncertainty from the \cPqb~quark fragmentation function is assessed from the largest deviation when varying the Bowler--Lund function within its uncertainties~\cite{Bowler:1981sb} and repeating the analysis with the Peterson model for \cPqb~quark fragmentation~\cite{PhysRevD.27.105}. An uncertainty from the semileptonic BF of \cPqb~hadrons is estimated by correcting the \ttbar\ simulation to match the BF in Ref.~\cite{PhysRevD.98.030001}.

The uncertainty from the PDFs is assessed from the standard deviation of the result when using the replicas of the NNPDF3.0 PDF set in the signal simulation~\cite{bib:NNPDF,Butterworth:2015oua}. An additional uncertainty is derived by varying the $\alpS$ value within its uncertainty in the PDF set~\cite{Butterworth:2015oua}. The dependence of the measurement on the assumed \mt\ value is estimated by varying the chosen \mt\ in the default setup by ${\pm}1\GeV$ with respect to the default value of $172.5\GeV$.

Previous CMS studies have shown that the \pt distribution of the top quark measured from data is softer than that in the NLO simulation of \ttbar\ production~\cite{bib:TOP-11-013_paper,bib:TOP-12-028_paper,Khachatryan:2015fwh,bib:TOP-16-008,Sirunyan:2018ucr}.
This is understood to arise at least partly from the missing higher-order QCD terms~\cite{bib:difftop,bib:kidonakis_13TeV,Czakon:2017wor,Czakon:2018nun}.
The change in the measurement when reweighting the simulated \ttbar event sample to match the top quark \pt spectrum in data is taken as a two-sided systematic uncertainty associated with the signal modeling.

Since \ttbar\ events producing electrons or muons originating from the decay of $\tau$ leptons are considered a background, the measured differential cross sections are sensitive to the relative BFs of \PW~bosons decaying to $\tau$ leptons and electrons or muons, and the $\tau$ semileptonic BFs assumed in the simulation. An uncertainty of $2.5\%$ is assigned to the relative normalization of this background process. The shape and absolute normalization of this process is taken from data, as described in Section~\ref{sec:diffxsec}.
The normalizations of all other backgrounds are varied by ${\pm}30\%$~\cite{bib:TOP-15-003_paper,Sirunyan:2018ucr}.

\section{Results}
\label{sec:results}

\subsection{Normalized differential cross sections}
\label{sec:measdiffxsecs}

\begin{figure*}[!htpb]
\centering
\includegraphics[width=\unfoldedFigWidth]{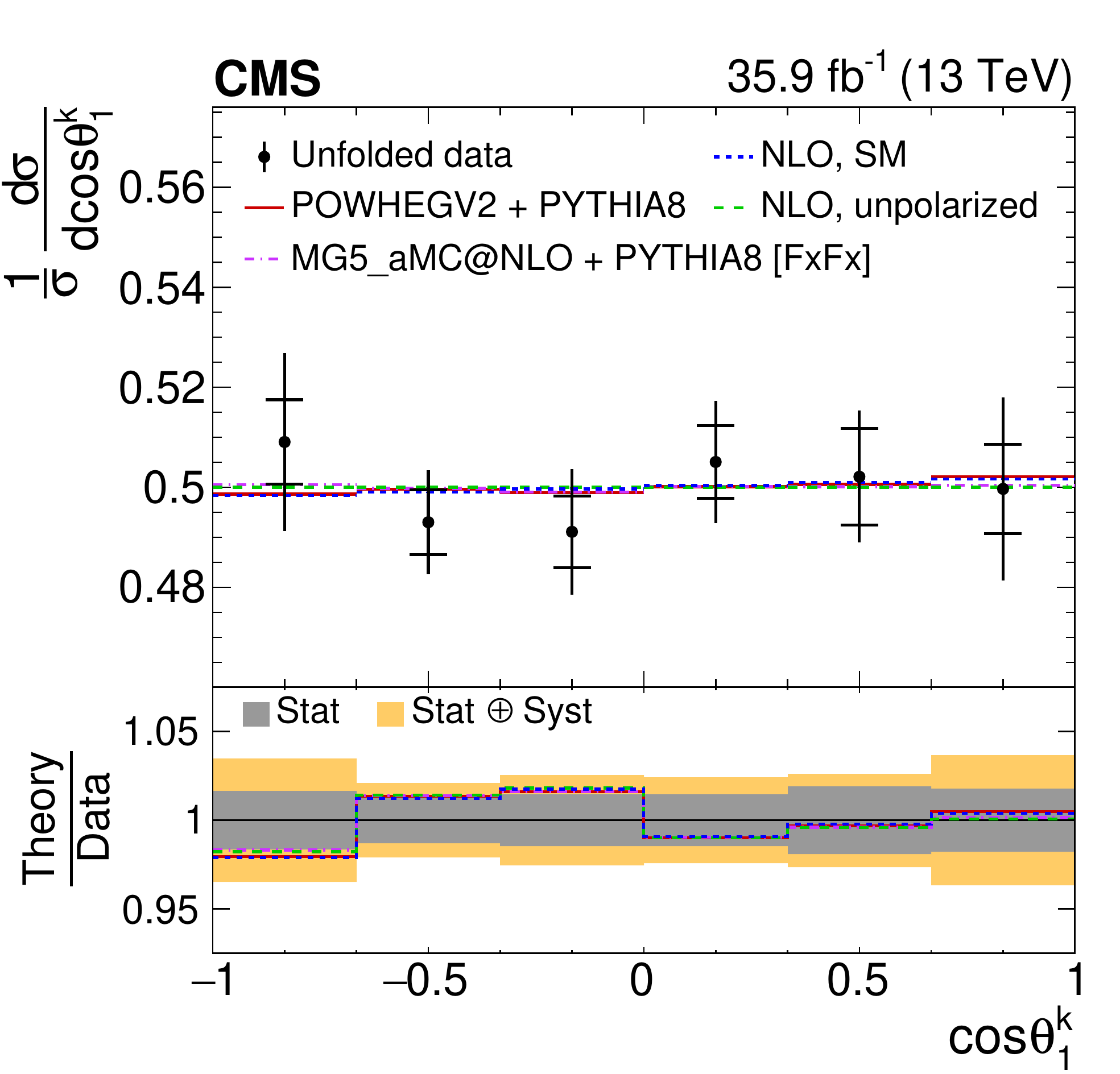}
	        \hfill
\includegraphics[width=\unfoldedFigWidth]{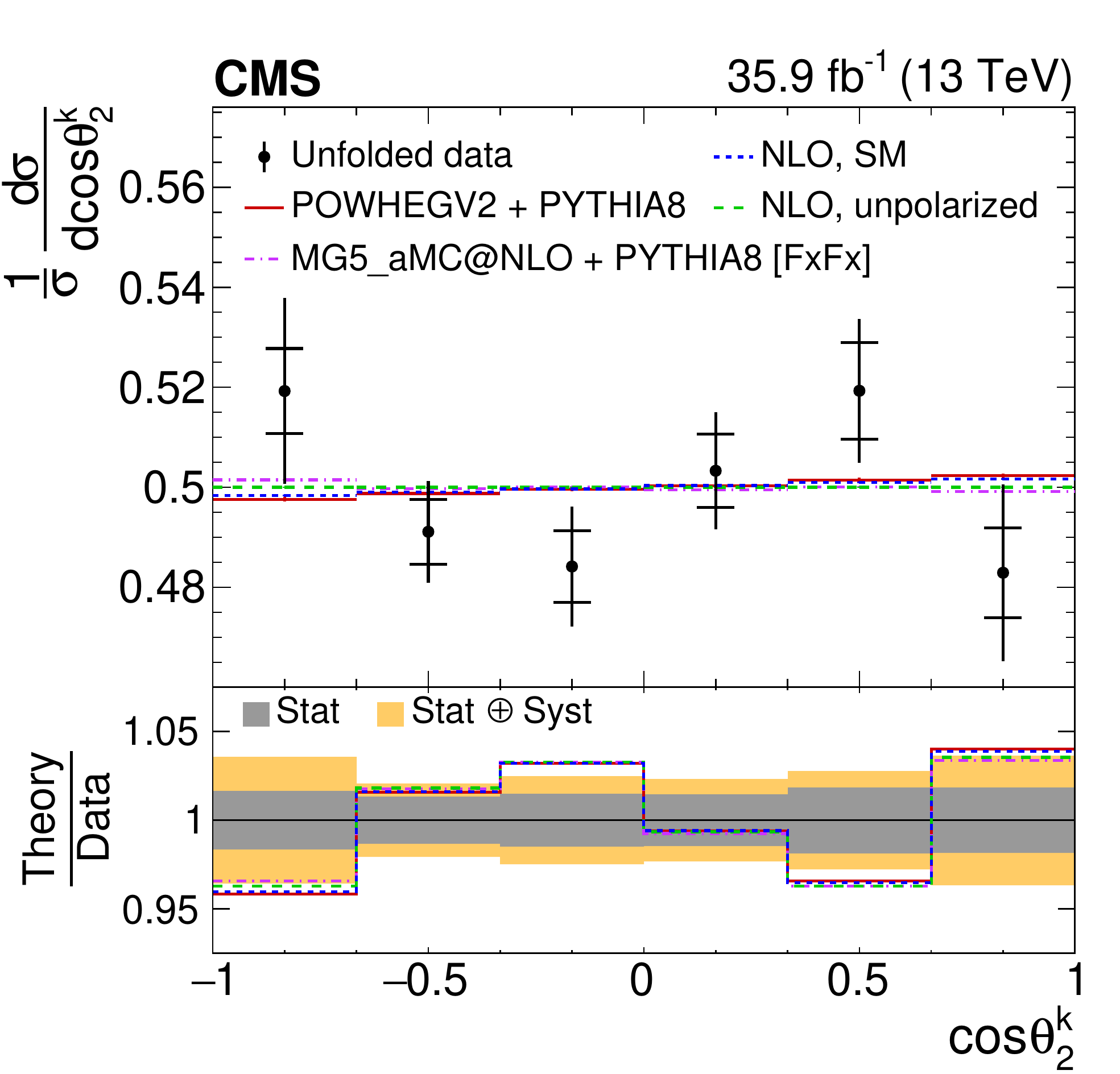} \\
\includegraphics[width=\unfoldedFigWidth]{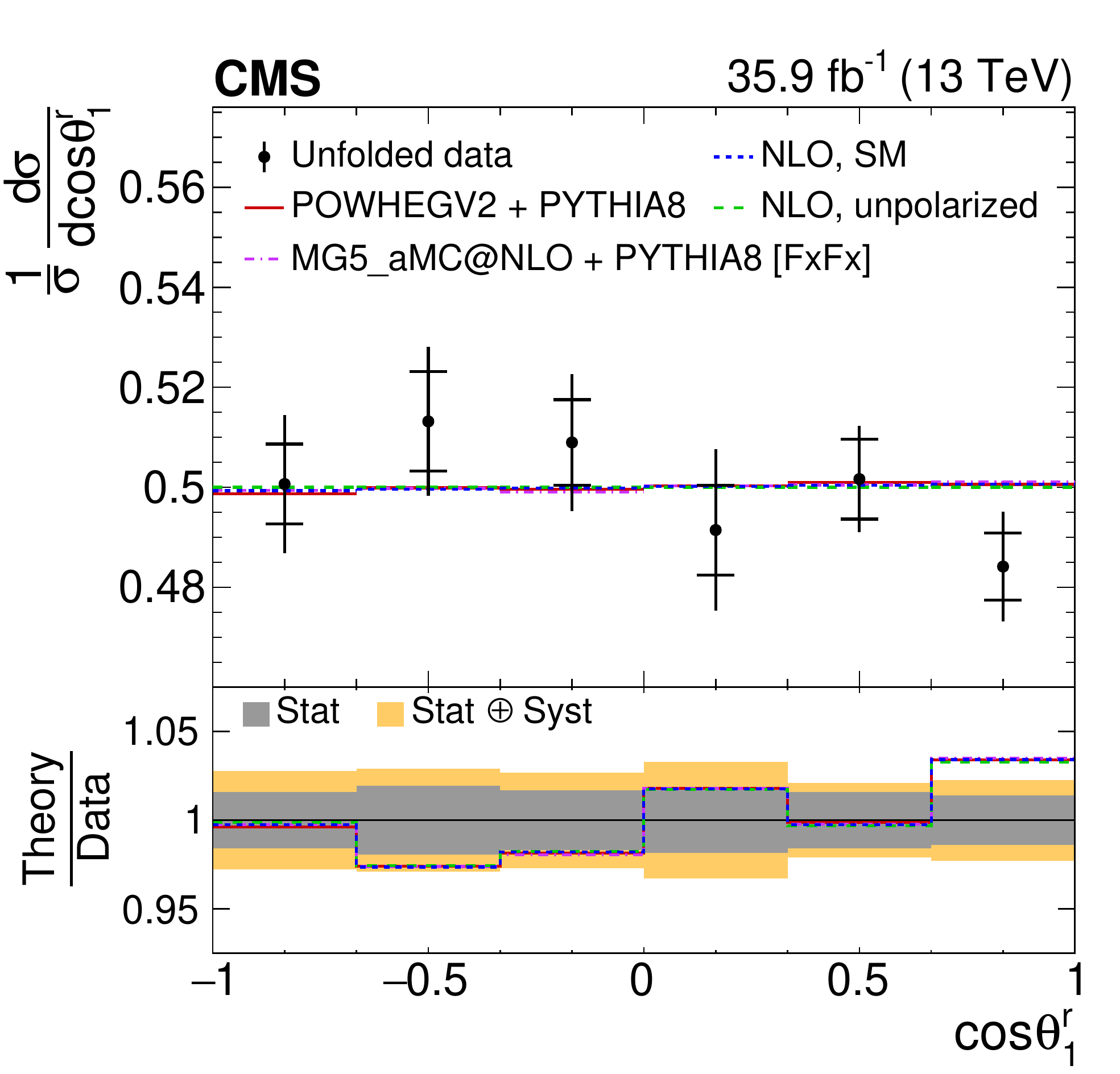}
	        \hfill
\includegraphics[width=\unfoldedFigWidth]{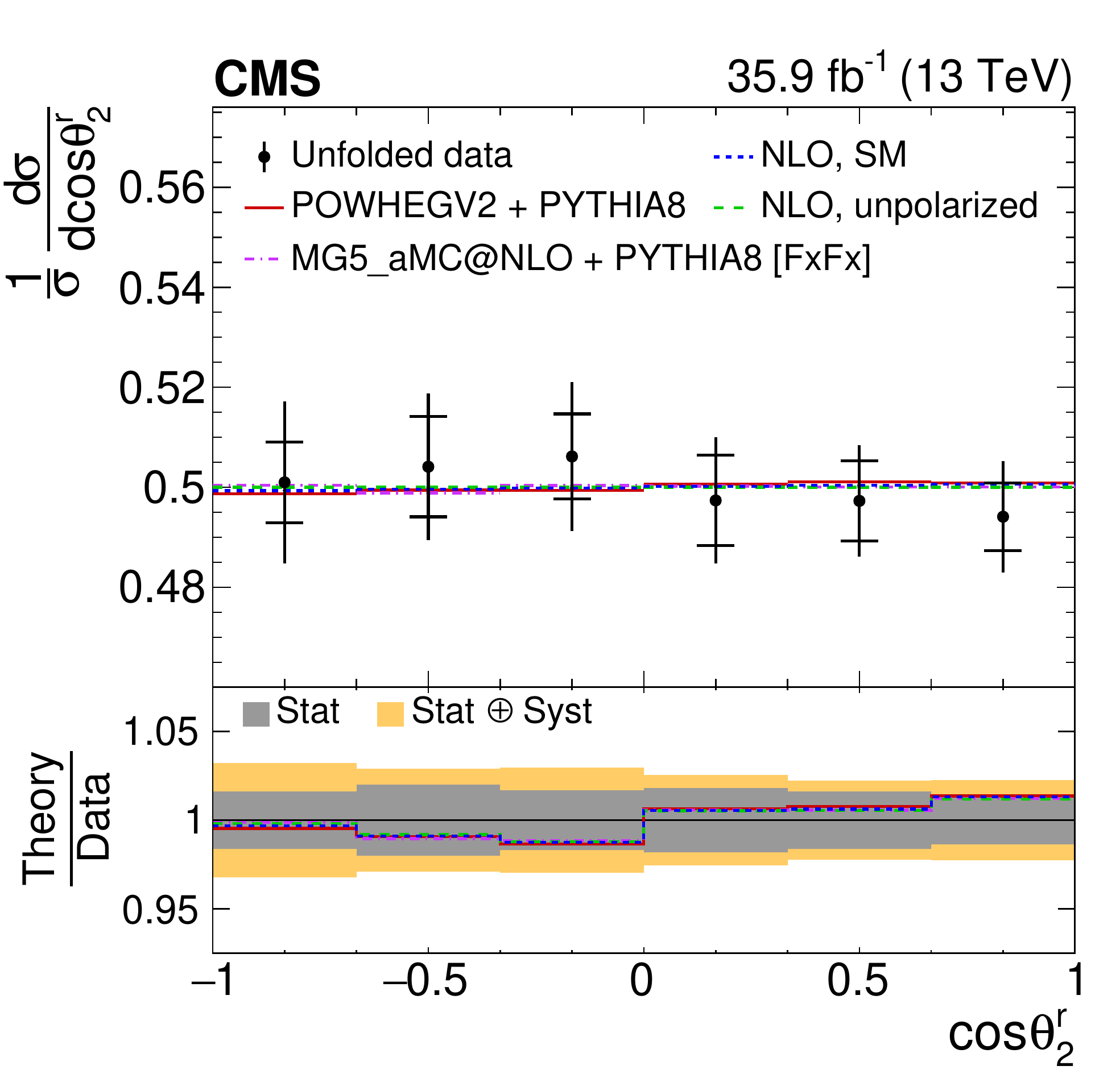} \\
\includegraphics[width=\unfoldedFigWidth]{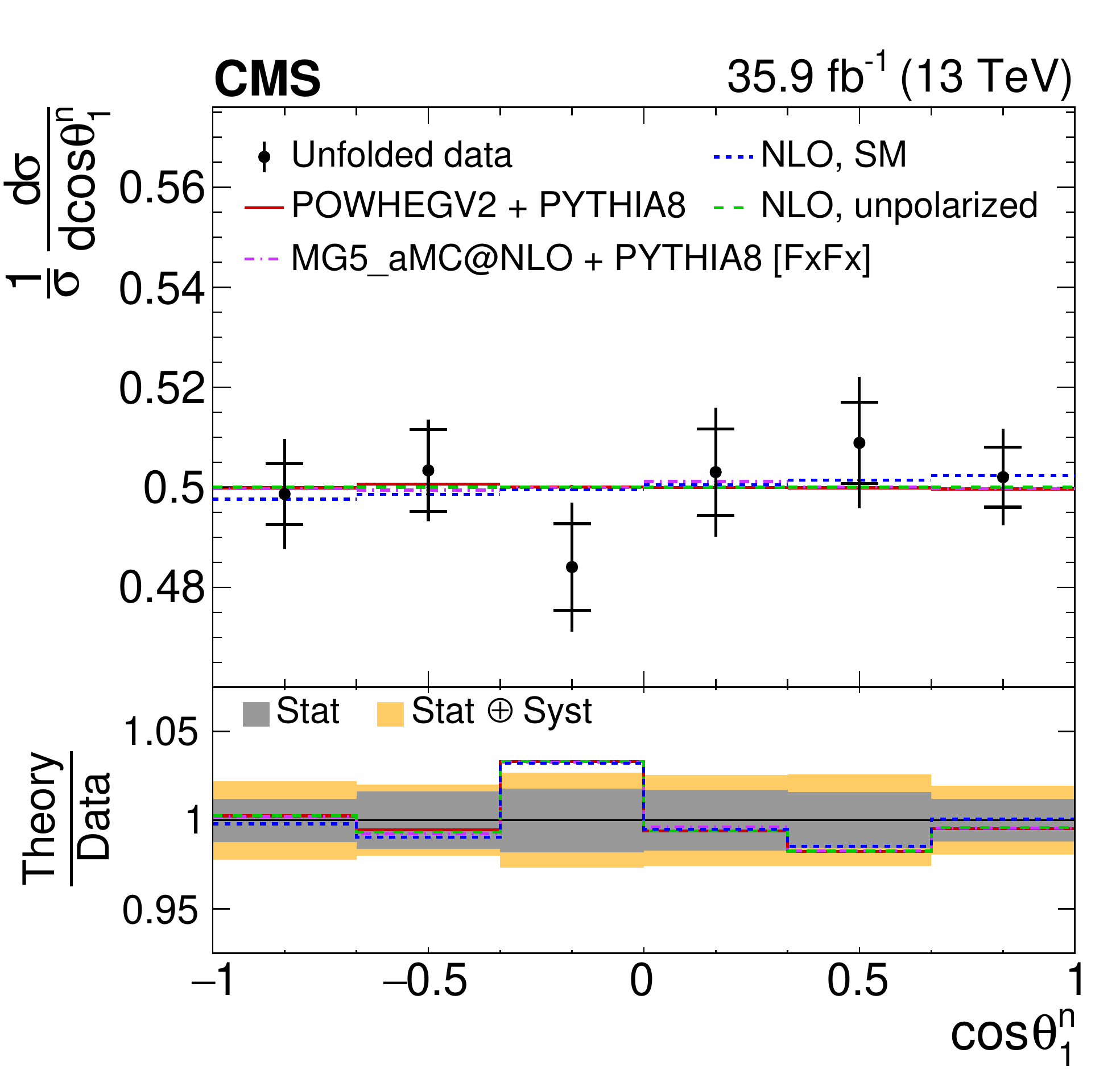}
	        \hfill 
\includegraphics[width=\unfoldedFigWidth]{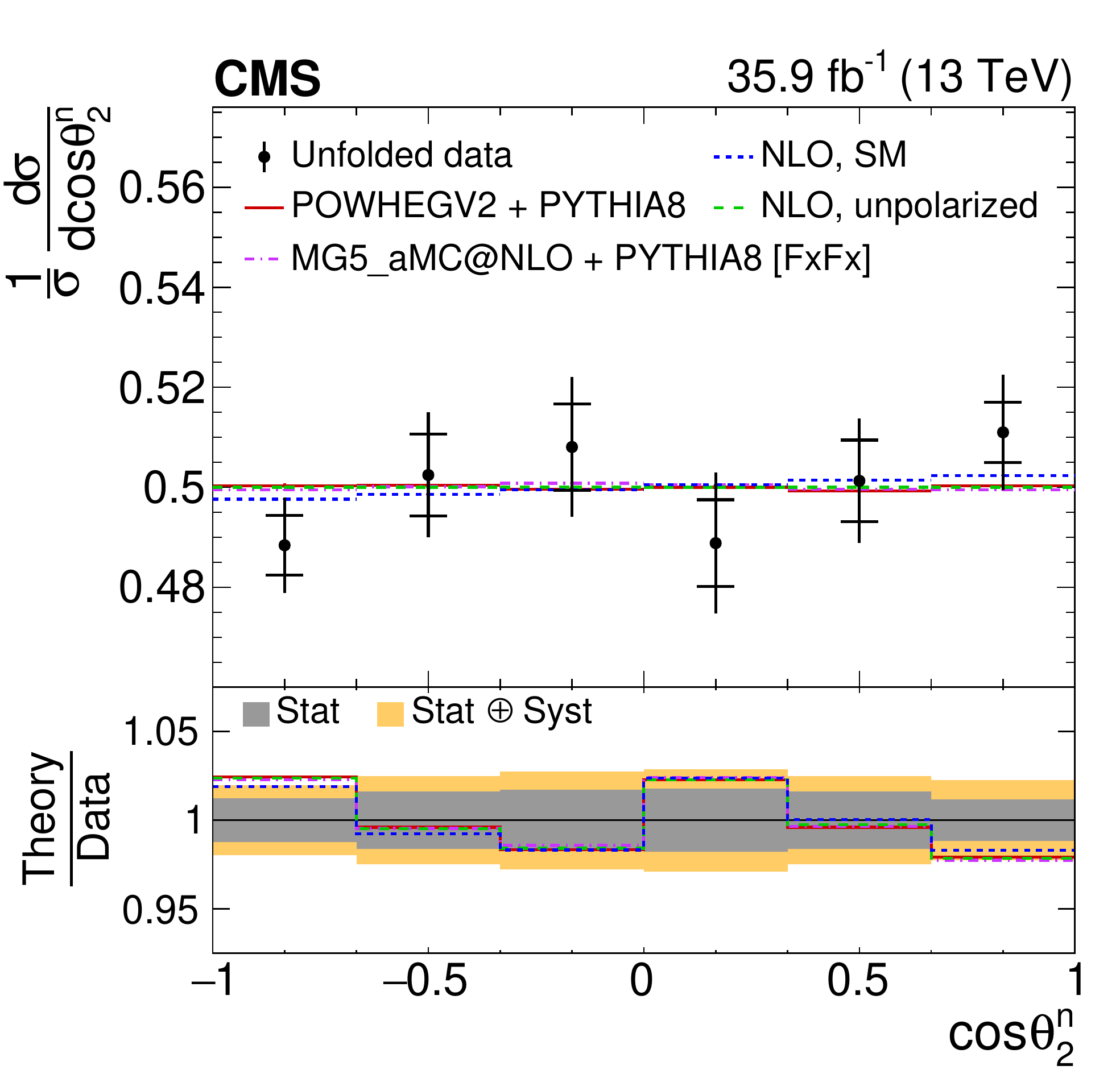} \\
\caption{\label{fig:UnfoldedNormXsecB}\protect
Unfolded data (points) and predicted (horizontal lines) normalized differential cross sections with respect
to $\cos\theta^i$ for top quarks (antiquarks) in the first (second) column,
probing polarization coefficients $B_{1}^{i}$ ($B_{2}^{i})$.
From top to bottom, the reference axis $i=\hat{k}$, $\hat{r}$, $\hat{n}$.
The vertical lines on the points represent the total uncertainties, with the statistical components indicated by horizontal bars.
The ratios of various predictions to the data are shown in the lower panels.
}
\end{figure*}

\begin{figure*}[!htpb]
\centering
\includegraphics[width=\unfoldedFigWidth]{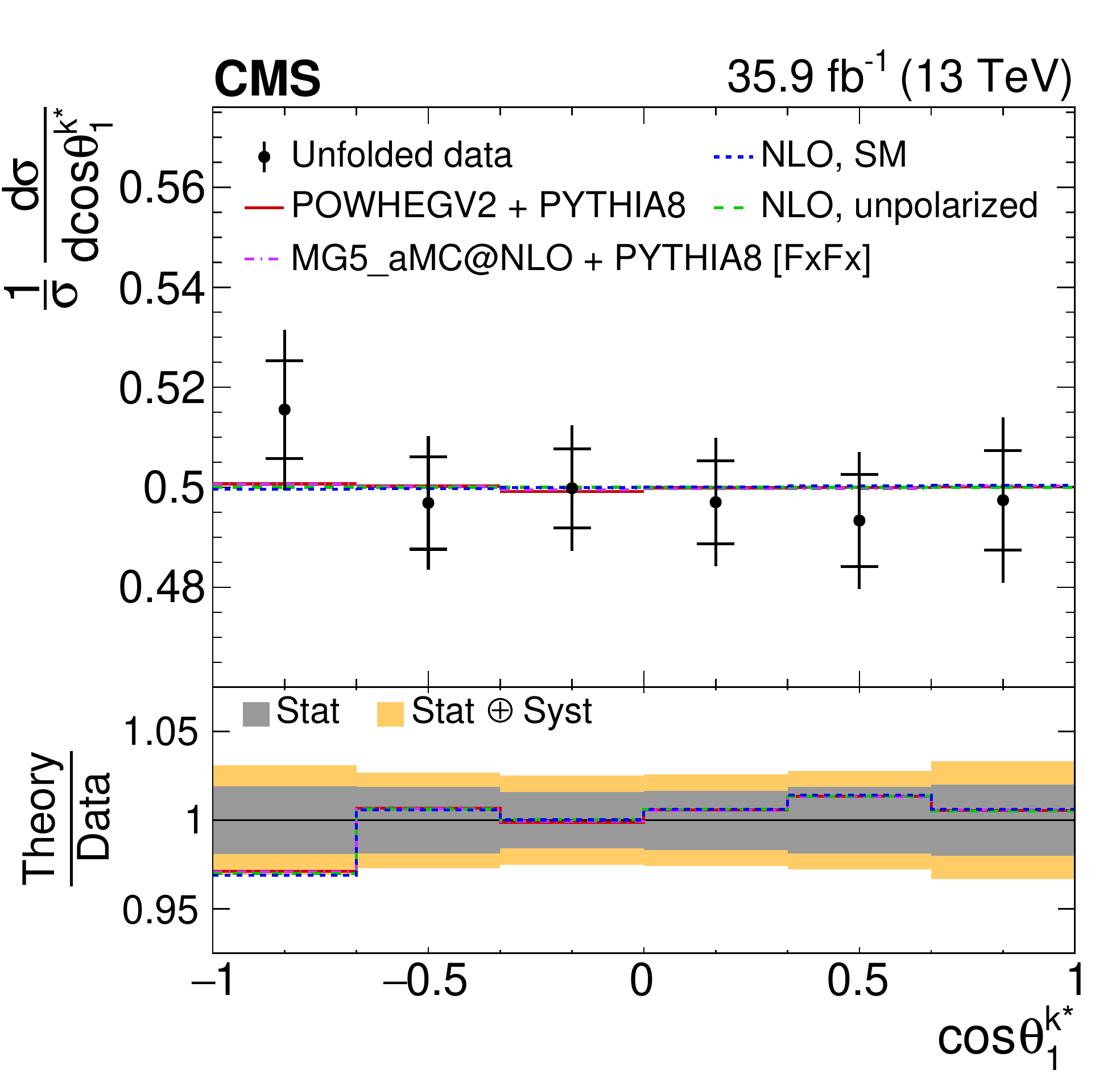}
	        \hfill
\includegraphics[width=\unfoldedFigWidth]{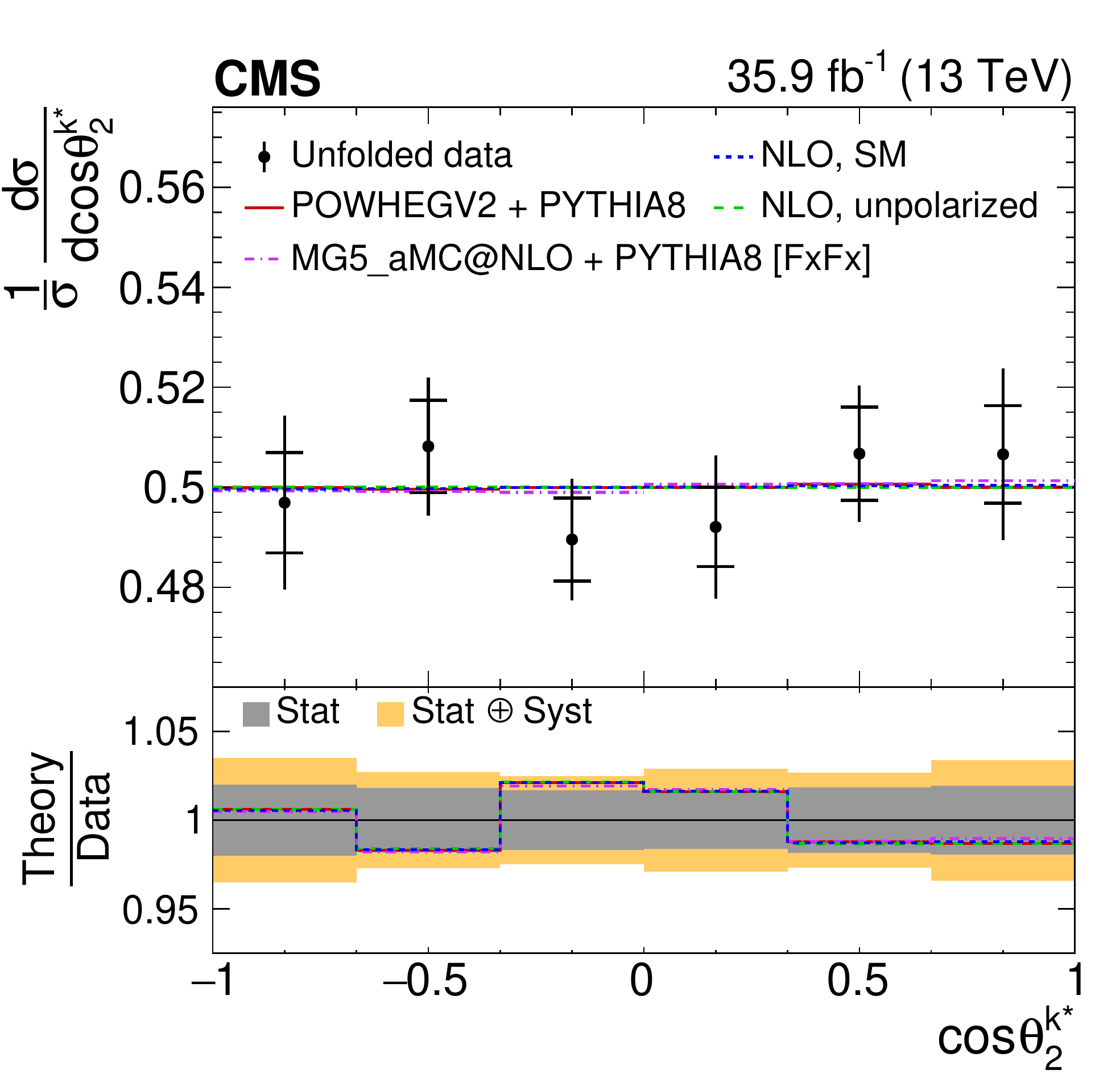} \\
\includegraphics[width=\unfoldedFigWidth]{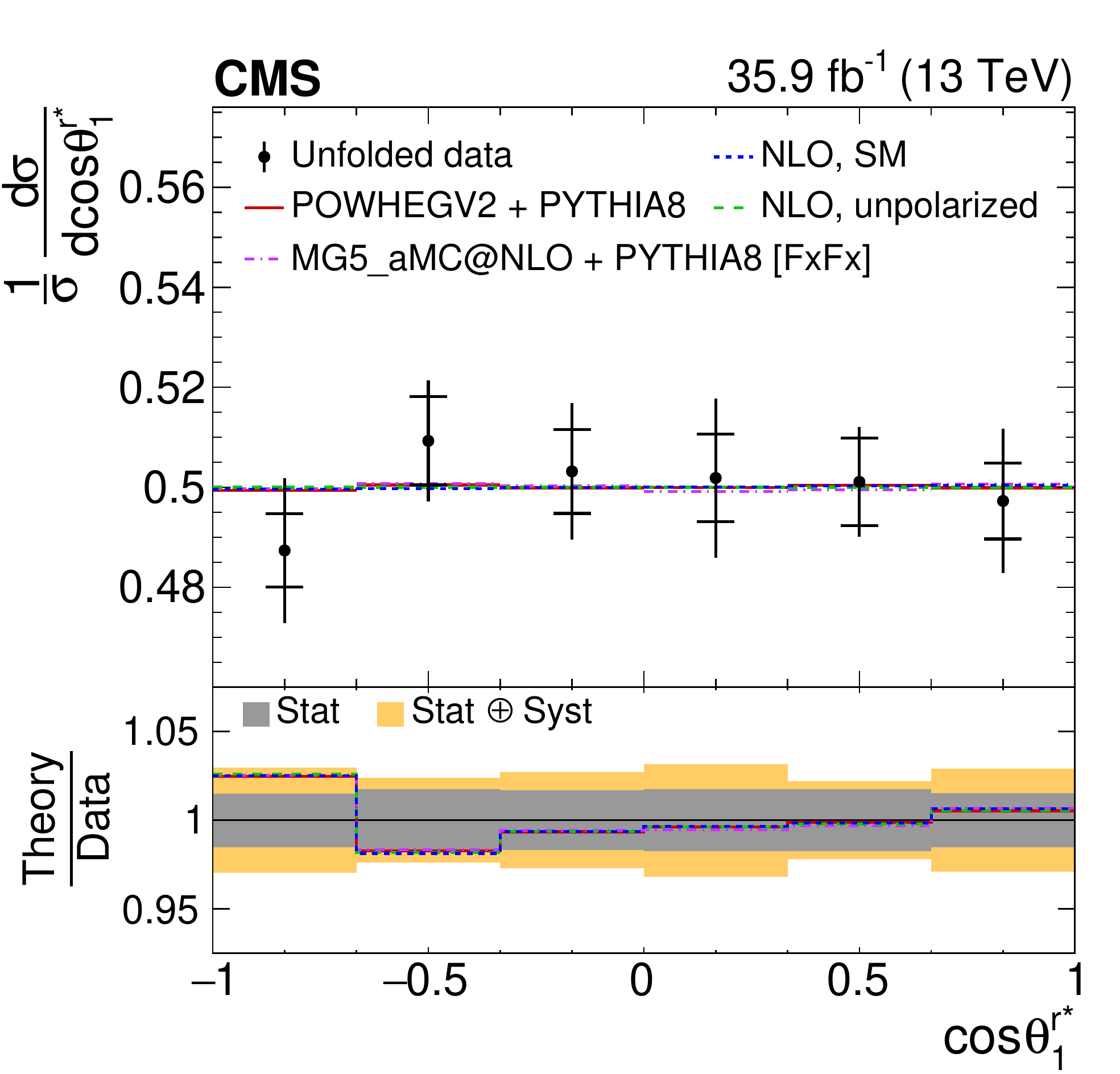}
	        \hfill
\includegraphics[width=\unfoldedFigWidth]{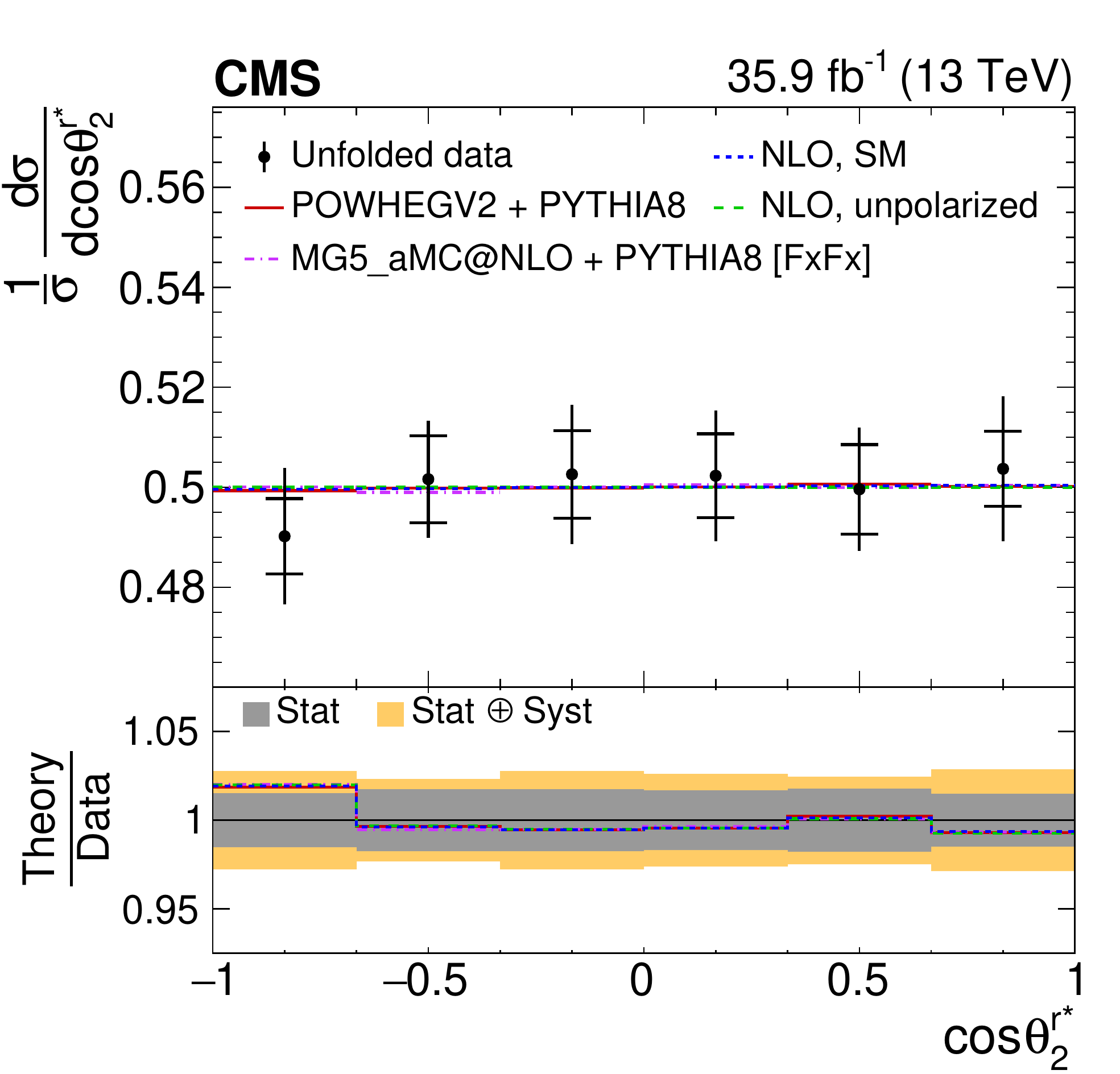} \\
\caption{\label{fig:UnfoldedNormXsecB2}\protect
Unfolded data (points) and predicted (horizontal lines) normalized differential cross sections with respect
to $\cos\theta^{i*}$ for top quarks (antiquarks) in the first (second) column,
probing polarization coefficients $B_{1}^{{i*}}$ ($B_{2}^{{i*}})$.
The reference axis $i^*=\hat{k}^*$ (top row) and $\hat{r}^*$ (bottom row).
The vertical lines on the points represent the total uncertainties, with the statistical components indicated by horizontal bars.
The ratios of various predictions to the data are shown in the lower panels.
}
\end{figure*}

\begin{figure*}[!htpb]
\centering
\includegraphics[width=\unfoldedFigWidth]{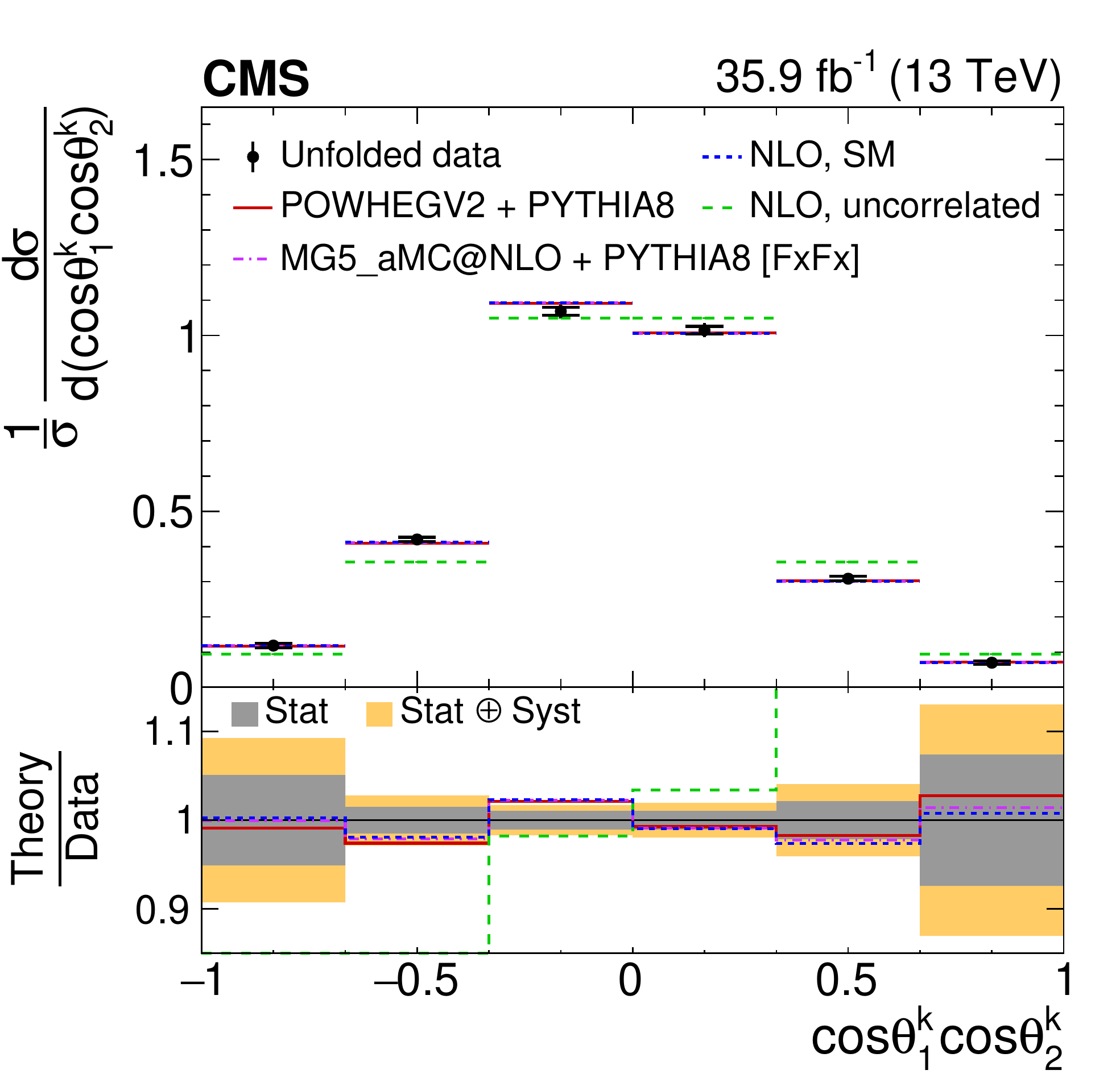}
	        \hfill
\includegraphics[width=\unfoldedFigWidth]{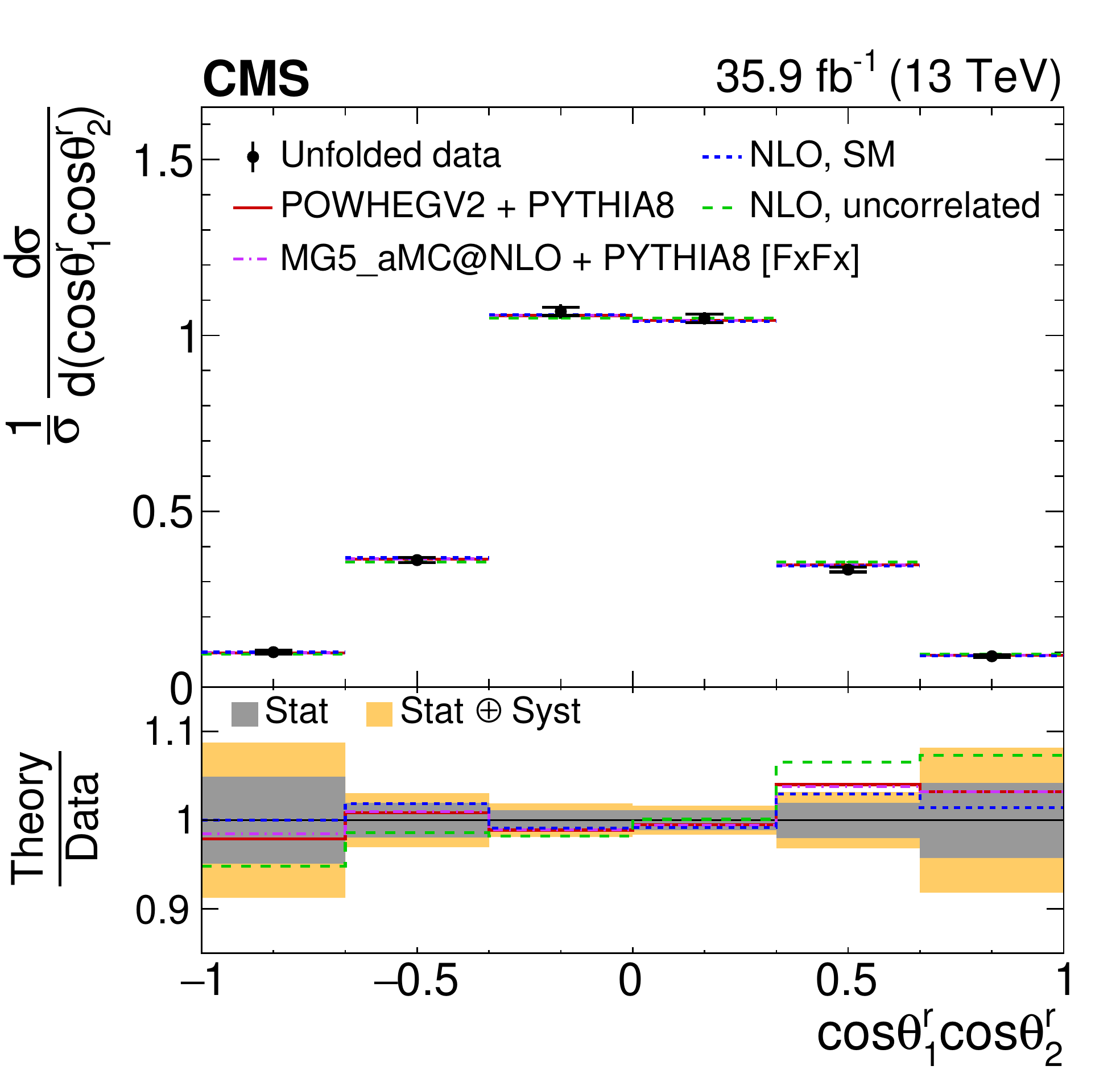} \\
\includegraphics[width=\unfoldedFigWidth]{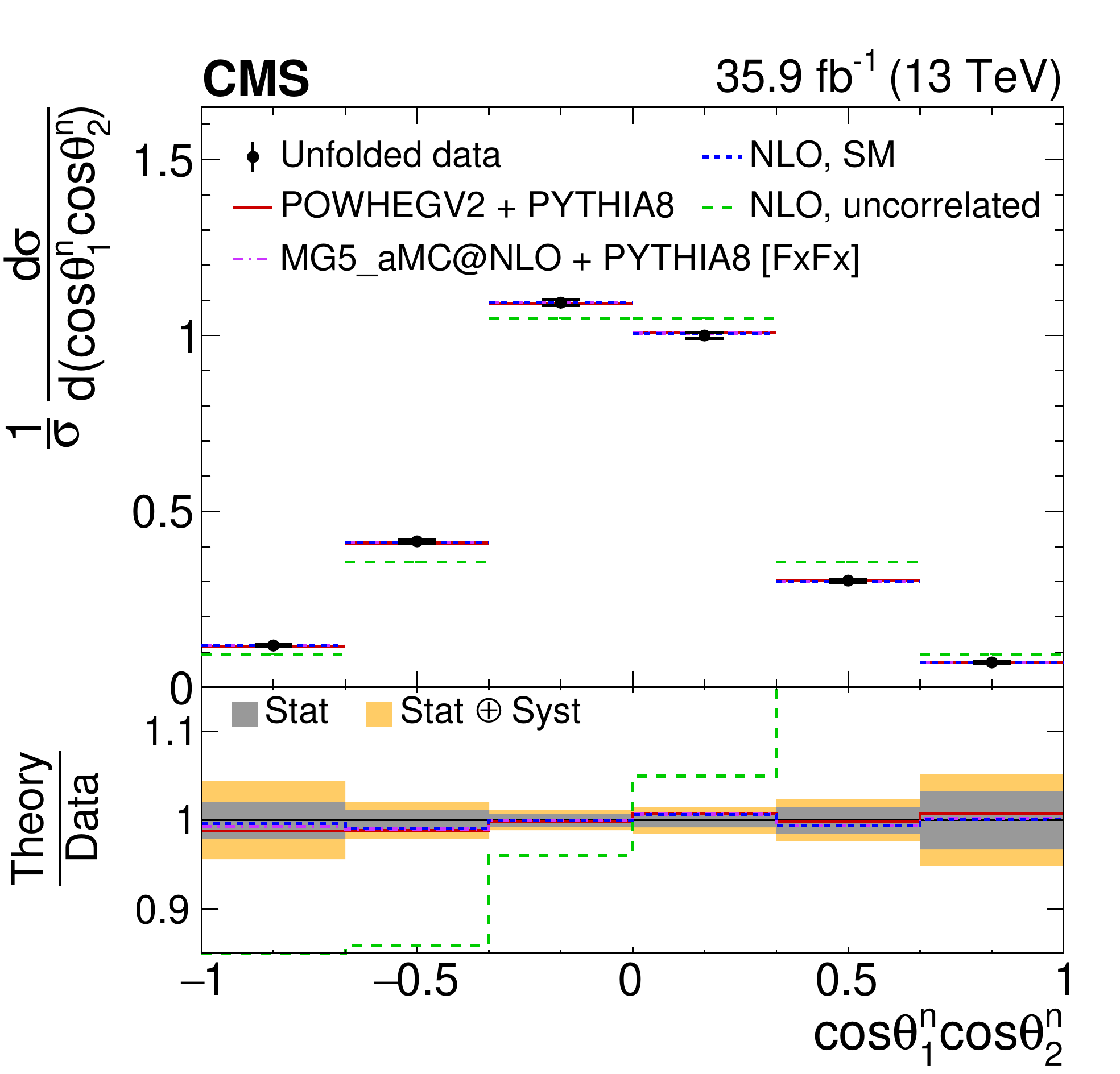}
	        \hfill
\includegraphics[width=\unfoldedFigWidth]{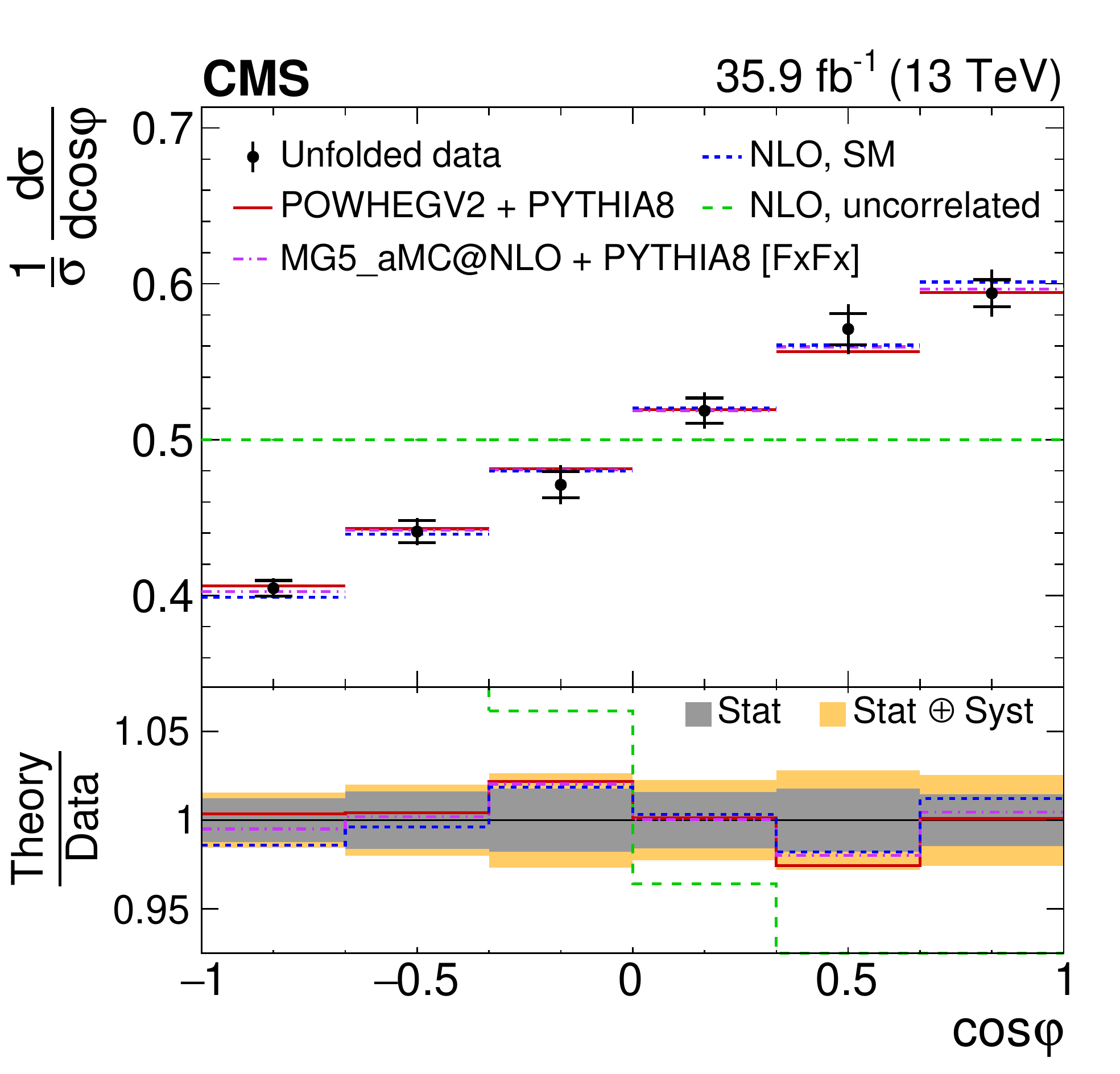} \\
\includegraphics[width=\unfoldedFigWidth]{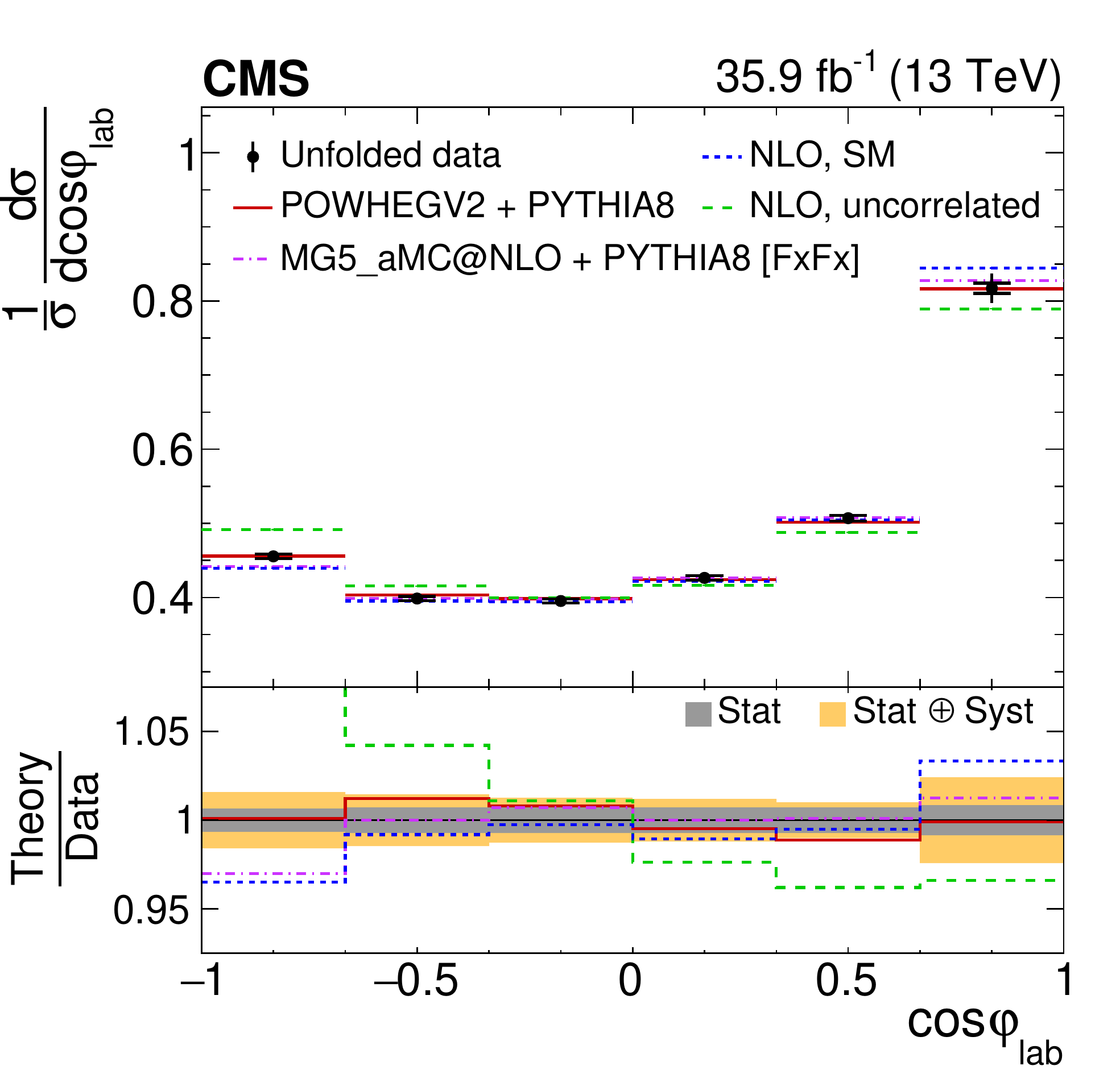}
	        \hfill
\includegraphics[width=\unfoldedFigWidth]{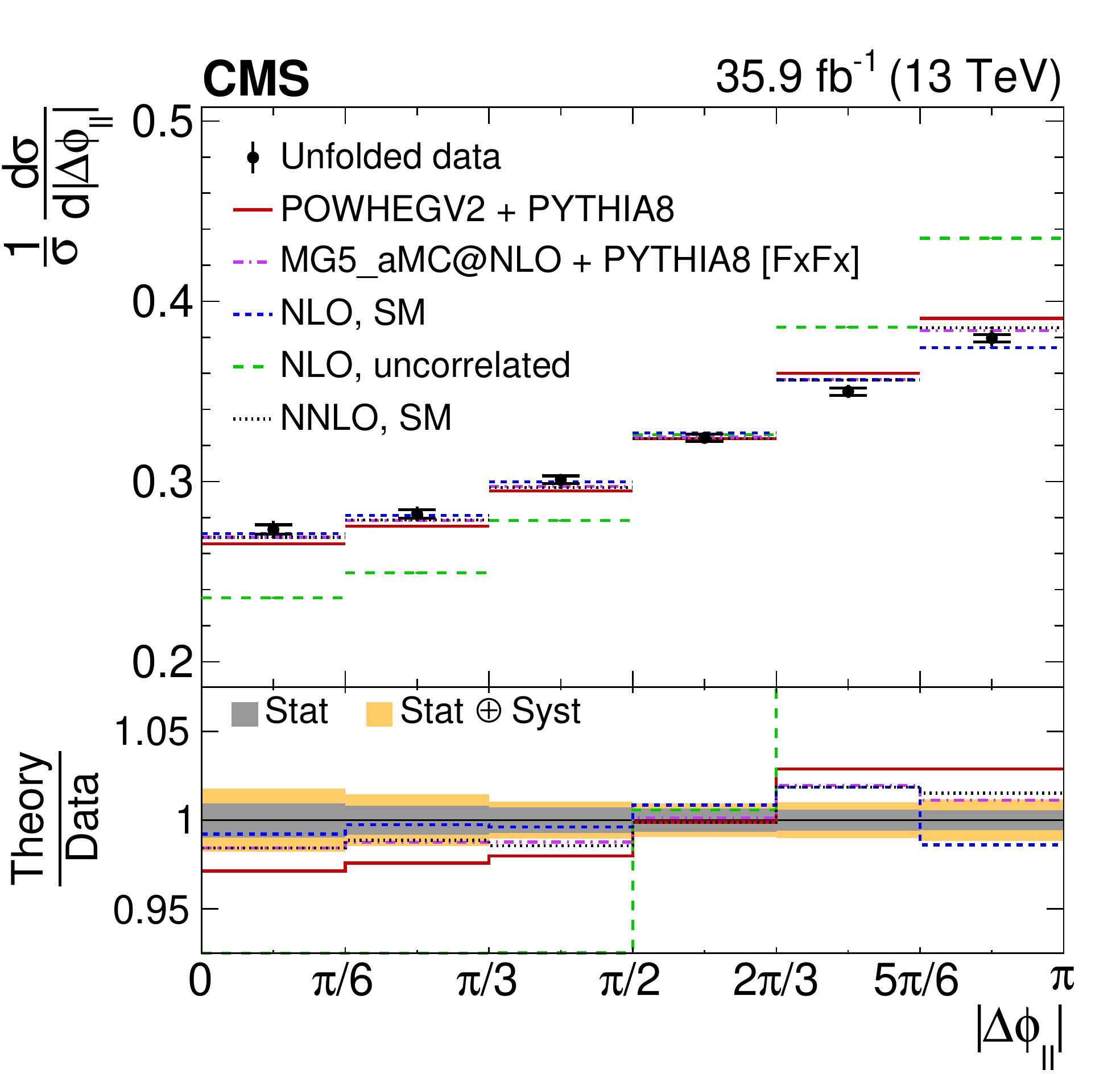} \\
\caption{\label{fig:UnfoldedNormXsecC}\protect
Unfolded data (points) and predicted (horizontal lines) normalized differential cross sections for the diagonal spin correlation observables (first two rows) and the laboratory-frame observables (bottom row). The vertical lines on the points represent the total uncertainties, with the statistical components indicated by horizontal bars.
The ratios of various predictions to the data are shown in the lower panels.
}
\end{figure*}

\begin{figure*}[!htpb]
\centering
\includegraphics[width=\unfoldedFigWidth]{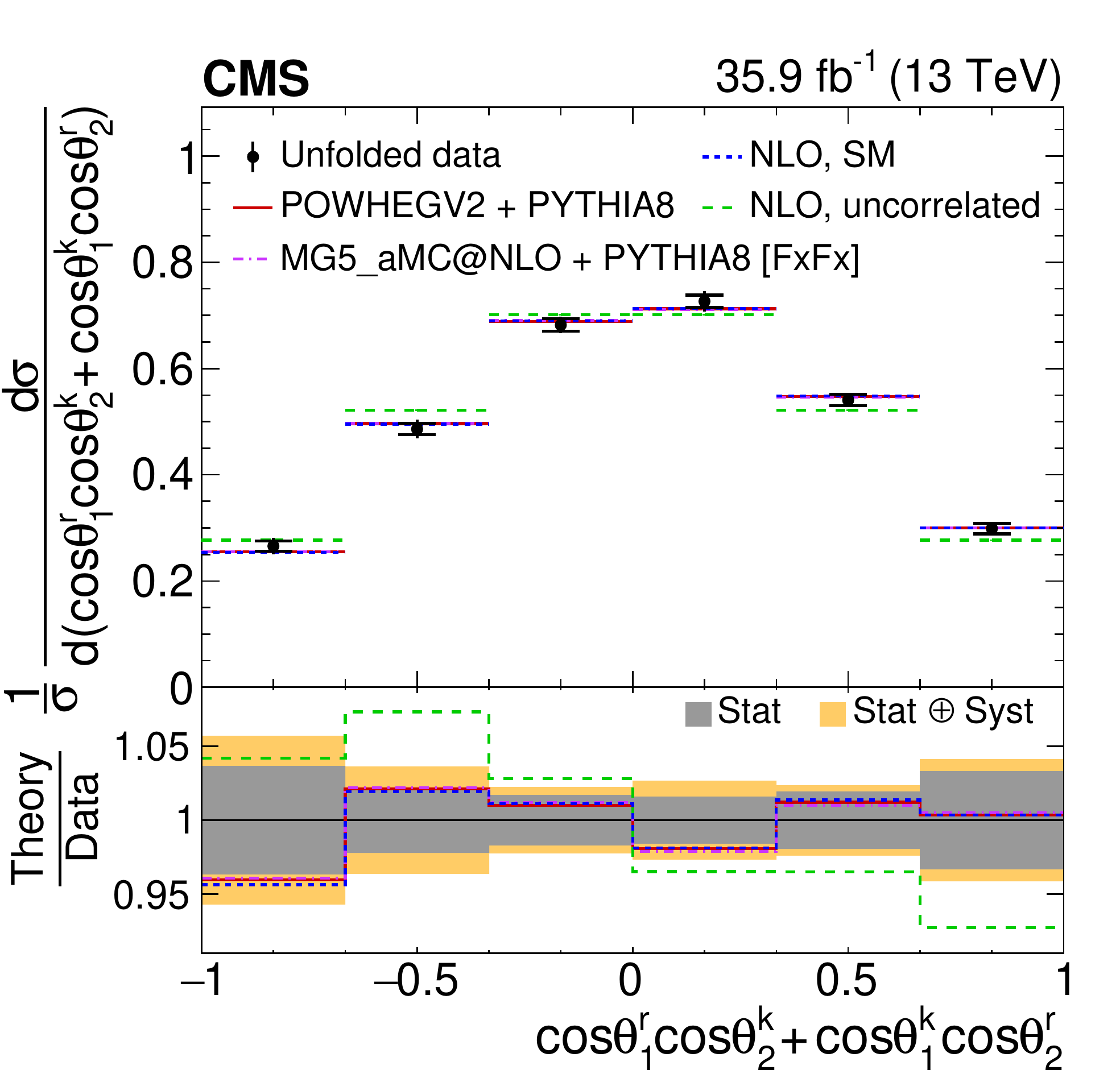}
	        \hfill
\includegraphics[width=\unfoldedFigWidth]{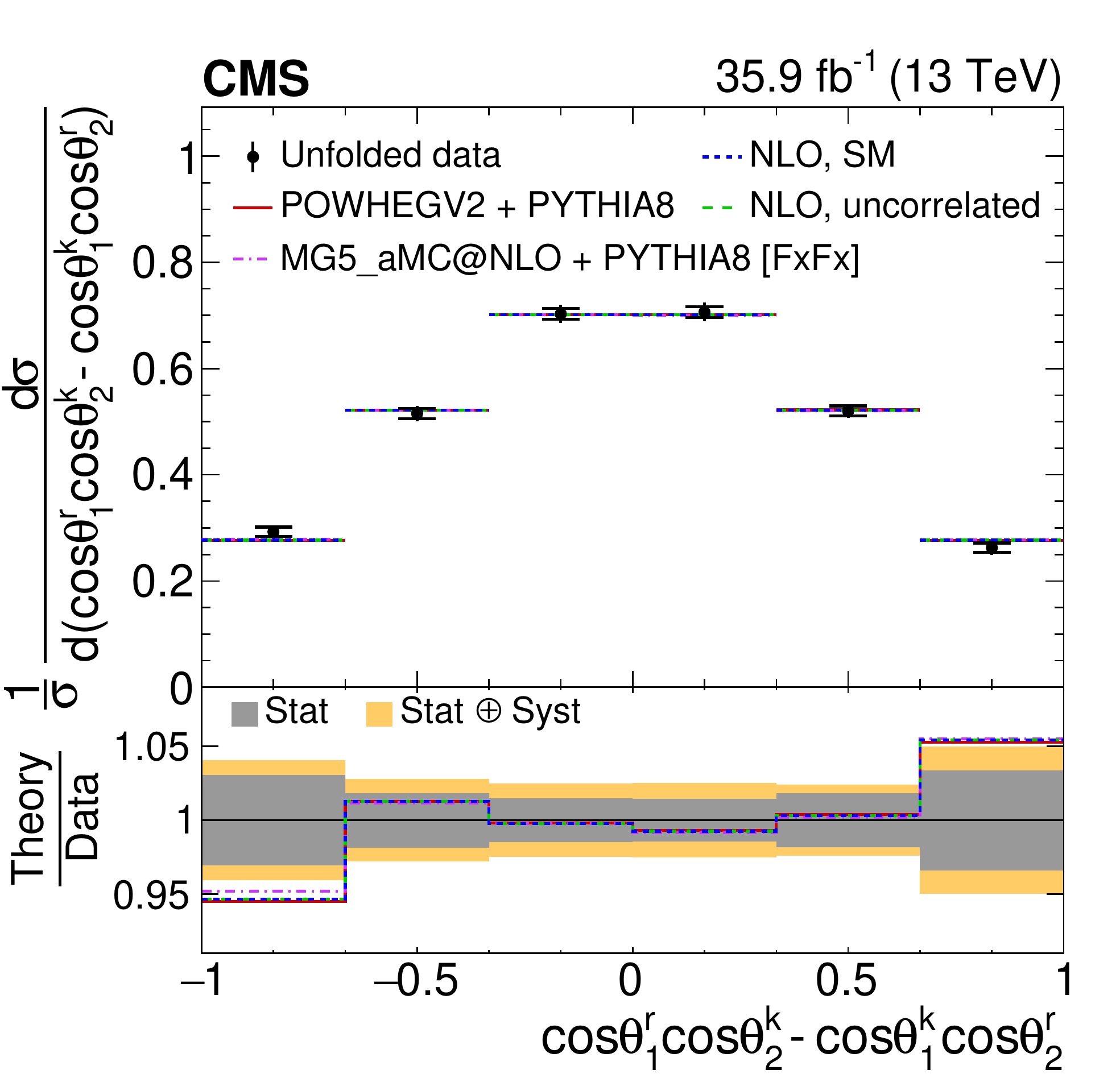} \\
\includegraphics[width=\unfoldedFigWidth]{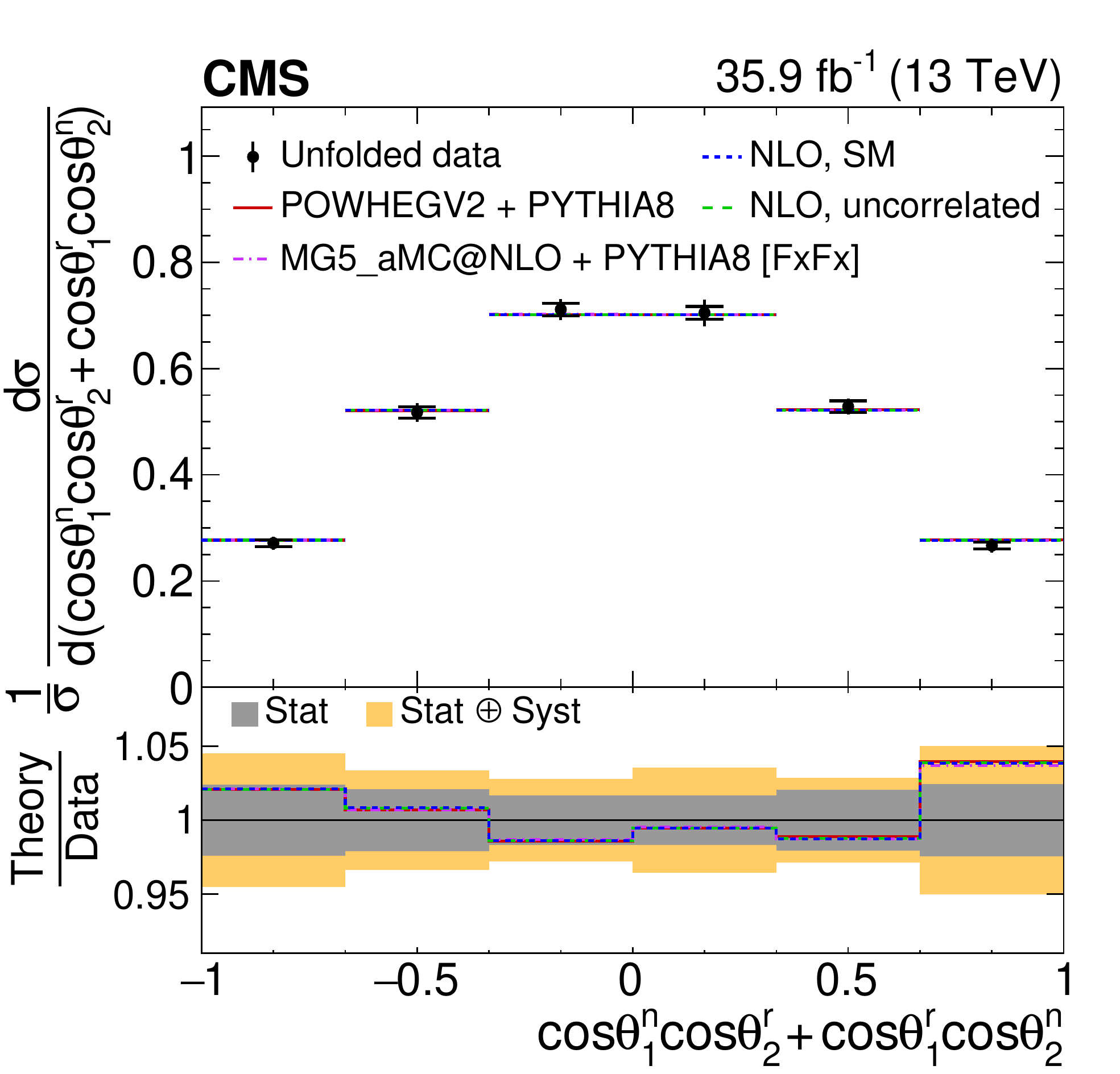}
	        \hfill
\includegraphics[width=\unfoldedFigWidth]{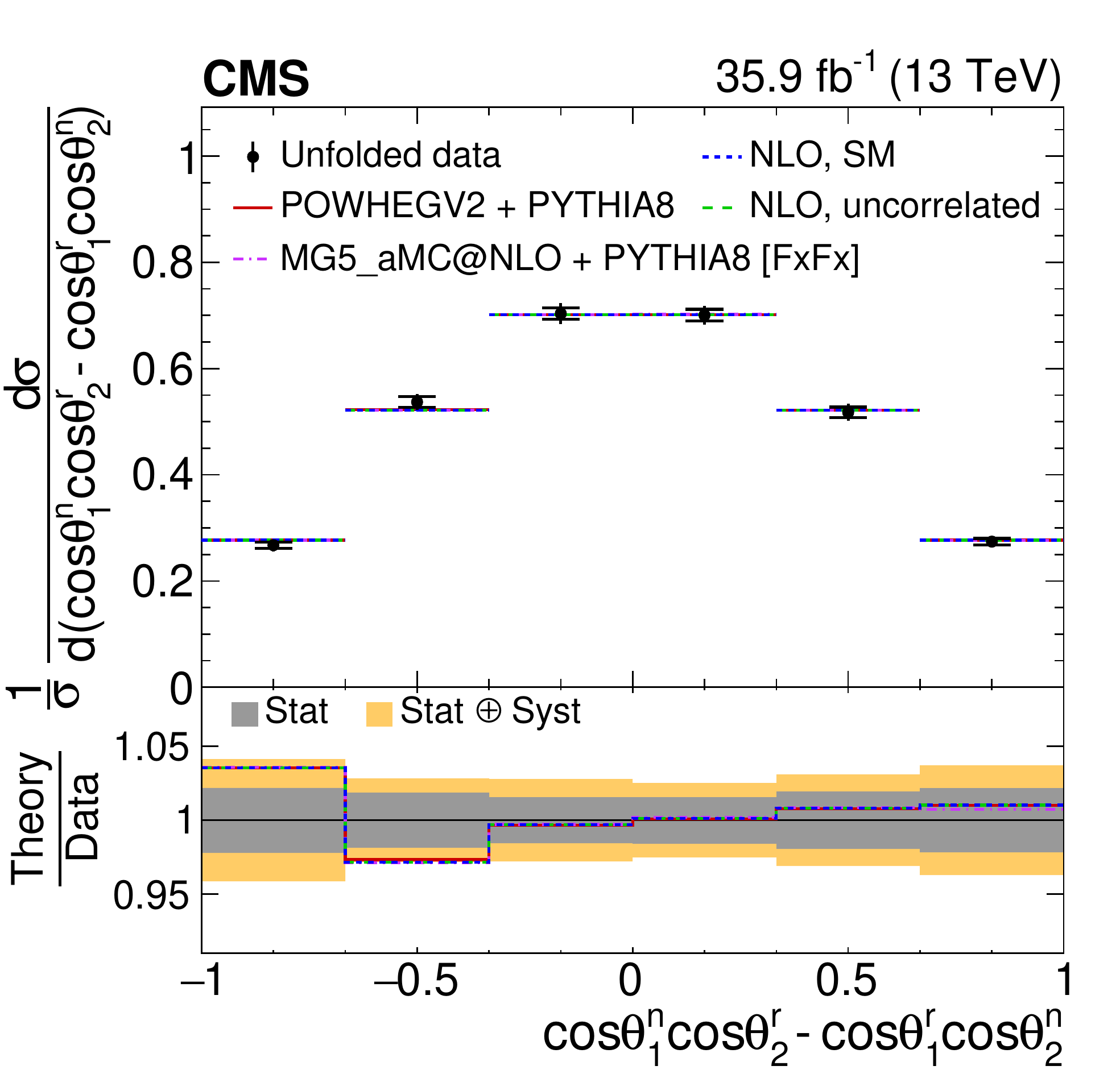} \\
\includegraphics[width=\unfoldedFigWidth]{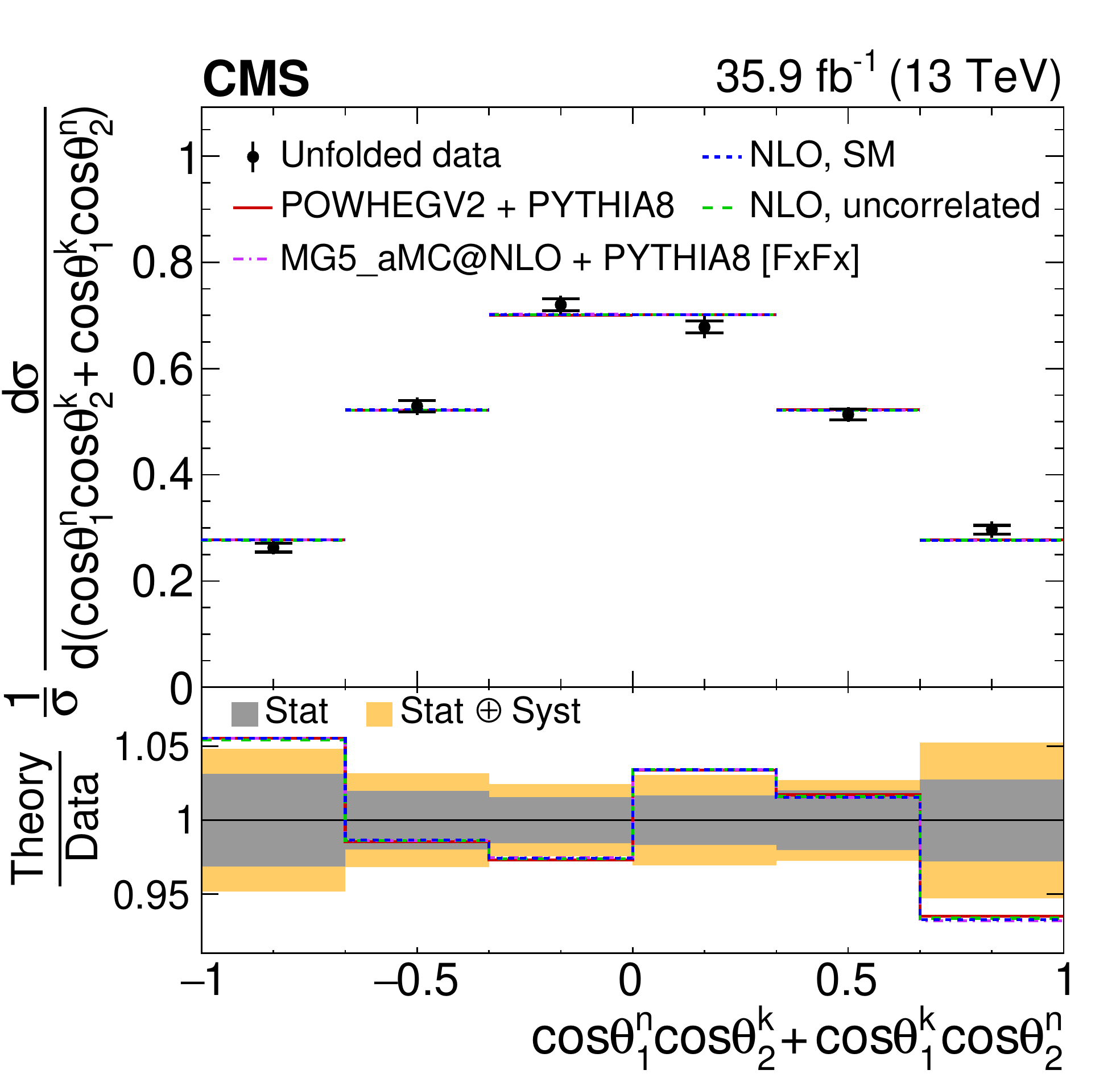}
	        \hfill
\includegraphics[width=\unfoldedFigWidth]{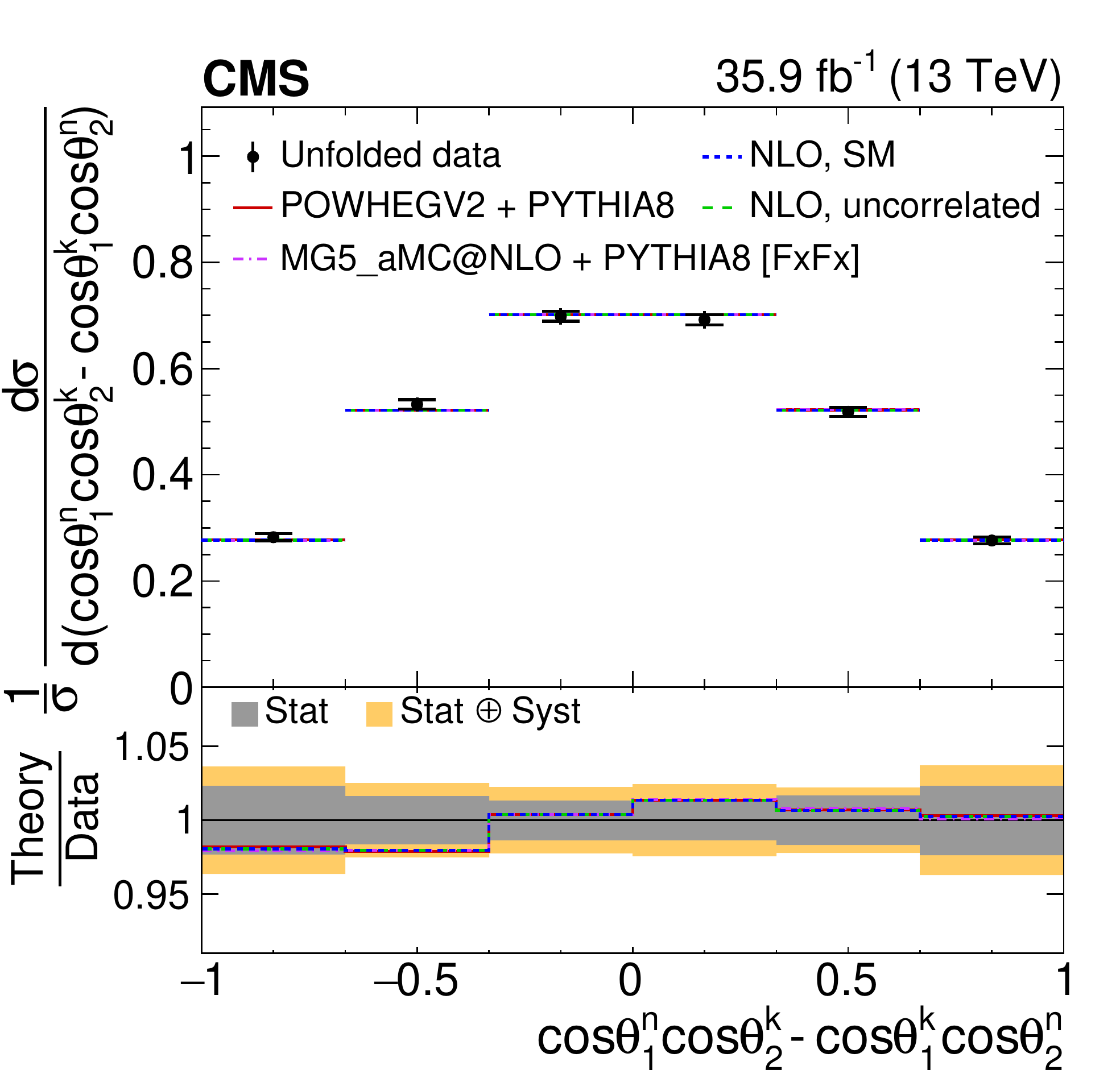} \\
\caption{\label{fig:UnfoldedNormXsecC2}\protect
Unfolded data (points) and predicted (horizontal lines) normalized differential cross sections for the cross spin correlation observables. The vertical lines on the points represent the total uncertainties, with the statistical components indicated by horizontal bars.
The ratios of various predictions to the data are shown in the lower panels.
}
\end{figure*}

Normalized differential cross sections at the parton level are measured for the 22 observables introduced in Section~\ref{sec:observables}.
For the top quark polarization observables measured using the nominal ($\hat{k}$, $\hat{r}$ and $\hat{n}$) and modified ($\hat{k}^*$ and $\hat{r}^*$) reference axes, the results are shown in Figs.~\ref{fig:UnfoldedNormXsecB} and~\ref{fig:UnfoldedNormXsecB2}, respectively.
The results for the observables that probe the diagonal \ttbar\ spin correlation coefficients and for
the laboratory-frame spin correlation observables are shown in Fig.~\ref{fig:UnfoldedNormXsecC}.
For the cross spin correlation observables, the results are shown in Fig.~\ref{fig:UnfoldedNormXsecC2}.
The measured distributions are compared with predictions from the \Powhegvtwo\ and \MGvATNLO\ simulations and with calculations for \ttbar\ production at NLO in QCD with EW corrections~\cite{Bernreuther:2013aga,Bernreuther:2015yna}, as well as similar calculations in the absence of top quark polarization or spin correlations.
For the observables measured in the top quark rest frame, the latter are equivalent to the predictions of Eqs.~(\ref{eq:theodist1})--(\ref{eq:theodist3}), with the coefficients set to zero.
For the laboratory-frame observables, dedicated calculations were made using the computational setup described in Refs.~\cite{Bernreuther2010,Bernreuther:2013aga}.
In addition, the only NNLO QCD prediction~\cite{Behring:2019iiv} is shown for the $\dphi$ distribution in Fig.~\ref{fig:UnfoldedNormXsecC}.

The effect of spin correlations is most clearly visible in the $\cosphi$ distribution in Fig.~\ref{fig:UnfoldedNormXsecC},
where the data strongly favor the predictions with spin correlations compared to the uncorrelated prediction.
The presence of spin correlations can also be seen in the other distributions sensitive to P- and CP-even spin correlations:
the three $\cos\theta_{1}^{i}\cos\theta_{2}^{i}$ distributions and
the two laboratory-frame distributions ($\cosphilab$ and $\dphi$) in Fig.~\ref{fig:UnfoldedNormXsecC},
and the $\cosPrk$ distribution in Fig.~\ref{fig:UnfoldedNormXsecC2}.
However, the measurements are not sensitive to the small level of top quark polarization predicted in the SM, and do not significantly disfavor the unpolarized predictions in Figs.~\ref{fig:UnfoldedNormXsecB} and~\ref{fig:UnfoldedNormXsecB2}.

The statistical and systematic correlation matrices for the normalized differential cross sections are determined simultaneously for all 132 measured bins to allow the fitting of multiple distributions,
and are shown in Fig.~\ref{fig:StatCorrMatrixAllVarsNorm}.
The statistical correlations are estimated using a bootstrap resampling of the data~\cite{10.2307/2958830}, and the systematic correlations are estimated by simultaneously evaluating the systematic variations described in Section~\ref{sec:errors} for all measured bins.
The statistical correlations among bins from the same distribution exhibit a typical pattern of correlation and anticorrelation arising from the unfolding.
The statistical correlations between bins from different distributions are typically small,
but the relationships between some of the distributions result in stronger correlations (for example, $\cos\theta_{1,2}^{i*}$ and $\cos\theta_{1,2}^{i}$ are the same up to a sign).
The systematic correlations are in general much stronger,
and the pattern of positive and negative correlations reflects the relative changes in shape of the different distributions in response to the systematic variations.

\ifthenelse{\boolean{cms@external}}
{
\begin{figure*}[!htpb]
\centering
\includegraphics[width=0.49\linewidth]{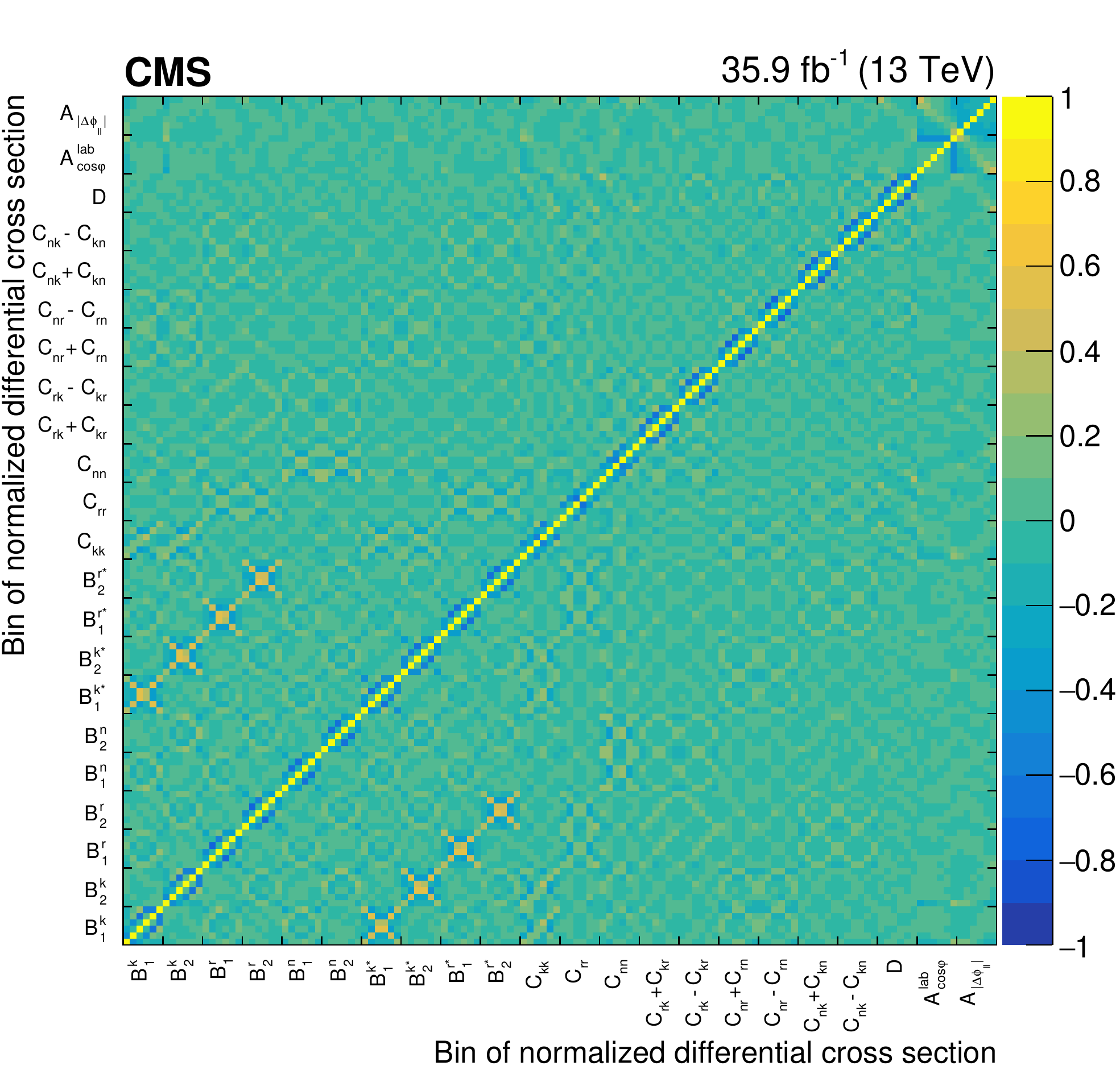}
\hfill
\includegraphics[width=0.49\linewidth]{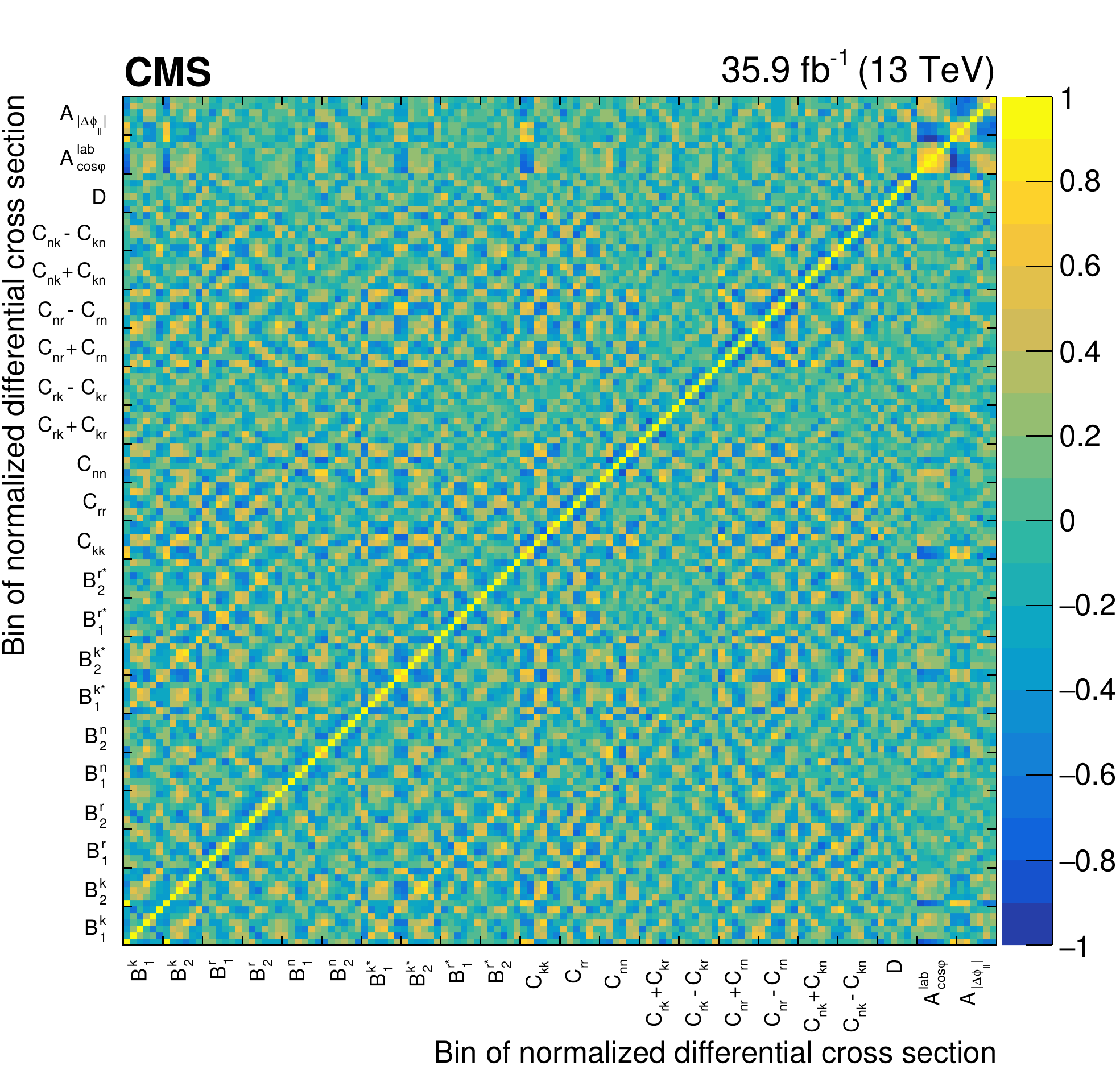}
\caption{\label{fig:StatCorrMatrixAllVarsNorm}\protect
Values (gray scale) of the total statistical (\cmsLeft) and systematic (\cmsRight) correlation matrices for all
measured bins of the normalized differential cross sections.
Each group of six bins along each axis corresponds to a measured distribution, and for conciseness is labeled by the name of the associated coefficient (as defined in Table~\ref{tab:obscoef}).
}
\end{figure*}
}
{
\begin{figure*}[!htpb]
\centering
\vspace{-0.4cm}
\includegraphics[width=0.72\linewidth]{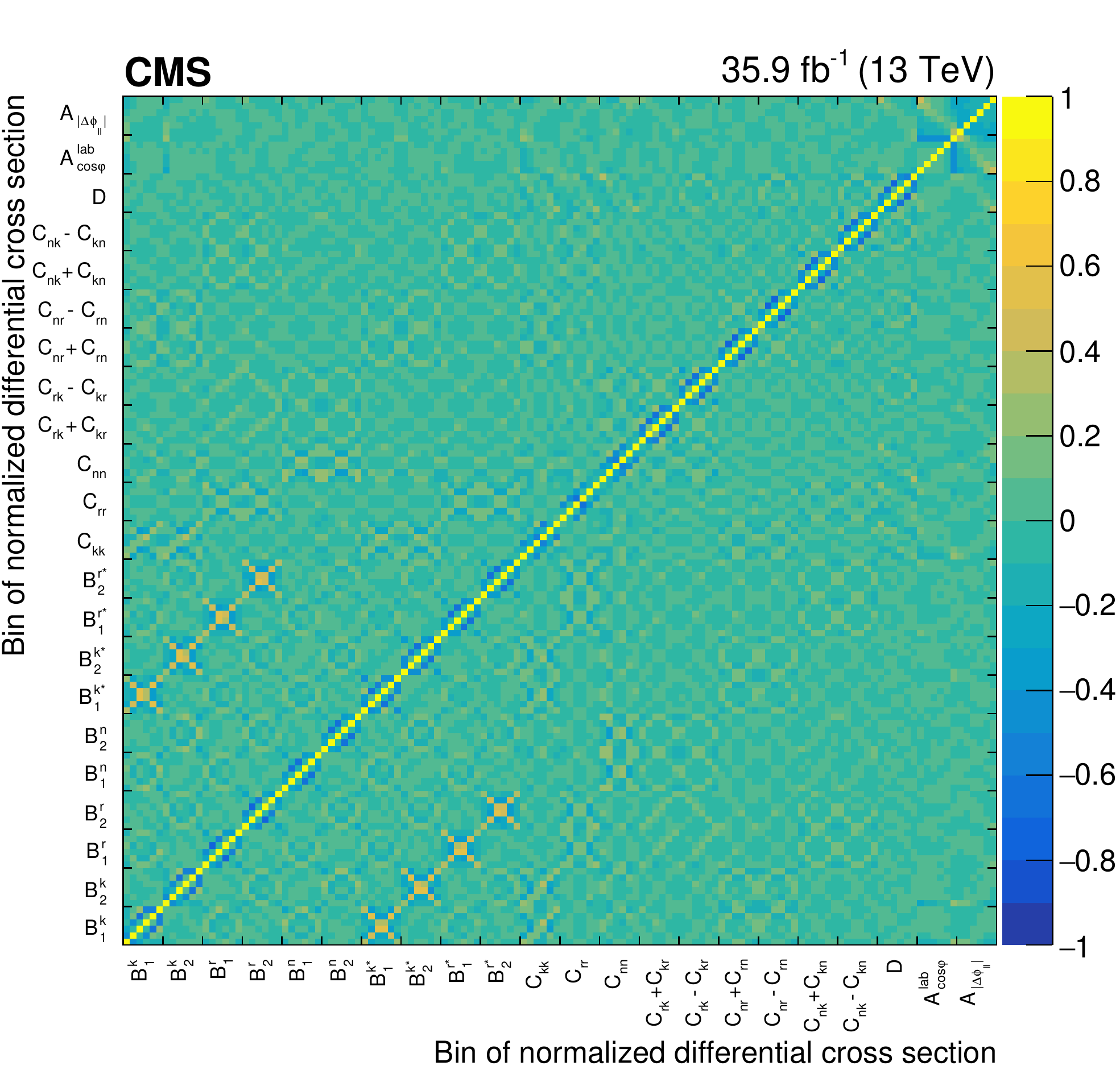}
\includegraphics[width=0.72\linewidth]{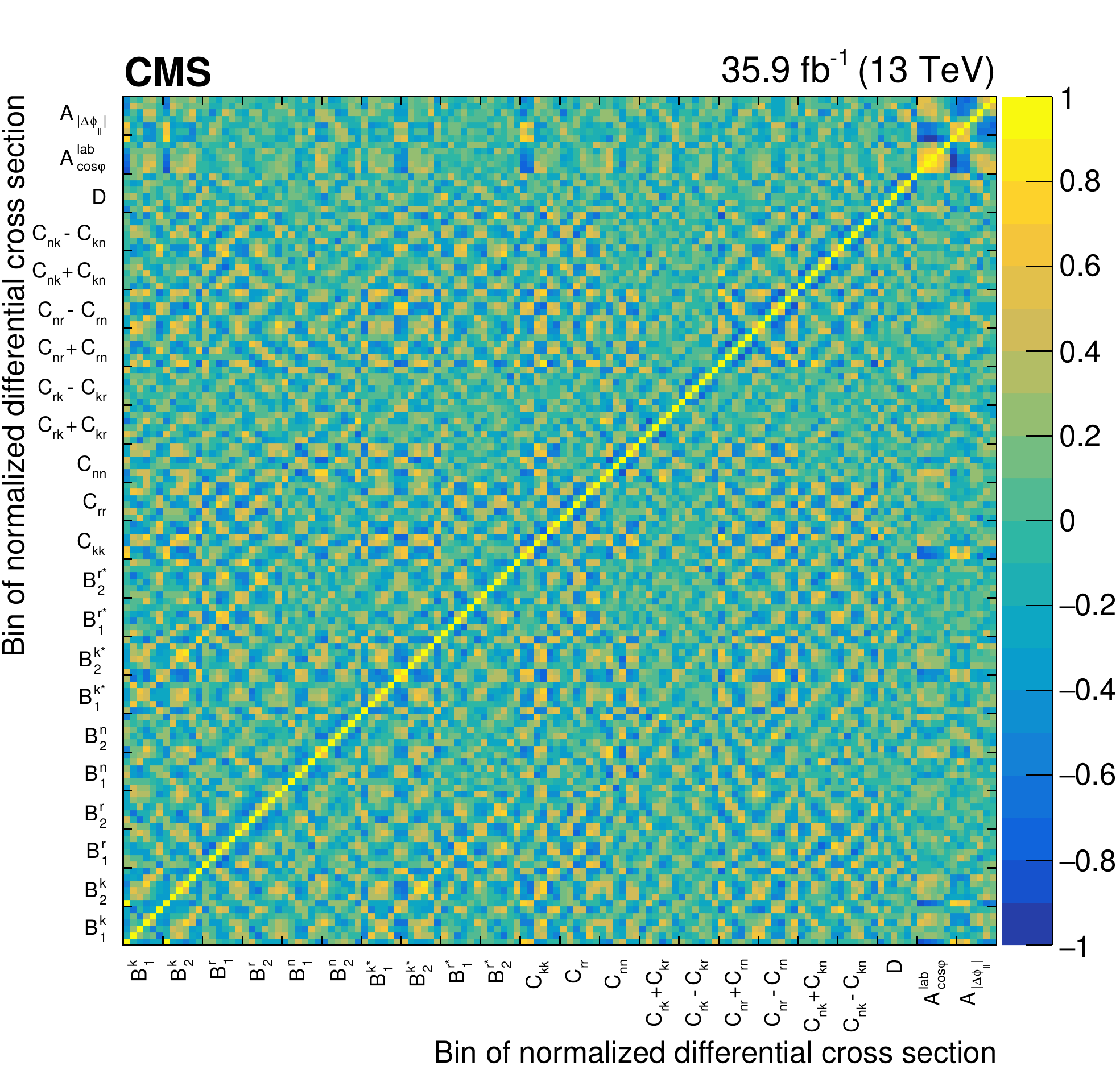}
\caption{\label{fig:StatCorrMatrixAllVarsNorm}\protect
Values (gray scale) of the total statistical (\cmsLeft) and systematic (\cmsRight) correlation matrices for all
measured bins of the normalized differential cross sections.
Each group of six bins along each axis corresponds to a measured distribution, and for conciseness is labeled by the name of the associated coefficient (as defined in Table~\ref{tab:obscoef}).
}
\end{figure*}
}

\ifthenelse{\boolean{cms@external}}{}{\pagebreak}

The agreement between the measured distributions and the four theoretical predictions shown in Figs.~\ref{fig:UnfoldedNormXsecB}--\ref{fig:UnfoldedNormXsecC2} is quantified by evaluating the $\chi^2$, taking the uncertainties from the sum of the measured statistical and systematic covariance matrices (and not including any uncertainties in the prediction). The results are shown in Table~\ref{tab:chi2Total}.
For the observables measured in the top quark rest frame, there is generally good agreement between the measured distributions and all the predictions in the presence of spin correlations.
For the two observables measured in the laboratory frame, there is greater variation between the predictions.
The \Powhegvtwo\ prediction best describes the data for $\cosphilab$, while for $\dphi$ the \MGvATNLO\ prediction provides the best agreement.
The NNLO QCD prediction shown in Fig.~\ref{fig:UnfoldedNormXsecC} also describes the observed $\dphi$ distribution well, with a $\chi^{2} / \mathrm{dof}$ of $4.3/5$,
where dof is the number of degrees of freedom.

\ifthenelse{\boolean{cms@external}}
{\begin{table*}[!htpb]
\centering
 \topcaption{\label{tab:chi2Total} The $\chi^2$ between the data and the predictions for all measured normalized differential cross sections (Figs.~\ref{fig:UnfoldedNormXsecB}--\ref{fig:UnfoldedNormXsecC2}). The $\chi^2$ values are evaluated using the sum of the measured statistical and systematic covariance matrices. The number of degrees of freedom (dof) is 5 for all observables. In the last row, the $\chi^2$ values are given for the set of all measured bins.}
 \begin{scotch}{l Y{4.0} Y{4.0} Y{4.0} Y{4.0} }
  \multirow{ 2}{*}{{Observable}}  & \multicolumn{4}{c}{{${\chi^2}$ between data and prediction (dof = 5)} } \\ [0.3ex]
  										 & \multicolumn{1}{c}{\Powhegvtwo} & \multicolumn{1}{c}{\MGaMCatNLO} & \multicolumn{1}{c}{NLO calculation} & \multicolumn{1}{c}{No spin correlation/polarization} \\ [0.3ex]
}
{\begin{table*}[!htpb]
\centering
 \topcaption{\label{tab:chi2Total} The $\chi^2$ between the data and the predictions for all measured normalized differential cross sections (Figs.~\ref{fig:UnfoldedNormXsecB}--\ref{fig:UnfoldedNormXsecC2}). The last column refers to the prediction in the case of no spin correlation (SC) or polarization (pol). The $\chi^2$ values are evaluated using the sum of the measured statistical and systematic covariance matrices. The number of degrees of freedom (dof) is 5 for all observables. In the last row, the $\chi^2$ values are given for the set of all measured bins.}
 \begin{scotch}{l Y{4.0} Y{4.0} Y{4.0} Y{4.0} }
 \Trule
  \multirow{ 2}{*}{{Observable}}  & \multicolumn{4}{c}{{${\chi^2}$ between data and prediction (dof = 5)} } \\ [0.3ex]
  										 & \multicolumn{1}{c}{\Powhegvtwo} & \multicolumn{1}{c}{\MGaMCatNLO} & \multicolumn{1}{c}{NLO calculation} & \multicolumn{1}{c}{No SC/pol} \\ [0.3ex]
}
 \hline
 \Trule
$\cospk$  & $1.1$ & $1.0$ & $1.1$ & $1.1$   \\ [0.3ex]
$\cosmk$  & $5.2$ & $5.0$ & $5.2$ & $5.2$   \\ [0.3ex]
$\cospr$  & $4.3$ & $4.4$ & $4.2$ & $3.9$   \\ [0.3ex]
$\cosmr$  & $0.7$ & $0.5$ & $0.6$ & $0.5$   \\ [0.3ex]
$\cospn$  & $1.9$ & $1.8$ & $1.8$ & $1.9$   \\ [0.3ex]
$\cosmn$  & $3.2$ & $3.1$ & $2.1$ & $3.1$   \\ [0.3ex]
$\cospks$  & $1.3$ & $1.3$ & $1.4$ & $1.3$   \\ [0.3ex]
$\cosmks$  & $1.8$ & $1.6$ & $1.7$ & $1.8$   \\ [0.3ex]
$\cosprs$  & $1.5$ & $1.5$ & $1.6$ & $1.6$   \\ [0.3ex]
$\cosmrs$  & $0.5$ & $0.6$ & $0.6$ & $0.6$   \\ [0.3ex]
$\coskk$  & $3.1$ & $3.2$ & $3.5$ & $66.7$   \\ [0.3ex]
$\cosrr$  & $2.0$ & $1.7$ & $1.1$ & $7.4$   \\ [0.3ex]
$\cosnn$  & $0.6$ & $0.3$ & $0.3$ & $267.0$   \\ [0.3ex]
$\cosPrk$  & $1.5$ & $1.6$ & $1.7$ & $12.3$   \\ [0.3ex]
$\cosMrk$  & $3.6$ & $3.1$ & $3.6$ & $3.6$   \\ [0.3ex]
$\cosPnr$  & $1.7$ & $1.7$ & $1.8$ & $1.8$   \\ [0.3ex]
$\cosMnr$  & $1.8$ & $1.9$ & $1.9$ & $1.9$   \\ [0.3ex]
$\cosPnk$  & $3.8$ & $4.0$ & $4.0$ & $3.9$   \\ [0.3ex]
$\cosMnk$  & $2.3$ & $2.4$ & $2.2$ & $2.2$   \\ [0.3ex]
$\cosphi$  & $1.5$ & $0.7$ & $1.4$ & $496.2$   \\ [0.3ex]
$\cosphilab$  & $3.9$ & $7.6$ & $7.0$ & $66.5$   \\ [0.3ex]
$\dphi$  & $10.8$ & $4.0$ & $9.2$ & $190.4$   \\ [\cmsTabSkip]
All (dof = 110) & $88.4$ & $89.7$ & $88.6$ & $2119.8$  \\ [0.1ex]
 \end{scotch}
 \end{table*}

\ifthenelse{\boolean{cms@external}}{}{\pagebreak}

\subsection{Coefficients}
\label{sec:coefficients}

From each measured normalized differential cross section, we extract the corresponding coefficient, 
using the functional forms of Eqs.~(\ref{eq:theodist1})--(\ref{eq:theodist3}) and combining the information from the measured bins in a way that minimizes the uncertainty in the coefficient.
For the laboratory-frame observables, the shapes of the distributions are instead quantified by the asymmetries defined in Eq.~(\ref{eq:asym}).
The results for all quantities are shown with their total uncertainties in Table~\ref{tab:coefficients},
where they are compared with predictions from \Powhegvtwo\ and \MGvATNLO simulations and the NLO calculations~\cite{Bernreuther:2013aga,Bernreuther:2015yna}.
The uncertainties in the NLO calculations come from varying $\mu_\mathrm{R}$ and $\mu_\mathrm{F}$ simultaneously up and down by a factor of 2 from their nominal value of $\mt$.
The NNLO QCD prediction for $\Adphi$ is $0.115\,^{+0.005}_{-0.001}$~\cite{Behring:2019iiv}, where the uncertainties are taken from the largest deviations when varying $\mu_\mathrm{R}$ and $\mu_\mathrm{F}$ individually and simultaneously up and down by a factor of 2 from the nominal choice of $(\mT^{\cPqt}+\mT^{\cPaqt})/4$.
The results are also shown in Figs.~\ref{fig:summaryB}--\ref{fig:summaryCcross}.

\begin{table*}[!htpb]
\centering
\topcaption{\label{tab:coefficients} Measured coefficients and asymmetries and their total uncertainties. Predicted values from simulation are quoted with a combination of statistical and scale uncertainties, while the 
NLO calculated values are quoted with their scale uncertainties~\cite{Bernreuther:2013aga,Bernreuther:2015yna}.
The NNLO QCD prediction for $\Adphi$, with scale uncertainties, is $0.115\,^{+0.005}_{-0.001}$~\cite{Behring:2019iiv}.
}

\begin{scotch}{l  X{6.6} r r r }
{Coefficient} & \multicolumn{1}{c}{{Measured}} & {\Powhegvtwo} & {\MGaMCatNLO} & {NLO calculation}  \\
\hline
\Trule
$\bpk$ & 0.005 , 0.023 & $0.004\,^{+0.001}_{-0.001}$ & $0.000\,^{+0.001}_{-0.001}\:\:\:\:\:\:$  &  $4.0\,^{+1.7}_{-1.2} \times 10^{-3} \:\:$  \\ [0.3ex]  
$\bmk$ & 0.007 , 0.023 & $0.006\,^{+0.001}_{-0.001}$ &  $-0.002\,^{+0.001}_{-0.001}\:\:\:\:\:\:$  &  $4.0\,^{+1.7}_{-1.2} \times 10^{-3} \:\:$  \\ [0.3ex]  
$\bpr$ & -0.023 , 0.017 & $0.002\,^{+0.001}_{-0.001}$ & $0.002\,^{+0.001}_{-0.001}\:\:\:\:\:\:$  &  $1.6\,^{+1.2}_{-0.9} \times 10^{-3} \:\:$  \\ [0.3ex]  
$\bmr$ & -0.010 , 0.020 & $0.003\,^{+0.001}_{-0.001}$ & $0.000\,^{+0.001}_{-0.001}\:\:\:\:\:\:$  &  $1.6\,^{+1.2}_{-0.9} \times 10^{-3} \:\:$  \\ [0.3ex]  
$\bpn$ & 0.006 , 0.013 & $-0.001\,^{+0.001}_{-0.001}$ & $0.001\,^{+0.001}_{-0.001}\:\:\:\:\:\:$  &  $5.7\,^{+0.5}_{-0.4} \times 10^{-3} \:\:$  \\ [0.3ex]  
$\bmn$ & 0.017 , 0.013 & $-0.001\,^{+0.001}_{-0.001}$ & $0.000\,^{+0.001}_{-0.001}\:\:\:\:\:\:$  &  $5.7\,^{+0.5}_{-0.4} \times 10^{-3} \:\:$  \\ [0.3ex]  
$\bpks$ & -0.016 , 0.018 & $-0.001\,^{+0.001}_{-0.001}$ & $0.000\,^{+0.001}_{-0.001}\:\:\:\:\:\:$  &  \multicolumn{1}{c}{${<}10^{-3}$}  \\ [0.3ex]  
$\bmks$ & 0.007 , 0.019 & $0.001\,^{+0.001}_{-0.001}$ & $0.003\,^{+0.002}_{-0.001}\:\:\:\:\:\:$  &  \multicolumn{1}{c}{${<}10^{-3}$}  \\ [0.3ex]  
$\bprs$ & 0.001 , 0.017 & $0.000\,^{+0.001}_{-0.001}$ & $0.000\,^{+0.001}_{-0.001}\:\:\:\:\:\:$  &  \multicolumn{1}{c}{${<}10^{-3}$}  \\ [0.3ex]  
$\bmrs$ & 0.010 , 0.017 & $0.001\,^{+0.001}_{-0.001}$ & $0.001\,^{+0.001}_{-0.001}\:\:\:\:\:\:$  &  \multicolumn{1}{c}{${<}10^{-3}$}  \\ [0.3ex]  
$\ckk$ & 0.300 , 0.038 & $0.314\,^{+0.005}_{-0.004}$ & $0.325\,^{+0.011}_{-0.006}\:\:\:\:\:\:$  &  $0.331\,^{+0.002}_{-0.002} \:\:\:\:\:\:$  \\ [0.3ex]  
$\crr$ & 0.081 , 0.032 & $0.048\,^{+0.007}_{-0.006}$ & $0.052\,^{+0.007}_{-0.005}\:\:\:\:\:\:$  &  $0.071\,^{+0.008}_{-0.006} \:\:\:\:\:\:$  \\ [0.3ex]  
$\cnn$ & 0.329 , 0.020 & $0.317\,^{+0.001}_{-0.001}$ & $0.324\,^{+0.002}_{-0.002}\:\:\:\:\:\:$  &  $0.326\,^{+0.002}_{-0.002} \:\:\:\:\:\:$  \\ [0.3ex]  
$\cPrk$ & -0.193 , 0.064 & $-0.201\,^{+0.004}_{-0.003}$ &  $-0.198\,^{+0.004}_{-0.005}\:\:\:\:\:\:$  &  $-0.206\,^{+0.002}_{-0.002} \:\:\:\:\:\:$  \\ [0.3ex]  
$\cMrk$ & 0.057 , 0.046 & $-0.001\,^{+0.002}_{-0.002}$ & $0.004\,^{+0.002}_{-0.002}\:\:\:\:\:\:$  &  \multicolumn{1}{c}{$0$}  \\ [0.3ex]  
$\cPnr$ & -0.004 , 0.037 & $-0.003\,^{+0.002}_{-0.002}$ & $0.001\,^{+0.002}_{-0.002}\:\:\:\:\:\:$  &  $1.06\,^{+0.01}_{-0.01} \times 10^{-3}$  \\ [0.3ex]  
$\cMnr$ & -0.001 , 0.038 & $0.002\,^{+0.002}_{-0.002}$ & $0.001\,^{+0.003}_{-0.002}\:\:\:\:\:\:$  &  \multicolumn{1}{c}{$0$}  \\ [0.3ex]  
$\cPnk$ & -0.043 , 0.041 & $-0.002\,^{+0.002}_{-0.002}$ & $0.003\,^{+0.002}_{-0.002}\:\:\:\:\:\:$  &  $2.15\,^{+0.04}_{-0.07} \times 10^{-3}$  \\ [0.3ex]  
$\cMnk$ & 0.040 , 0.029 & $-0.001\,^{+0.002}_{-0.002}$ &  $-0.001\,^{+0.002}_{-0.002}\:\:\:\:\:\:$  &  \multicolumn{1}{c}{$0$}  \\ [0.3ex]  
$D$ & -0.237 , 0.011 & $-0.226\,^{+0.003}_{-0.004}$ &  $-0.233\,^{+0.004}_{-0.006}\:\:\:\:\:\:$  &  $-0.243\,^{+0.003}_{-0.003} \:\:\:\:\:\:$  \\ [0.3ex]  
$\Aphilab$ & 0.167 , 0.010 & $0.161\,^{+0.002}_{-0.002}$ & $0.174\,^{+0.004}_{-0.003}\:\:\:\:\:\:$  &  $0.181\,^{+0.004}_{-0.003} \:\:\:\:\:\:$ \\ [0.3ex]  
$\Adphi$ & 0.103 , 0.008 & $0.125\,^{+0.004}_{-0.005}$ & $0.115\,^{+0.003}_{-0.005}\:\:\:\:\:\:$  &  $0.108\,^{+0.009}_{-0.012} \:\:\:\:\:\:$ \\ [0.4ex]
\end{scotch}
\end{table*}

\begin{figure}[!htpb]
\centering
\includegraphics[width=0.99\linewidth]{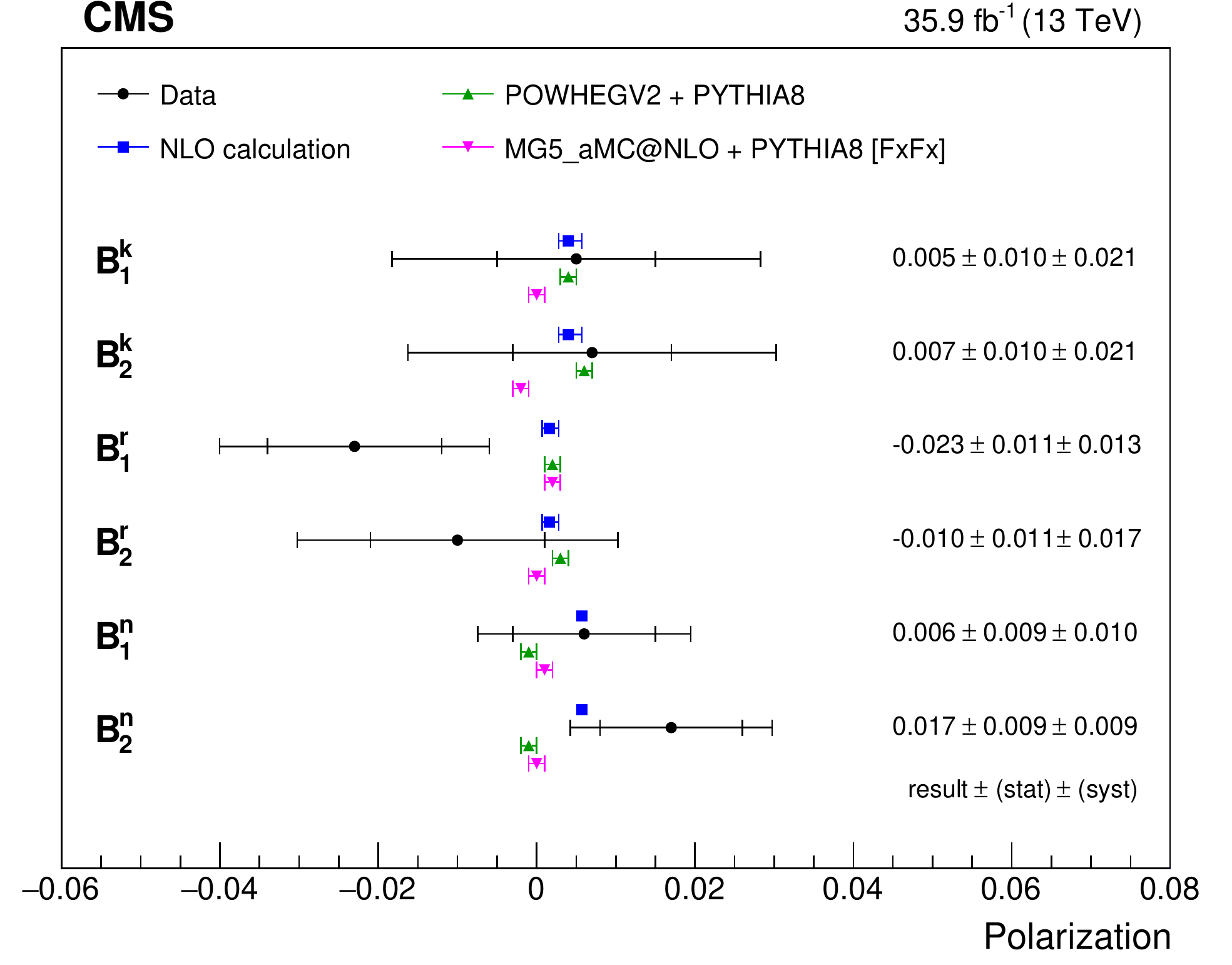}
\caption{\label{fig:summaryB}\protect
Measured values of the polarization coefficients (circles) and the predictions from \Powhegvtwo\ (triangles), \MGvATNLO\ (inverted triangles), and the NLO calculation~\cite{Bernreuther:2015yna} (squares). The inner vertical bars on the circles give the statistical uncertainty in the data and the outer bars the total uncertainty. The numerical measured values with their statistical and systematic uncertainties are given on the right. The vertical bars on the values from simulation represent the combination of statistical and scale uncertainties, while for the calculated values they represent the scale uncertainties.
}
\end{figure}

\begin{figure}[!htpb]
\centering
\includegraphics[width=0.99\linewidth]{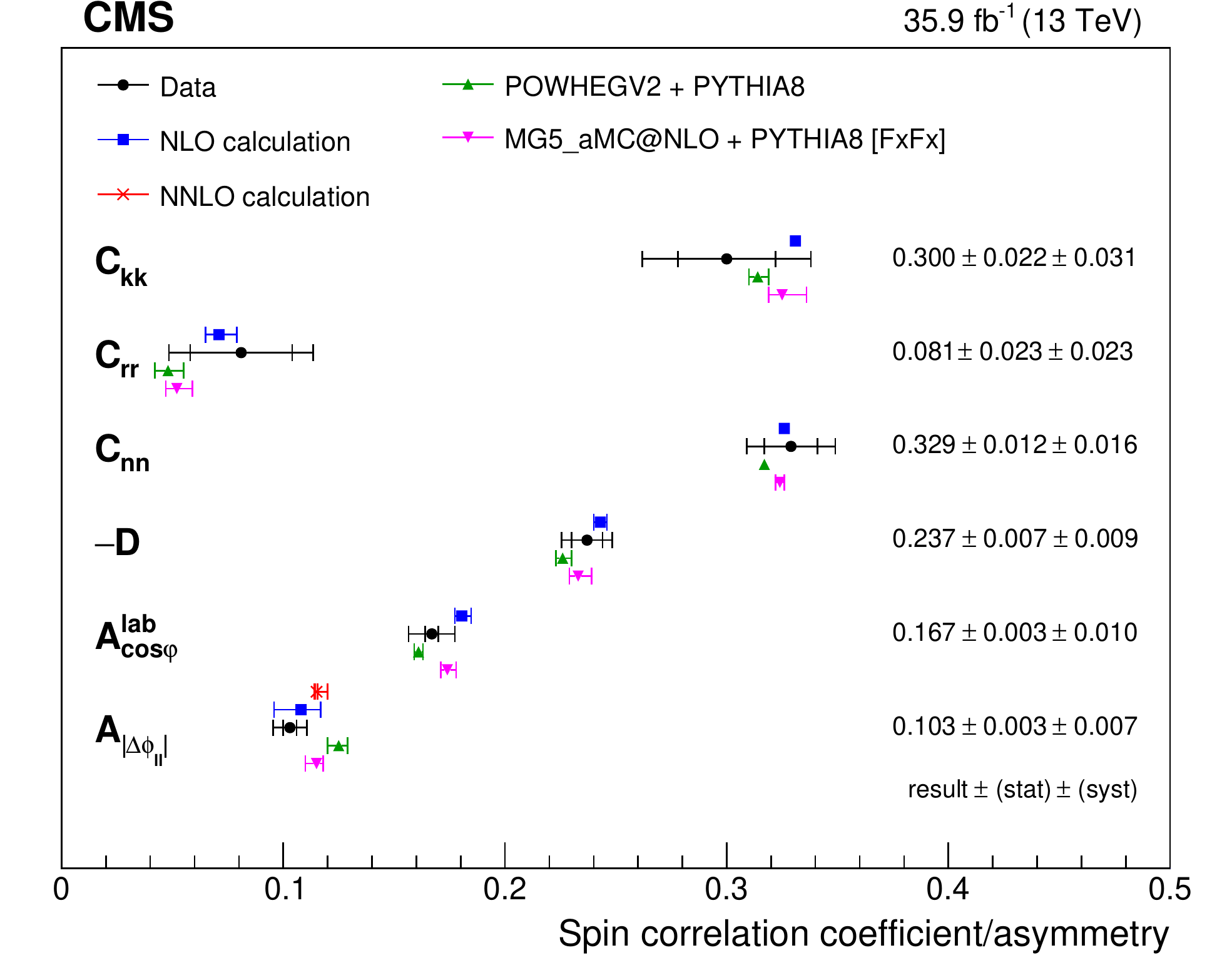}
\caption{\label{fig:summaryC}\protect
Measured values of the spin correlation coefficients and asymmetries (circles) and the predictions from \Powhegvtwo\ (triangles), \MGvATNLO\ (inverted triangles), the NLO calculation~\cite{Bernreuther:2013aga,Bernreuther:2015yna} (squares), and the NNLO calculation~\cite{Behring:2019iiv} (cross). The inner vertical bars on the circles give the statistical uncertainty in the data and the outer bars the total uncertainty. The numerical measured values with their statistical and systematic uncertainties are given on the right. The vertical bars on the values from simulation represent the combination of statistical and scale uncertainties, while for the calculated values they represent the scale uncertainties.
}
\end{figure}

\begin{figure}[!htpb]
\centering
\includegraphics[width=0.99\linewidth]{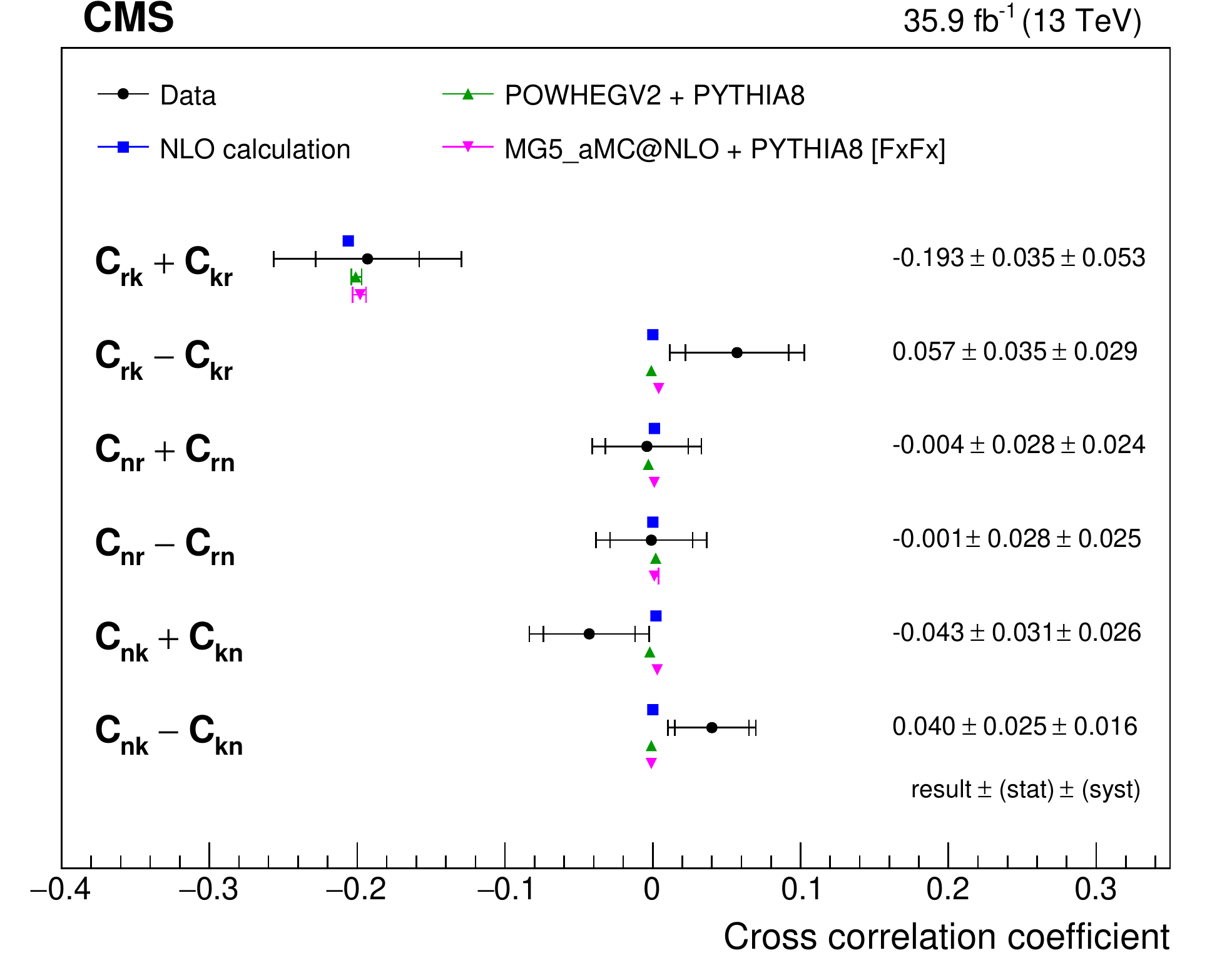}
\caption{\label{fig:summaryCcross}\protect
Measured values of the cross spin correlation coefficients (circles) and the predictions from \Powhegvtwo\ (triangles), \MGvATNLO\ (inverted triangles), and the NLO calculation~\cite{Bernreuther:2015yna} (squares). The inner vertical bars on the circles give the statistical uncertainty in the data and the outer bars the total uncertainty. The numerical measured values with their statistical and systematic uncertainties are given on the right. The vertical bars on the values from simulation represent the combination of statistical and scale uncertainties, while for the calculated values they represent the scale uncertainties.
}
\end{figure}

There is good agreement between the measured coefficients and all the SM predictions,
while substantial variation is seen in the predicted laboratory-frame asymmetries, 
which have sizable scale uncertainties.
In the fixed-order calculations of Refs.~\cite{Bernreuther:2013aga,Bernreuther:2015yna},
the numerator and denominator of the normalized differential cross section are computed
at NLO QCD including EW corrections, and then the
ratio is consistently expanded to NLO. On the other hand, in the
computations from simulation the ratio is not expanded. This leads
to differences that are nominally of order $\alpS^2$, in addition to the
EW corrections (which are not included in the simulation).
However, for $\Adphi$ the EW corrections are found to be
only at the level of $2\%$
using the computational setup of Ref.~\cite{Bernreuther:2013aga}.
In the NNLO QCD calculation, the ratio is not expanded~\cite{Behring:2019iiv}.

The breakdown of the systematic and statistical uncertainties in the polarization and spin correlation measurements is given in Tables~\ref{tab:systematicsTotalBcombined} and~\ref{tab:systematicsTotalCcombined}, respectively.
The systematic and statistical uncertainties are of comparable size for most of the measured coefficients. The exception is $B_{1,2}^{k}$, because the JES and \cPqb~quark fragmentation uncertainties have a large effect on the reconstructed top quark momentum in the \ttbar CM frame. The laboratory-frame asymmetries have statistical uncertainties smaller than their systematic uncertainties. The excellent reconstruction resolution results in little dilution of the statistical precision of the measured asymmetries. There is also a large background uncertainty in $\Aphilab$, owing to the large \Zjets contribution near $\cosphilab = 1$, and a large top quark \pt modeling uncertainty in $\Adphi$.

\ifthenelse{\boolean{cms@external}}
{
\begin{table*}[!htpb]
}
{
\begin{sidewaystable*}
}
\centering

    \topcaption{
    Summary of the systematic, statistical, and total uncertainties in the extracted top quark polarization coefficients. \adash (\NA) is shown where the values are ${<}0.0005$.
    }
    \label{tab:systematicsTotalBcombined}

\begin{scotch}{lcccccccccc}
{Source} & \multicolumn{10}{c}{{Uncertainty}} \\
 & $\bpk$ & $\bmk$ & $\bpr$ & $\bmr$ & $\bpn$ & $\bmn$ & $\bpks$ & $\bmks$ & $\bprs$ & $\bmrs$ \\ [0.3ex]
\hline
Trigger  & 0.001 & 0.001 & 0.001 & 0.001 & \NA & \NA & 0.001 & 0.001 & 0.002 & 0.002 \\
Lepton ident./isolation  & 0.001 & \NA & \NA & \NA & \NA & \NA & \NA & \NA & \NA & \NA \\
Kinematic reconstruction  & \NA & \NA & \NA & \NA & \NA & \NA & \NA & \NA & \NA & \NA \\
Pileup  & \NA & \NA & 0.002 & 0.002 & \NA & 0.001 & 0.001 & 0.001 & \NA & \NA \\
\cPqb~tagging  & 0.003 & 0.004 & 0.003 & 0.003 & \NA & \NA & 0.002 & 0.002 & 0.001 & 0.001 \\
JES  & 0.011 & 0.012 & 0.007 & 0.009 & 0.003 & 0.003 & 0.009 & 0.008 & 0.007 & 0.007 \\
Unclustered energy  & 0.001 & 0.002 & 0.001 & 0.001 & \NA & 0.001 & 0.001 & \NA & 0.001 & 0.002 \\
JER  & 0.001 & 0.002 & 0.001 & 0.001 & 0.001 & 0.001 & \NA & 0.001 & 0.001 & 0.001 \\[\cmsTabSkip]
Scales  & 0.005 & 0.004 & 0.004 & 0.009 & 0.003 & 0.004 & 0.003 & 0.004 & 0.006 & 0.005 \\
ME/PS matching  & 0.006 & 0.006 & 0.004 & 0.001 & 0.003 & 0.004 & 0.003 & 0.003 & 0.004 & 0.004 \\
Color reconnection  & 0.005 & 0.003 & 0.003 & 0.004 & 0.008 & 0.005 & 0.006 & 0.008 & 0.006 & 0.008 \\
Underlying event  & 0.001 & 0.003 & 0.001 & 0.003 & 0.002 & 0.003 & 0.003 & 0.002 & 0.004 & 0.004 \\
\cPqb~quark fragmentation  & 0.009 & 0.009 & 0.004 & 0.005 & \NA & 0.001 & 0.001 & 0.001 & 0.001 & 0.001 \\
\cPqb~hadron semilept. decays  & 0.001 & 0.001 & \NA & \NA & \NA & \NA & \NA & \NA & \NA & \NA \\
PDF  & 0.001 & 0.001 & \NA & \NA & \NA & \NA & 0.001 & 0.001 & 0.001 & 0.001 \\
Top quark mass  & 0.007 & 0.007 & \NA & 0.001 & 0.001 & 0.002 & 0.002 & 0.001 & 0.002 & 0.002 \\
Top quark \pt  & 0.003 & 0.004 & 0.001 & 0.001 & \NA & \NA & 0.001 & 0.001 & \NA & \NA \\
Background  & 0.008 & 0.008 & 0.005 & 0.008 & 0.001 & 0.001 & 0.004 & 0.005 & 0.002 & 0.002 \\[\cmsTabSkip]
Total systematic  & 0.021 & 0.021 & 0.013 & 0.017 & 0.010 & 0.009 & 0.013 & 0.014 & 0.013 & 0.013 \\[\cmsTabSkip]
Data statistical  & 0.009 & 0.008 & 0.009 & 0.009 & 0.007 & 0.008 & 0.010 & 0.010 & 0.010 & 0.009 \\
Signal simulation statistical  & 0.003 & 0.003 & 0.003 & 0.003 & 0.003 & 0.003 & 0.004 & 0.004 & 0.004 & 0.003 \\
Background sim. statistical  & 0.005 & 0.005 & 0.005 & 0.005 & 0.004 & 0.004 & 0.006 & 0.006 & 0.005 & 0.005 \\[\cmsTabSkip]
Total statistical  & 0.010 & 0.010 & 0.011 & 0.011 & 0.009 & 0.009 & 0.012 & 0.012 & 0.012 & 0.011 \\[\cmsTabSkip]
Total  & 0.023 & 0.023 & 0.017 & 0.020 & 0.013 & 0.013 & 0.018 & 0.019 & 0.017 & 0.017 \\
\end{scotch}

\ifthenelse{\boolean{cms@external}}
{
\end{table*}
}
{
\end{sidewaystable*}
}

\ifthenelse{\boolean{cms@external}}
{
\begin{table*}[!htpb]
\centering

    \topcaption{
    Summary of the systematic, statistical, and total uncertainties in the extracted \ttbar spin correlation coefficients and asymmetries. \adash (\NA) is shown where the values are ${<}0.0005$.
    }
    \label{tab:systematicsTotalCcombined}

\begin{scotch}{lcccccccccccc}
{Source} & \multicolumn{12}{c}{{Uncertainty}} \\
 & $\ckk$ & $\crr$ & $\cnn$ & $\cPrk$ & $\cMrk$ & $\cPnr$ & $\cMnr$ & $\cPnk$ & $\cMnk$ & $D$ & $\Aphilab$ & $\Adphi$ \\ [0.3ex]
\hline
Trigger  & 0.001 & 0.001 & \NA & 0.002 & \NA & \NA & \NA & \NA & \NA & \NA & 0.001 & \NA \\
Lepton ident./iso.  & 0.001 & 0.001 & \NA & 0.001 & \NA & \NA & \NA & \NA & \NA & \NA & \NA & \NA \\
Kinematic reco.  & \NA & \NA & \NA & \NA & \NA & \NA & \NA & \NA & \NA & \NA & \NA & \NA \\
Pileup  & 0.002 & \NA & 0.001 & 0.004 & 0.001 & 0.001 & 0.002 & 0.001 & 0.001 & 0.001 & \NA & 0.001 \\
\cPqb~tagging  & 0.004 & 0.001 & 0.002 & 0.005 & 0.001 & 0.001 & 0.001 & 0.001 & 0.001 & 0.001 & \NA & \NA \\
JES  & 0.012 & 0.009 & 0.005 & 0.022 & 0.011 & 0.011 & 0.009 & 0.012 & 0.007 & 0.002 & \NA & 0.001 \\
Unclust. energy  & 0.001 & 0.001 & 0.001 & 0.004 & 0.001 & 0.001 & 0.002 & 0.001 & 0.001 & \NA & \NA & 0.001 \\
JER  & 0.001 & 0.002 & 0.001 & 0.004 & 0.002 & 0.001 & 0.001 & 0.003 & 0.001 & \NA & \NA & \NA \\[\cmsTabSkip]
Scales  & 0.012 & 0.006 & 0.007 & 0.026 & 0.011 & 0.007 & 0.014 & 0.011 & 0.007 & 0.003 & 0.002 & 0.003 \\
ME/PS matching  & 0.004 & 0.003 & 0.001 & 0.009 & 0.016 & 0.011 & 0.001 & 0.012 & 0.009 & 0.002 & 0.002 & 0.004 \\
Color reconnect.  & 0.005 & 0.013 & 0.006 & 0.013 & 0.011 & 0.014 & 0.017 & 0.009 & 0.008 & 0.002 & 0.001 & 0.001 \\
Underlying event  & 0.008 & 0.002 & 0.002 & 0.004 & 0.010 & 0.007 & 0.005 & 0.007 & 0.002 & 0.003 & 0.001 & 0.001 \\
\cPqb~quark fragment.  & 0.014 & 0.002 & 0.005 & 0.017 & 0.001 & 0.001 & 0.001 & 0.002 & 0.001 & 0.003 & \NA & 0.001 \\
\cPqb~had. semilep. d.  & \NA & 0.001 & 0.001 & 0.002 & \NA & 0.001 & \NA & \NA & \NA & 0.001 & \NA & \NA \\
PDF  & 0.002 & 0.002 & 0.001 & 0.002 & \NA & \NA & \NA & \NA & \NA & 0.001 & 0.003 & 0.001 \\
Top quark mass  & 0.001 & 0.002 & 0.006 & 0.006 & 0.009 & 0.002 & 0.002 & 0.009 & 0.001 & 0.002 & 0.001 & \NA \\
Top quark \pt  & 0.008 & 0.011 & 0.005 & 0.019 & \NA & 0.001 & \NA & 0.001 & \NA & 0.004 & 0.003 & 0.005 \\
Background  & 0.017 & 0.009 & 0.008 & 0.025 & 0.006 & 0.004 & 0.004 & 0.007 & 0.003 & 0.004 & 0.008 & 0.002 \\[\cmsTabSkip]
Total systematic  & 0.031 & 0.023 & 0.016 & 0.053 & 0.029 & 0.024 & 0.025 & 0.026 & 0.016 & 0.009 & 0.010 & 0.007 \\[\cmsTabSkip]
Data statistical  & 0.018 & 0.019 & 0.010 & 0.029 & 0.029 & 0.024 & 0.025 & 0.025 & 0.020 & 0.006 & 0.003 & 0.003 \\
Signal sim. stat.  & 0.007 & 0.007 & 0.004 & 0.011 & 0.011 & 0.009 & 0.009 & 0.010 & 0.008 & 0.002 & 0.001 & 0.001 \\
Bkg. sim. stat.  & 0.010 & 0.010 & 0.005 & 0.018 & 0.017 & 0.012 & 0.010 & 0.015 & 0.012 & 0.003 & 0.002 & 0.002 \\[\cmsTabSkip]
Total statistical  & 0.022 & 0.023 & 0.012 & 0.035 & 0.035 & 0.028 & 0.028 & 0.031 & 0.025 & 0.007 & 0.003 & 0.003 \\[\cmsTabSkip]
Total  & 0.038 & 0.032 & 0.020 & 0.064 & 0.046 & 0.037 & 0.038 & 0.041 & 0.029 & 0.011 & 0.010 & 0.008 \\
\end{scotch}

\end{table*}
}
{
\begin{sidewaystable*}
\centering

    \topcaption{
    Summary of the systematic, statistical, and total uncertainties in the extracted \ttbar spin correlation coefficients and asymmetries. \adash (\NA) is shown where the values are ${<}0.0005$.
    }
    \label{tab:systematicsTotalCcombined}

\cmsTable{
\begin{scotch}{lcccccccccccc}
{Source} & \multicolumn{12}{c}{{Uncertainty}} \\
 & $\ckk$ & $\crr$ & $\cnn$ & $\cPrk$ & $\cMrk$ & $\cPnr$ & $\cMnr$ & $\cPnk$ & $\cMnk$ & $D$ & $\Aphilab$ & $\Adphi$ \\ [0.3ex]
\hline
Trigger  & 0.001 & 0.001 & \NA & 0.002 & \NA & \NA & \NA & \NA & \NA & \NA & 0.001 & \NA \\
Lepton ident./isolation  & 0.001 & 0.001 & \NA & 0.001 & \NA & \NA & \NA & \NA & \NA & \NA & \NA & \NA \\
Kinematic reconstruction  & \NA & \NA & \NA & \NA & \NA & \NA & \NA & \NA & \NA & \NA & \NA & \NA \\
Pileup  & 0.002 & \NA & 0.001 & 0.004 & 0.001 & 0.001 & 0.002 & 0.001 & 0.001 & 0.001 & \NA & 0.001 \\
\cPqb~tagging  & 0.004 & 0.001 & 0.002 & 0.005 & 0.001 & 0.001 & 0.001 & 0.001 & 0.001 & 0.001 & \NA & \NA \\
JES  & 0.012 & 0.009 & 0.005 & 0.022 & 0.011 & 0.011 & 0.009 & 0.012 & 0.007 & 0.002 & \NA & 0.001 \\
Unclustered energy  & 0.001 & 0.001 & 0.001 & 0.004 & 0.001 & 0.001 & 0.002 & 0.001 & 0.001 & \NA & \NA & 0.001 \\
JER  & 0.001 & 0.002 & 0.001 & 0.004 & 0.002 & 0.001 & 0.001 & 0.003 & 0.001 & \NA & \NA & \NA \\[\cmsTabSkip]
Scales  & 0.012 & 0.006 & 0.007 & 0.026 & 0.011 & 0.007 & 0.014 & 0.011 & 0.007 & 0.003 & 0.002 & 0.003 \\
ME/PS matching  & 0.004 & 0.003 & 0.001 & 0.009 & 0.016 & 0.011 & 0.001 & 0.012 & 0.009 & 0.002 & 0.002 & 0.004 \\
Color reconnection  & 0.005 & 0.013 & 0.006 & 0.013 & 0.011 & 0.014 & 0.017 & 0.009 & 0.008 & 0.002 & 0.001 & 0.001 \\
Underlying event  & 0.008 & 0.002 & 0.002 & 0.004 & 0.010 & 0.007 & 0.005 & 0.007 & 0.002 & 0.003 & 0.001 & 0.001 \\
\cPqb~quark fragmentation  & 0.014 & 0.002 & 0.005 & 0.017 & 0.001 & 0.001 & 0.001 & 0.002 & 0.001 & 0.003 & \NA & 0.001 \\
\cPqb~hadron semilept. decays  & \NA & 0.001 & 0.001 & 0.002 & \NA & 0.001 & \NA & \NA & \NA & 0.001 & \NA & \NA \\
PDF  & 0.002 & 0.002 & 0.001 & 0.002 & \NA & \NA & \NA & \NA & \NA & 0.001 & 0.003 & 0.001 \\
Top quark mass  & 0.001 & 0.002 & 0.006 & 0.006 & 0.009 & 0.002 & 0.002 & 0.009 & 0.001 & 0.002 & 0.001 & \NA \\
Top quark \pt  & 0.008 & 0.011 & 0.005 & 0.019 & \NA & 0.001 & \NA & 0.001 & \NA & 0.004 & 0.003 & 0.005 \\
Background  & 0.017 & 0.009 & 0.008 & 0.025 & 0.006 & 0.004 & 0.004 & 0.007 & 0.003 & 0.004 & 0.008 & 0.002 \\[\cmsTabSkip]
Total systematic  & 0.031 & 0.023 & 0.016 & 0.053 & 0.029 & 0.024 & 0.025 & 0.026 & 0.016 & 0.009 & 0.010 & 0.007 \\[\cmsTabSkip]
Data statistical  & 0.018 & 0.019 & 0.010 & 0.029 & 0.029 & 0.024 & 0.025 & 0.025 & 0.020 & 0.006 & 0.003 & 0.003 \\
Signal sim. statistical  & 0.007 & 0.007 & 0.004 & 0.011 & 0.011 & 0.009 & 0.009 & 0.010 & 0.008 & 0.002 & 0.001 & 0.001 \\
Background sim. statistical  & 0.010 & 0.010 & 0.005 & 0.018 & 0.017 & 0.012 & 0.010 & 0.015 & 0.012 & 0.003 & 0.002 & 0.002 \\[\cmsTabSkip]
Total statistical  & 0.022 & 0.023 & 0.012 & 0.035 & 0.035 & 0.028 & 0.028 & 0.031 & 0.025 & 0.007 & 0.003 & 0.003 \\[\cmsTabSkip]
Total  & 0.038 & 0.032 & 0.020 & 0.064 & 0.046 & 0.037 & 0.038 & 0.041 & 0.029 & 0.011 & 0.010 & 0.008 \\
\end{scotch}
}
\end{sidewaystable*}
}

The statistical and systematic correlation matrices for the measured coefficients are shown in Fig.~\ref{fig:StatCorrMatrixAllVarsCoeff}.
The coefficients are largely statistically uncorrelated, as expected for the measurement of independent quantities.
The expected statistical correlations between the related $D$ and diagonal $C$ coefficients are clear, as are the correlations between $D$ and the two related laboratory-frame asymmetries.
The systematic correlations are in general much stronger. In particular, strong correlations are evident for the polarization measurements with positively and negatively charged leptons (except for the $B_{i}^{n}$, where the largest sources of systematic uncertainty have a substantial statistical uncertainty from the simulation). The coefficients with significant statistical correlations naturally have significant systematic correlations as well.

\ifthenelse{\boolean{cms@external}}
{
\begin{figure*}[!htpb]
\centering
\includegraphics[width=0.49\linewidth]{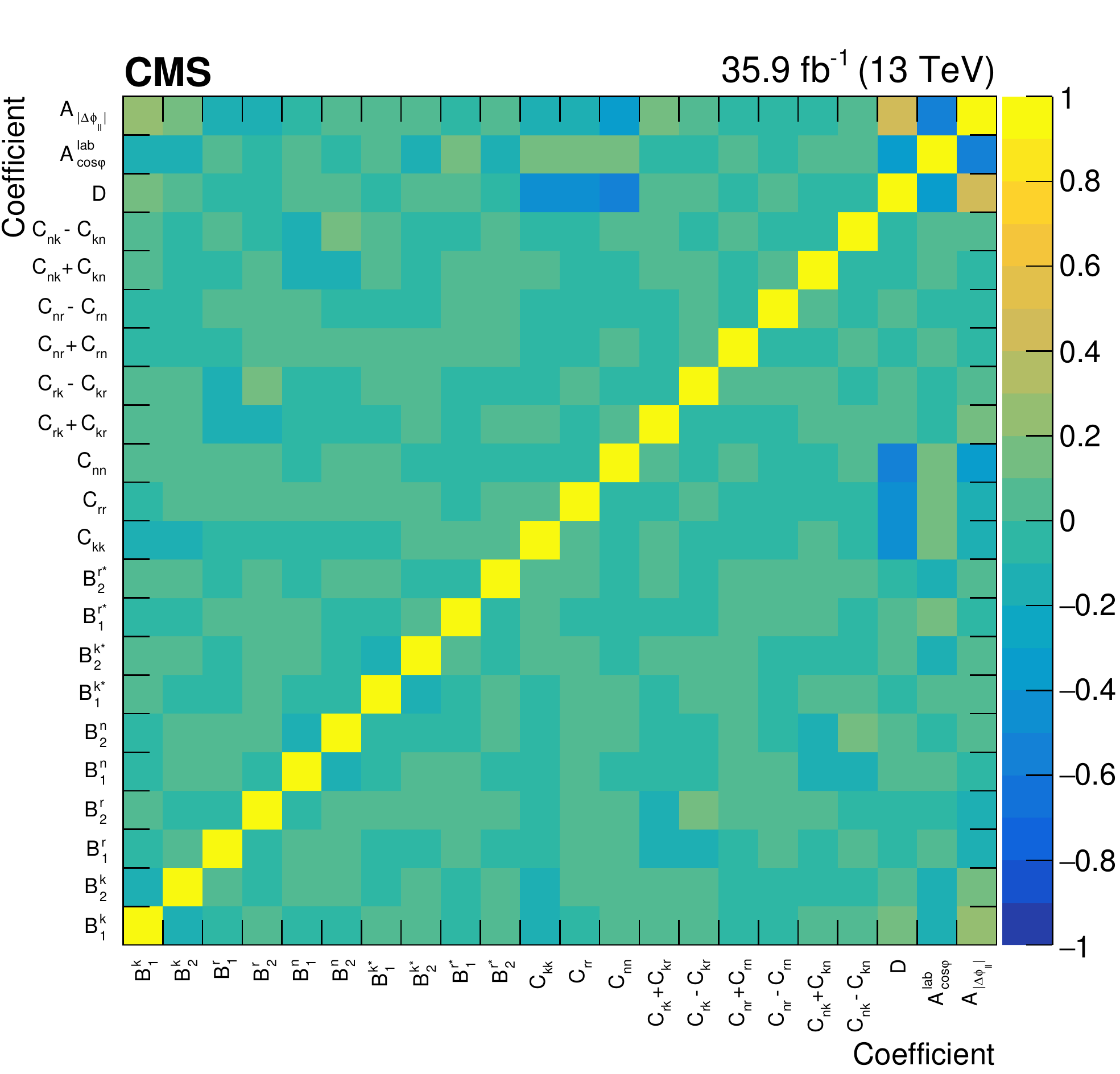}
\hfill
\includegraphics[width=0.49\linewidth]{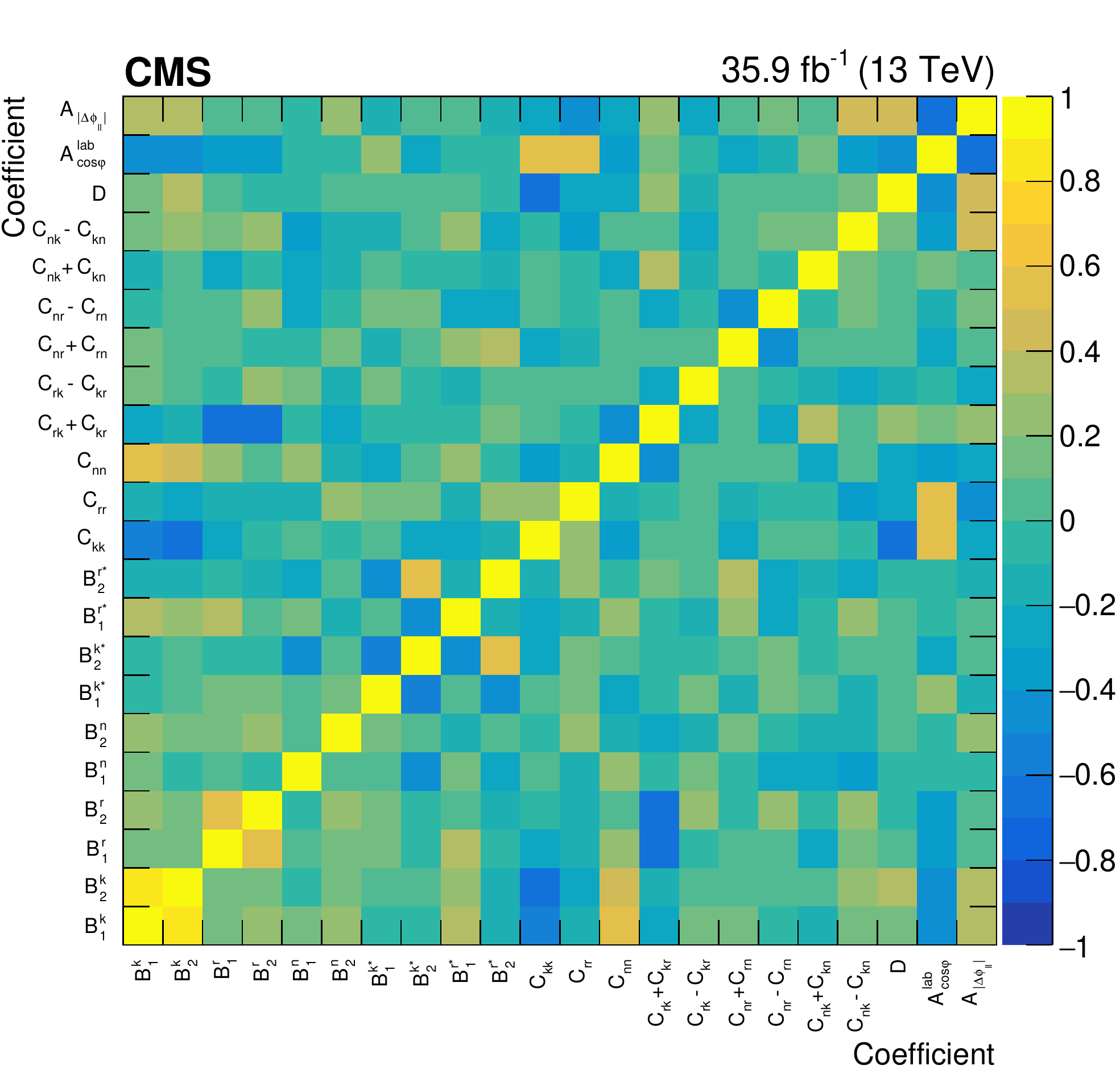}
\caption{\label{fig:StatCorrMatrixAllVarsCoeff}\protect
Values (gray scale) of the total statistical (\cmsLeft) and systematic (\cmsRight) correlation matrices for all
measured coefficients and laboratory-frame asymmetries.
}
\end{figure*}
}
{
\begin{figure*}[!htpb]
\centering
\includegraphics[width=0.72\linewidth]{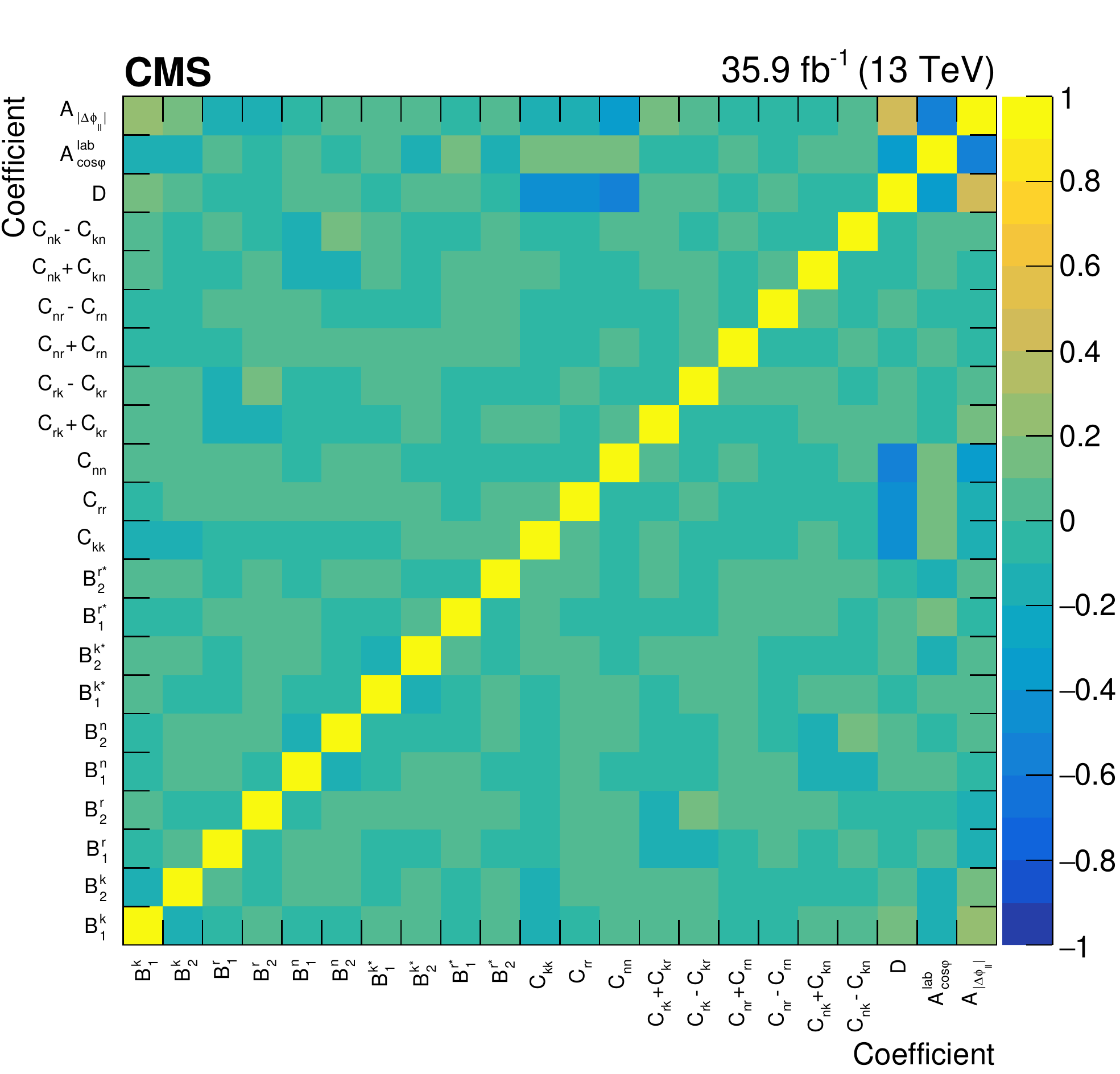}
\includegraphics[width=0.72\linewidth]{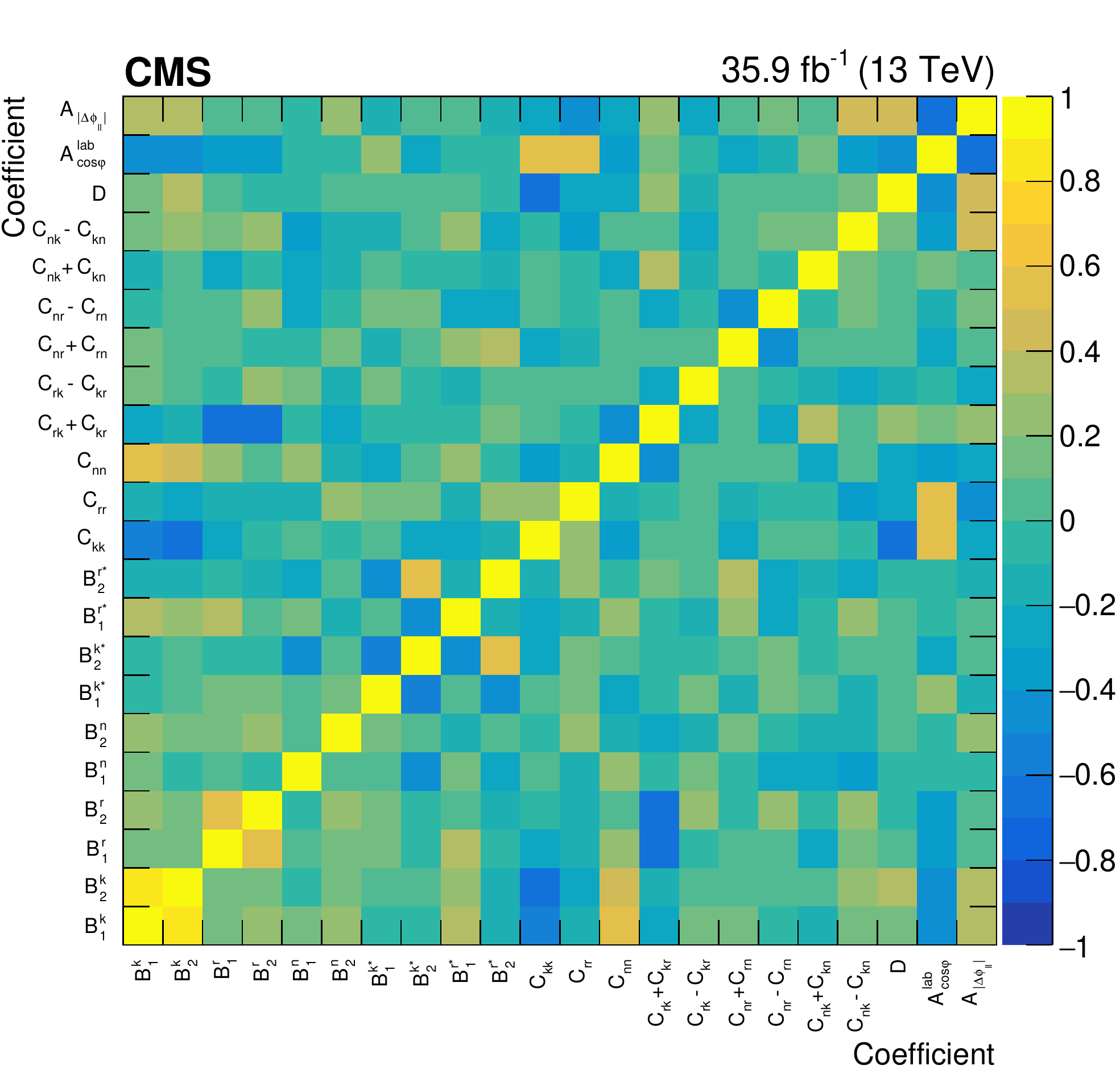}
\caption{\label{fig:StatCorrMatrixAllVarsCoeff}\protect
Values (gray scale) of the total statistical (\cmsLeft) and systematic (\cmsRight) correlation matrices for all
measured coefficients and laboratory-frame asymmetries.
}
\end{figure*}
}

The sums and differences of the pairs of $B$ coefficients are of interest, as they correspond to the CP-even and CP-odd components of the polarization. The results obtained using the measured coefficients and their covariance matrices are given in Table~\ref{tab:Bcoefficients}, and are consistent with the SM predictions.

\begin{table}[!htpb]
\centering
\topcaption{\label{tab:Bcoefficients} Measured sums and differences of the $B$ coefficients and their statistical and systematic uncertainties. 
The NLO calculated coefficients are quoted with their scale uncertainties~\cite{Bernreuther:2015yna}.}
\begin{scotch}{l  X{6.12} c }
{Coefficient} & \multicolumn{1}{c}{{$\text{Measured} \pm \text{(stat)} \pm \text{(syst)}$}} & {NLO calculation}  \\
\hline
\Trule
$\bpk+\bmk$	&	0.012		,	0.013	\pm	0.040 & $8.0\,^{+3.4}_{-2.4}  \times 10^{-3}$  \\ [0.3ex]
$\bpk-\bmk$	&	-0.002		,	0.015	\pm	0.011 & $0$ \\ [0.3ex]    
$\bpr+\bmr$	&	-0.033		,	0.015	\pm	0.026 & $3.2\,^{+2.3}_{-1.7}  \times 10^{-3}$  \\ [0.3ex]
$\bpr-\bmr$	&	-0.012		,	0.016	\pm	0.014 & $0$ \\ [0.3ex]    
$\bpn+\bmn$	&	0.024		,	0.012	\pm	0.013 & $11.3\,^{+0.9}_{-0.7} \times 10^{-3}$  \\ [0.3ex]
$\bpn-\bmn$	&	-0.011		,	0.014	\pm	0.013 & $0$ \\ [0.3ex]    
$\bpks+\bmks$	&	-0.010		,	0.016	\pm	0.012 & ${<}10^{-3}$  \\ [0.3ex]  
$\bpks-\bmks$	&	-0.023		,	0.018	\pm	0.024 & $0$ \\ [0.3ex]    
$\bprs+\bmrs$	&	0.011		,	0.016	\pm	0.018 & ${<}10^{-3}$  \\ [0.3ex]  
$\bprs-\bmrs$	&	-0.008		,	0.016	\pm	0.020 & $0$ \\ [0.3ex]    
\end{scotch}
\end{table}

\begin{table}[!tpb]
\centering
\topcaption{\label{tab:fSM}
Values of $f_{\mathrm{SM}}$, the strength of the measured spin correlations relative to the SM prediction, derived from the measurements in Table~\ref{tab:coefficients}.
The uncertainties shown are statistical, systematic, and theoretical, respectively. Their sum in quadrature is shown in the last column.
}
\begin{scotch}{l X{7.18} X{-1}  }
  {Coefficient}
& \multicolumn{1}{c}{{${f_{\mathrm{SM}}} \pm \text{(stat)} \pm \text{(syst)} \pm \text{(theo)}$}}
& \multicolumn{1}{c}{{Total uncertainty}} \\
\hline
\Trule
$\ckk$	&	0.90		,	0.07	\pm	0.09	\pm	0.01	&	\multicolumn{1}{c}{$\,\pm\;\!	0.11	$}	\\	[0.3ex]					
$\crr$	&	1.13		,	0.32	\pm	0.32	\:^{+\:	0.10	}_{-\:	0.13	}	&	\multicolumn{1}{c}{$^{+\:	0.46	}_{-\:	0.47	}$}	\\	[0.3ex]
$\cnn$	&	1.01		,	0.04	\pm	0.05	\pm	0.01	&	\multicolumn{1}{c}{$\,\pm\;\!	0.06	$}	\\	[0.3ex]					
$\cPrk$	&	0.94		,	0.17	\pm	0.26	\pm	0.01	&	\multicolumn{1}{c}{$\,\pm\;\!	0.31	$}	\\	[0.3ex]					
$D$	&	0.98		,	0.03	\pm	0.04	\pm	0.01	&	\multicolumn{1}{c}{$\,\pm\;\!	0.05	$}	\\	[0.3ex]		
$\Aphilab$	&	0.74		,	0.07	\pm	0.19	\:^{+\:	0.06	}_{-\:	0.08	}	&	\multicolumn{1}{c}{$^{+\:	0.21	}_{-\:	0.22	}$}	\\	[0.3ex]
$\Adphi$	&	1.05		,	0.03	\pm	0.08	\:^{+\:	0.09	}_{-\:	0.12	}	&	\multicolumn{1}{c}{$^{+\:	0.13	}_{-\:	0.15	}$}	\\	[0.4ex]
\end{scotch}
\end{table}

\begin{figure*}[!tpb]
\centering
\includegraphics[width=\fSMWidth]{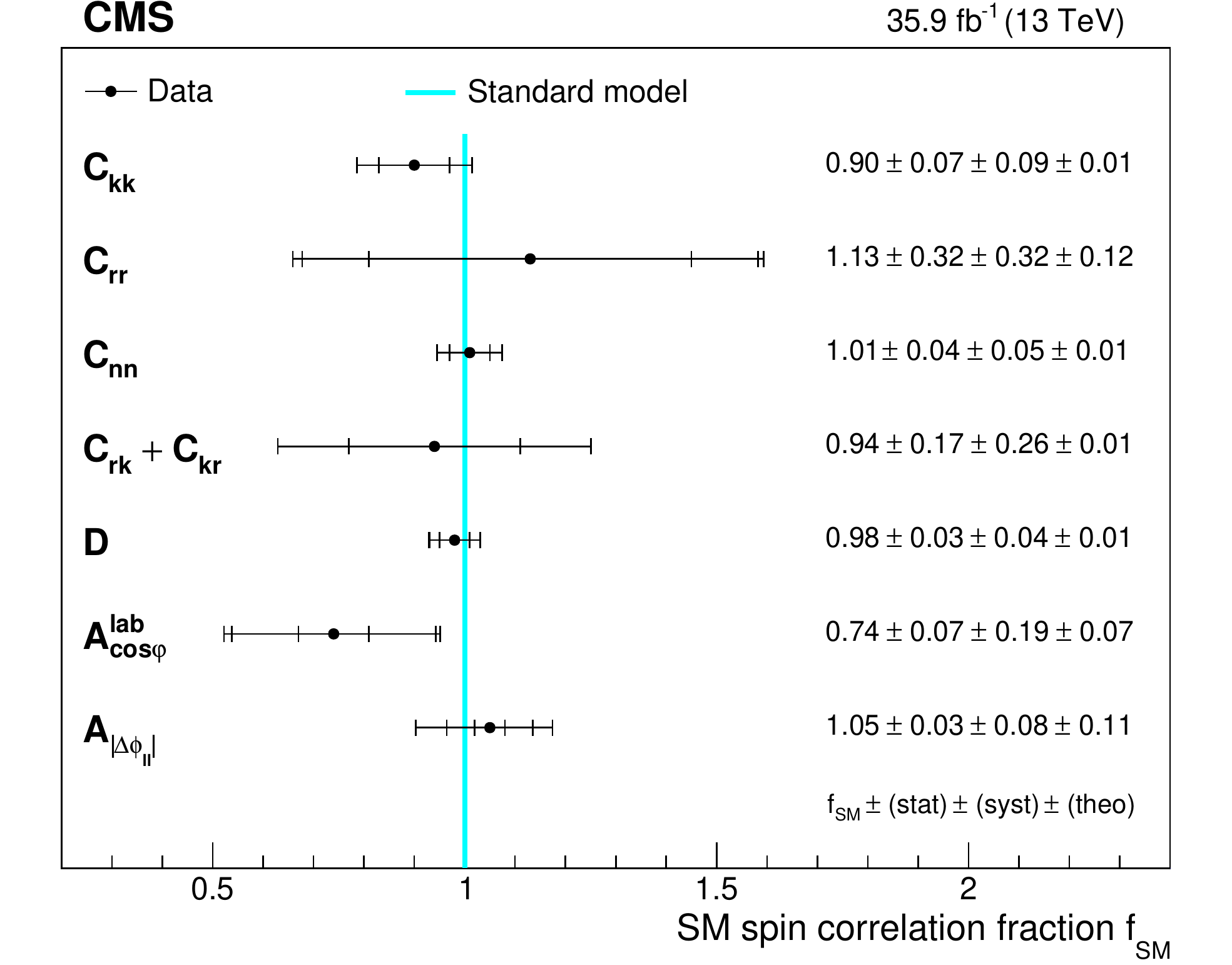}
\caption{\label{fig:summaryf}\protect
Measured values of $f_{\mathrm{SM}}$, the strength of the measured spin correlations relative to the SM prediction.
The inner vertical bars give the statistical uncertainty, the middle bars the total experimental uncertainty (statistical and systematic), and the outer bars the total uncertainty.
The numerical measured values with their uncertainties are given on the right.
}
\end{figure*}

For the coefficients in Table~\ref{tab:coefficients} sensitive to P- and CP-even spin correlations (which are substantial in the SM), we use the NLO calculations 
to transform the measurements into determinations of $f_{\mathrm{SM}}$, the strength of the given measure of spin correlations relative to the SM prediction.
A linear dependence of $f_{\mathrm{SM}}$ on the measured coefficient is defined, where $f_{\mathrm{SM}}=1$ and $f_{\mathrm{SM}}=0$ correspond to
measurements in agreement with the NLO calculations in the presence and absence of spin correlations, respectively.
The resulting measurements of $f_{\mathrm{SM}}$ are shown in Table~\ref{tab:fSM}, where the theoretical scale uncertainty from the transformation is shown as a separate uncertainty.
There is a potential correlation between the theoretical scale uncertainty and the scale component of the experimental systematic uncertainty. A similar correlation may exist with the top quark \pt systematic uncertainty, owing to its connection to missing higher-order QCD terms. These effects are neglected because their effect on the total uncertainty would be small.
The $f_{\mathrm{SM}}$ results are also shown in Fig.~\ref{fig:summaryf}.
The results are all consistent with unity, demonstrating the agreement of the measured spin correlation strengths with the SM predictions for all considered combinations of reference axes.

\ifthenelse{\boolean{cms@external}}{}{\pagebreak}

\section{Limits on higher-dimensional operators}
\label{sec:eft}

\subsection{Constraining the top quark CMDM}
\label{sec:CMDM}
Analogous to the magnetic dipole moment of an electrically charged
particle, the chromomagnetic dipole moment (CMDM) of a
color-charged particle in color fields can be defined. In the SM, the
intrinsic spin of the top quark and its color charge give it a small
CMDM~\cite{Bernreuther:2013aga,Bernreuther:2015yna}. Several BSM models,
such as two-Higgs-doublet models (e.g., supersymmetry),
technicolor, and top quark compositeness
models~\cite{bib:zhang,Martinez:2007qf}, predict an anomalous CMDM,
leading to modifications of the \ttbar\ production rate 
and spin structure.
As a consequence, the measurement of the
\ttbar\ production spin density matrix represents a powerful probe of the top
quark CMDM and can be used to search for BSM phenomena. 

As in Ref.~\cite{Sirunyan:2018ucr}, the effect of an anomalous
CMDM on \ttbar\ production is predicted using an EFT framework  
in which a fixed set of dimension-six operators is added to the SM Lagrangian
\cite{Buchmuller:1985jz,Grzadkowski:2010es}. The anomalous CMDM of the top quark is 
a consequence of the \otg\ operator~\cite{bib:zhang}, 
\begin{linenomath}
\begin{equation}
\otg = y_{\cPqt} g_{S} ({\overline Q} \sigma^{\mu \nu} T^{a} t) \tilde{\phi} G_{\mu \nu}^{a},
\end{equation}
\end{linenomath}
where $y_{\cPqt}$ denotes the Yukawa coupling of the top quark, $g_{S}$ the
strong coupling ($g_{S} = 2\sqrt{\smash[b]{\pi\alpS}}$), 
$Q$ the left-handed third-generation quark doublet, 
$\sigma^{\mu \nu}$ the Dirac matrices, 
$T^{a}$ the Gell--Mann matrices divided by 2, $t$ the right-handed
top quark singlet, $\tilde{\phi}$ the charge-conjugated Higgs doublet
field, and $G_{\mu \nu}^{a}$ the gluon field strength tensor. Besides
modifying the $\cPg\ttbar$ vertex, \otg\ also leads to a new
$\cPg\cPg\ttbar$ vertex.
The contribution due to \otg\ is parametrized by a dimensionless Wilson coefficient
divided by the square of the BSM scale ($\Lambda$), assumed
to be large compared to the scales typically probed at the LHC.
The real part of this Wilson coefficient is denoted as \ctg.
The imaginary part corresponds to a top quark chromoelectric dipole moment (CEDM), and is assumed to be zero in this section.
The top quark CEDM is constrained in Section~\ref{sec:simple}.

To produce predictions for the normalized \ttbar\ differential cross section,
the model of
Ref.~\cite{bib:zhang} is implemented in the \MGvATNLO\ generator at
NLO in QCD.
The setup is similar to that of the \MGvATNLO\ sample introduced in Section~\ref{sec:simulation},
but without extra partons at the ME level.
The \Rivet\ framework~\cite{Buckley:2010ar}
is used to apply the object definitions and calculate the spin
density matrix observables.

Four observables are chosen to constrain \ctgl,
corresponding to the four dimensions in Eq.~(\ref{eq:fulldist}), with the restriction that they are
independent from each other. For example, only two of the observables
$\cospk$, $\cospr$, and $\cospn$
are independent because they 
are the direction cosines of the $\{\hat{k},\hat{r},\hat{n}\}$ coordinate system.
The $\ckk$, $\cnn$, $\cPrk$, and $D$ coefficients are all directly sensitive to \ctgl~\cite{Bernreuther:2015yna},
and the corresponding four observables (as defined in Table~\ref{tab:obscoef}) are chosen.

A $\chi^{2}$ minimization technique is used to constrain \ctgl, with:
\begin{linenomath}
\ifthenelse{\boolean{cms@external}}
{
    \begin{multline}
	\chi^{2}(\ctgl)=\sum_{i=1}^{N}\sum_{j=1}^{N} \Big(\text{data}_{i}-\text{pred}_{i} (\ctgl)\Big) \\ 
	\Big(\text{data}_{j}-\text{pred}_{j} (\ctgl)\Big) \, \text{Cov}^{-1}_{ij},
    \end{multline}
}
{
    \begin{equation}
	\chi^{2}(\ctgl)=\sum_{i=1}^{N}\sum_{j=1}^{N} \Big(\text{data}_{i}-\text{pred}_{i} (\ctgl)\Big) \Big(\text{data}_{j}-\text{pred}_{j} (\ctgl)\Big) \, \text{Cov}^{-1}_{ij},
    \end{equation}
}
\end{linenomath}
where $\text{data}_{i}$ and $\text{pred}_{i}$(\ctgl) are the measured
and predicted normalized differential cross sections in the $i$th of the $N$ bins of the chosen observables,
and $\mathrm{Cov}^{-1}_{ij}$ is the ($i$th,
$j$th) element of the inverse of the data covariance matrix for those $N$ bins.
The covariance matrix, corresponding to a subset of the bins illustrated in Fig.~\ref{fig:StatCorrMatrixAllVarsNorm}, 
accounts for all systematic and statistical uncertainties, as well as the inter-bin correlations
introduced in the unfolding process.
In order to break the linear dependencies between the
bins of each distribution
after normalization to unit area,
one bin from each distribution is excluded from the fit, along with the rows and columns associated 
with it in the covariance matrix.
The fit result is independent of the choice of excluded bins.

The $\chi^{2}$ minimization procedure is performed twice: first 
including the full contribution from \ctgl\ to the \ttbar\ cross section,
and second including only the contribution that is linear in \ctgl, which describes the interference of the \otg\ amplitudes with those of the SM.
In both cases the best fit value of \ctgl\ is $0.06\TeV^{-2}$, corresponding to a $\chi^{2} / \mathrm{dof}$ of $8/19$. 
The difference between the two results is negligible, indicating that the value of \ctgl\ is small enough to justify the linear approximation.

Assuming Gaussian probability density functions for the
uncertainties in the unfolded data, constraints with confidence levels (\CL) can be
estimated from the values of \ctgl\ for which the $\Delta \chi^{2}$
reaches certain values. The $\Delta \chi^{2}$ is defined as the
change in $\chi^{2}$ from its minimum
value, and is shown as a function of \ctgl\ in Fig.~\ref{fig:eft_compare}.
Since the uncertainties in the theoretical predictions do not have a clear
frequentist interpretation, they are not included in the confidence
intervals. 
They are estimated separately in Fig.~\ref{fig:eft_compare} from the
maximally positive and negative
effects on the best fit value of \ctgl\ when 
changing $\mu_\mathrm{R}$ and $\mu_\mathrm{F}$
individually and simultaneously up and down by a factor of 2
in the \MGvATNLO\ predictions.

\begin{figure}[!htpb]
  \centering
        \includegraphics[width=\cmsFigWidth]{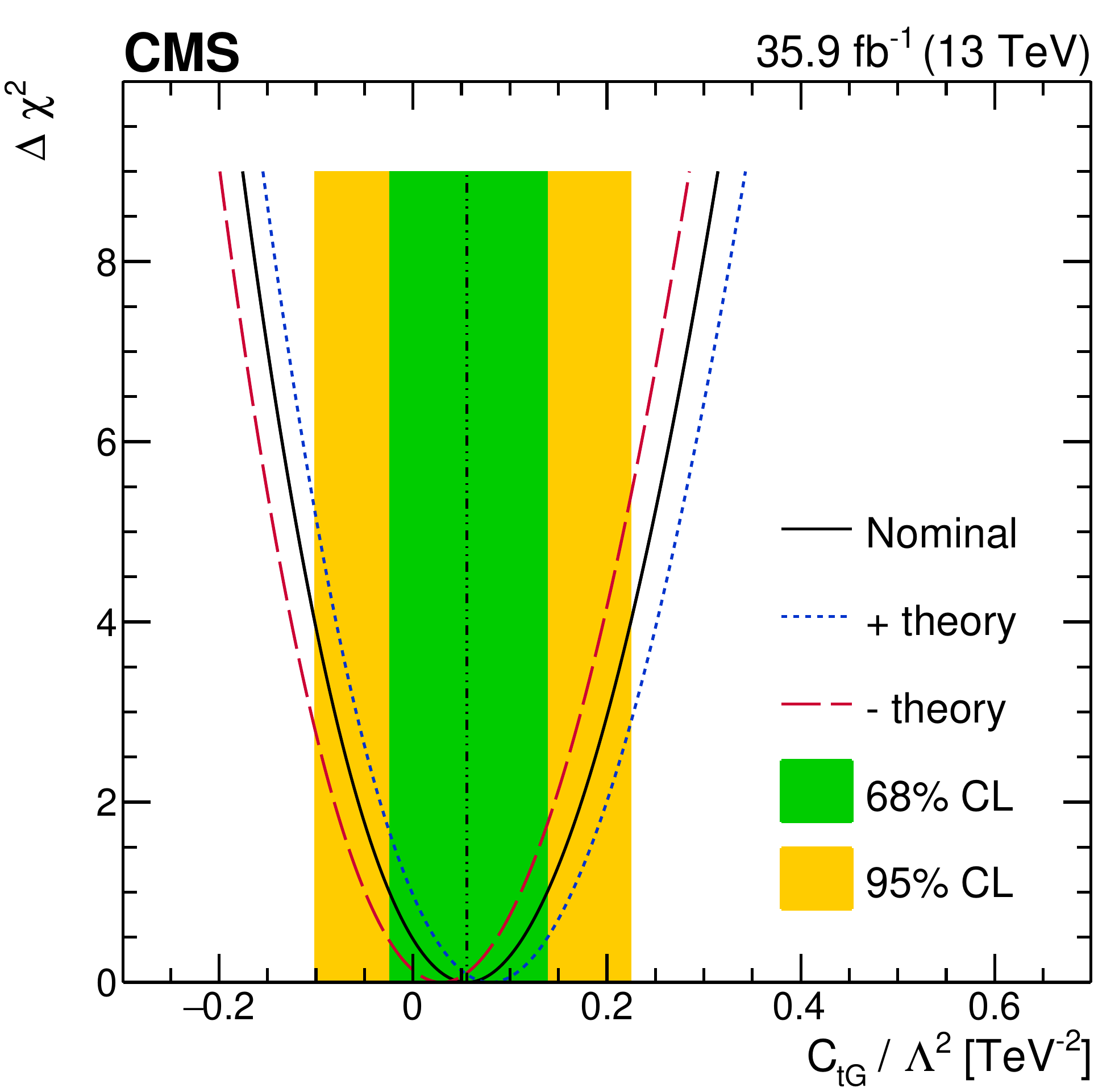}
\caption{The $\Delta \chi^{2}$ values from the fit to the data as a function of \ctgl.
The solid line is the result of the nominal fit, and the dotted and dashed lines show the most-positive and most-negative shifts in the best fit \ctgl, respectively, when the theoretical inputs are allowed to vary within their uncertainties.
The vertical line denotes the best fit value from the nominal fit,
and the inner and outer areas indicate the 68 and 95\% \CL, respectively.
} 

\label{fig:eft_compare}
\end{figure}

The resulting constraint at 95\% \CL is \ctgfitCIold.
In Ref.~\cite{bib:zhang}, a 95\% \CL constraint of $-0.42 < \ctgl\ < 0.30\TeV^{-2}$
was derived using NLO predictions for the contributions from \otg\ to the total \ttbar\ cross section,
combined with ATLAS and CMS measurements at $\sqrt{s} = 8\TeV$,
as well as $-0.32 < \ctgl\ < 0.73\TeV^{-2}$, using Fermilab Tevatron results.
From a measurement of the absolute \ttbar\ differential cross section as
a function of $\dphi$ at the particle level, 
CMS determined $-0.06 < \ctgl\ < 0.41\TeV^{-2}$ at 95\% \CL~\cite{Sirunyan:2018ucr}.
The results presented here are consistent with and improve on these previous limits.
Compared to Ref.~\cite{Sirunyan:2018ucr}, the sensitivity to \ctgl\ in this analysis 
is improved by 30\% and the theoretical uncertainties are substantially smaller.

\subsection{Constraining anomalous couplings}
\label{sec:simple}

The top quark anomalous CMDM operator is just one of the 11 independent dimension-six operators relevant for hadronic \ttbar production~\cite{Bernreuther:2015yna}.
The normalized differential cross sections measured in Section~\ref{sec:measdiffxsecs} are sensitive to ten of these operators,
and each can be constrained using a fit similar to that in Section~\ref{sec:CMDM}.
However, in the absence of a consistent simulation of all these operators compatible with the NLO QCD predictions in the \MGvATNLO\ generator~\cite{bib:zhang},
we instead use the known functional forms of Eqs.~(\ref{eq:theodist1})--(\ref{eq:theodist3}) to fit the data.
We use the calculations from Ref.~\cite{Bernreuther:2015yna} to determine the coefficients and their dependence on the contributions from the different operators.
The calculations for the NLO SM part are the same as those introduced in Section~\ref{sec:measdiffxsecs}.
For the contributions from the operators, only tree-level interference terms with the QCD amplitudes in the linear approximation are considered~\cite{Bernreuther:2015yna}.

The anomalous couplings associated with the 11 operators are listed in Table~\ref{tab:operators}, with a brief description of their properties. 
Unlike the Wilson coefficient \ctg\ considered in Section~\ref{sec:CMDM}, the anomalous couplings apply to operators in their form after spontaneous symmetry breaking~\cite{Bernreuther:2015yna}.
The couplings $\mut$ and $\dt$ represent the top quark anomalous CMDM and CEDM, respectively, and there are two further CP-odd operators involving two top quarks and up to three gluons (with couplings $\cmm$ and $\cmp$). The operators associated with the remaining couplings represent CP-even four-quark interactions, with weak isospin quantum numbers either 0 or 1.
The operators are described in detail in Ref.~\cite{Bernreuther:2015yna}.

\begin{table}[!htbp]
\centering
\topcaption{\label{tab:operators} Anomalous couplings associated with the dimension-six operators relevant for hadronic \ttbar production, the operator type of the effective interaction vertex they represent, and their P and CP symmetry properties.
It is not possible to combine the isospin-1 operators such that they have definite properties with respect to C and P~\cite{Bernreuther:2015yna}.
}
\begin{scotch}{l l Z{7.8} }
{Coupling} & {Operator type} &  \multicolumn{1}{c}{ Symmetry properties }  \\
\hline
\Trule

$	\mut	$            & 2 quarks plus gluon(s)     &  \text{P-even}, \text{CP-even}  \\
$	\dt	$            & 2 quarks plus gluon(s)     &  \text{P-odd}, \text{CP-odd}  \\
$	\cmm	$            & 2 quarks plus gluon(s)     &  \text{P-odd}, \text{CP-odd}  \\
$	\cmp	$            & 2 quarks plus gluon(s)     &  \text{P-even}, \text{CP-odd}  \\
$	\cVV	$            & 4 quarks (weak isospin 0)     &  \text{P-even}, \text{CP-even}  \\
$	\cVA	$            & 4 quarks (weak isospin 0)     &  \text{P-odd}, \text{CP-even}  \\
$	\cAV	$            & 4 quarks (weak isospin 0)     &  \text{P-odd}, \text{CP-even}  \\
$	\cAA	$            & 4 quarks (weak isospin 0)     &  \text{P-even}, \text{CP-even}  \\
$	\cone	$            & 4 quarks (weak isospin 1)     &  \multicolumn{1}{c}{CP-even}  \\
$	\ctwo	$            & 4 quarks (weak isospin 1)     &  \multicolumn{1}{c}{CP-even}  \\
$	\cthree	$            & 4 quarks (weak isospin 1)     &  \multicolumn{1}{c}{CP-even} \\ [0.3ex]
\end{scotch}
\end{table}

The normalized differential cross sections measured in Section~\ref{sec:measdiffxsecs} are sensitive to all the anomalous couplings given in Table~\ref{tab:operators} except $\cAA$, which is constrained by measurements of the \ttbar\ charge asymmetry~\cite{Bernreuther:2015yna}.
Using the same fitting procedure as in Section~\ref{sec:CMDM}, we set a limit on each coupling, setting the other couplings to zero. The 95\% \CL limits are given in Table~\ref{tab:fulllimits}, and the measured values and uncertainties are listed and displayed in Fig.~\ref{fig:summaryeft}. Theoretical uncertainties are estimated from the simultaneous variation of $\mu_\mathrm{R}$ and $\mu_\mathrm{F}$ up and down by a factor of 2.
Limits are given for the combination of couplings $\cone-\ctwo+\cthree$ rather than $\ctwo$ alone because this is the combination of couplings to which the measurements are directly sensitive~\cite{Bernreuther:2015yna}.
The strongest constraints are found for the operators probed in the $\cPg\cPg$ initial state. The four-quark operators with isospin 0 are more constrained than those with isospin 1, where contributions
from the up and down quark $\cPq\cPaq$ initial states
have opposite signs and similar magnitudes~\cite{Degrande2011,Bernreuther:2015yna}.

\begin{table*}[!htpb]
\centering
\topcaption{\label{tab:fulllimits}
The 95\% \CL limits on the anomalous couplings listed in Table~\ref{tab:operators}, derived by fitting the distributions measured in Section~\ref{sec:measdiffxsecs} and setting the other anomalous couplings to zero.
The confidence intervals include only the experimental uncertainties as in Section~\ref{sec:CMDM}.
The theoretical uncertainties, the $\chi^2$ values (dof = 19), and the distributions used in each fit are given in the last three columns.
For conciseness, the distributions are labeled by their associated coefficients (as defined in Table~\ref{tab:obscoef}).
\adash (\NA) is shown where the uncertainties are ${<}0.0005$.}
\cmsTable{
\begin{scotch}{l c X{0.5} c c }
\sTrule
  {Coupling} 
& \multicolumn{1}{c}{{95\% \CL}} 
& \multicolumn{1}{c}{{Theoretical unc.}}
& \multicolumn{1}{c}{ \chisqtext }
& \multicolumn{1}{c}{{Coefficients}} \\
\hline
\Trule
$	\mut	$ & 	\multicolumn{1}{c}{$	-0.014	<	\mut	<	0.004	$}	&	, 0.001	&	7 & $\ckk$,\, $\cnn$,\, $\cPrk$,\, $D$	\\	[0.3ex]
$	\dt	$ & 	\multicolumn{1}{c}{$	-0.020	<	\dt	<	0.012	$}	&	\multicolumn{1}{c}{\NA}	&	9 & $\bmr$,\, $\bpn$,\, $\cMnr$,\, $\cMnk$	\\	[0.3ex]
$	\cmm	$ & 	\multicolumn{1}{c}{$	-0.040	<	\cmm	<	0.006	$}	&	, 0.001	&	7 &	$\bmr$,\, $\bpn$,\, $\cMnr$,\, $\cMnk$ \\	[0.3ex]
$	\cmp	$ & 	\multicolumn{1}{c}{$	-0.009	<	\cmp	<	0.005	$}	&	\multicolumn{1}{c}{\NA}	&	11 & $\bpn$,\, $\bmn$,\, $\bprs$,\, $\cPnk$	\\	[0.3ex]
$	\cVV	$ & 	\multicolumn{1}{c}{$	-0.011	<	\cVV	<	0.042	$}	&	, 0.004	&	7 & $\ckk$,\, $\cnn$,\, $\cPrk$,\, $D$	\\	[0.3ex]
$	\cVA	$ & 	\multicolumn{1}{c}{$	-0.044	<	\cVA	<	0.027	$}	&	, 0.003	&	9 &	$\bmk$,\, $\bmr$,\, $\ckk$,\, $\cPnr$ \\	[0.3ex]
$	\cAV	$ & 	\multicolumn{1}{c}{$	-0.035	<	\cAV	<	0.032	$}	&	, 0.001	&	6 &	$\bpks$,\, $\bmks$,\, $\bprs$,\, $\bmrs$ \\	[0.3ex]
$	\cone	$ & 	\multicolumn{1}{c}{$	-0.09	<	\cone	<	0.34	$}	&	, 0.04	&	7 & $\ckk$,\, $\cnn$,\, $\cPrk$,\, $D$	\\	[0.3ex]
$	\cthree	$ & 	\multicolumn{1}{c}{$	-0.35	<	\cthree	<	0.21	$}	&	, 0.02	&	9 & $\bmk$,\, $\bmr$,\, $\ckk$,\, $\cPnr$ \\	[0.3ex]
$	\cone-\ctwo+\cthree	$ & 	\multicolumn{1}{c}{$	-0.17	<	\cone-\ctwo+\cthree	<	0.15	$}	&	, 0.01	&	6 & $\bpks$,\, $\bmks$,\, $\bprs$,\, $\bmrs$
\\	[0.3ex]
\end{scotch}
}
\end{table*}

\begin{figure}[!htpb]
\centering
\includegraphics[width=0.99\linewidth]{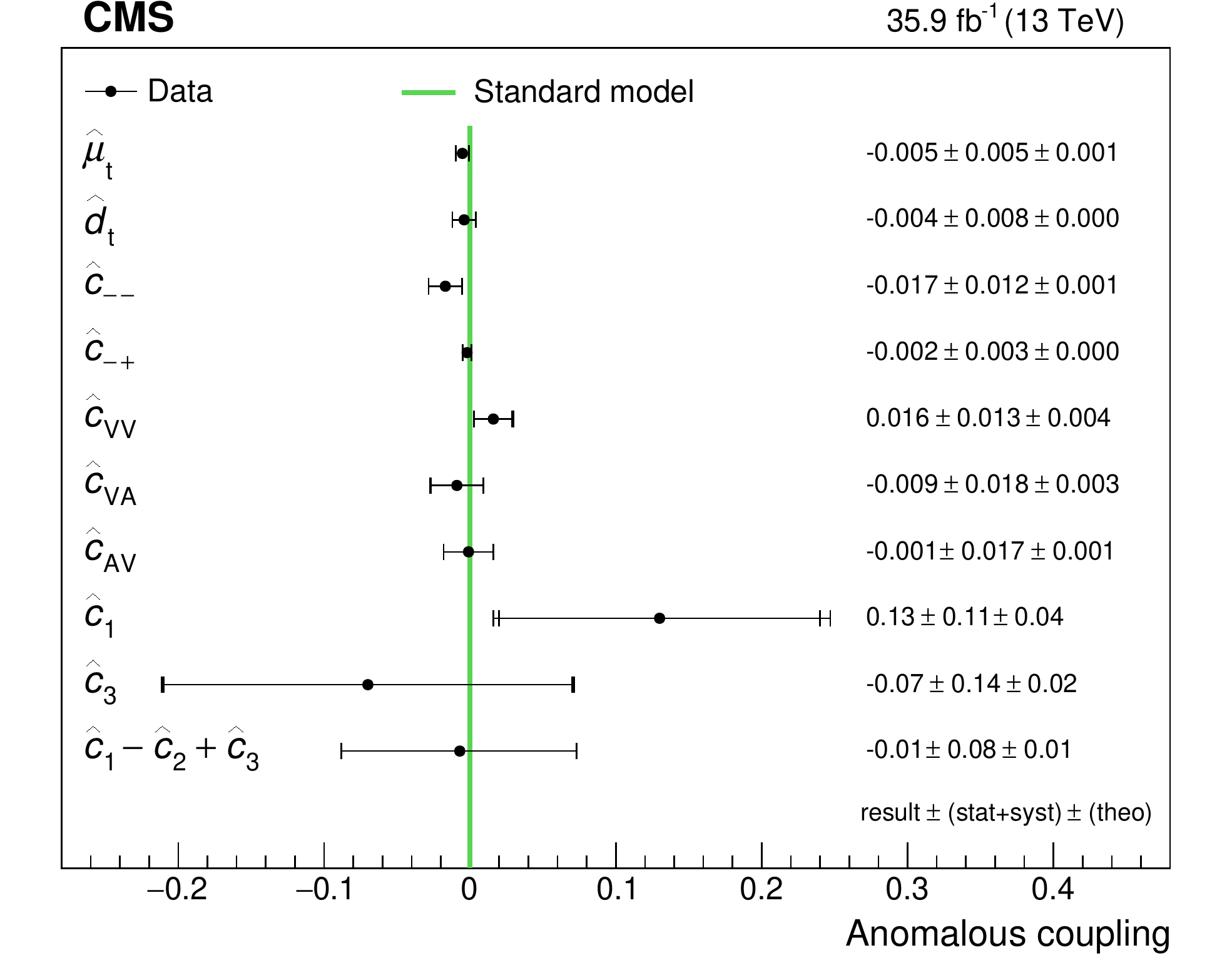}
\caption{\label{fig:summaryeft}\protect
Measured values of and uncertainties in the fitted anomalous couplings,
assuming other anomalous couplings to be zero.
The first and second quoted uncertainties are from experimental (statistical and systematic, at 68\% \CL) and theoretical sources, respectively, and are shown by the inner and outer vertical bars on the points.
The expected SM value is shown by the vertical line.
}
\end{figure}

We also consider the simultaneous fitting of multiple couplings. We find that the pairs of four-quark couplings $(\cVV,\cone)$, $(\cVA,\cthree)$, and $(\cAV,\cone-\ctwo+\cthree)$ cannot be simultaneously constrained because
their predicted effects on the measured distributions can approximately cancel each other.
The constraints on the other couplings are found to be independent, and therefore sufficiently characterized by the results of Table~\ref{tab:fulllimits}, with the exception of three combinations of couplings for which we derive two-dimensional 68 and 95\% \CL limits, shown in Fig.~\ref{fig:wbern_2D_ut_cvv_c1}.

\begin{figure*}[!htpb]
\centering
\includegraphics[width=0.49\linewidth]{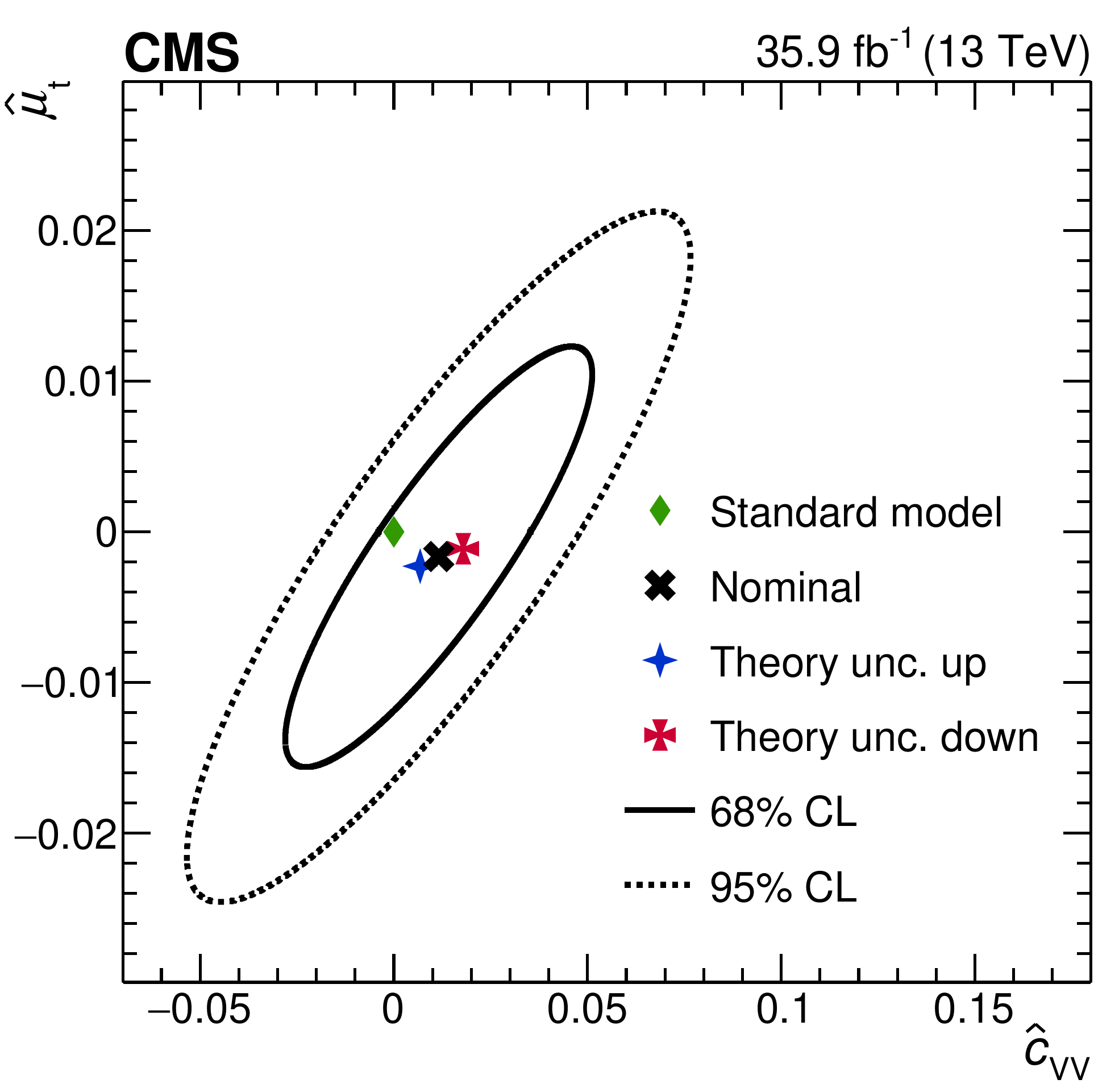}
        \hfill
\includegraphics[width=0.49\linewidth]{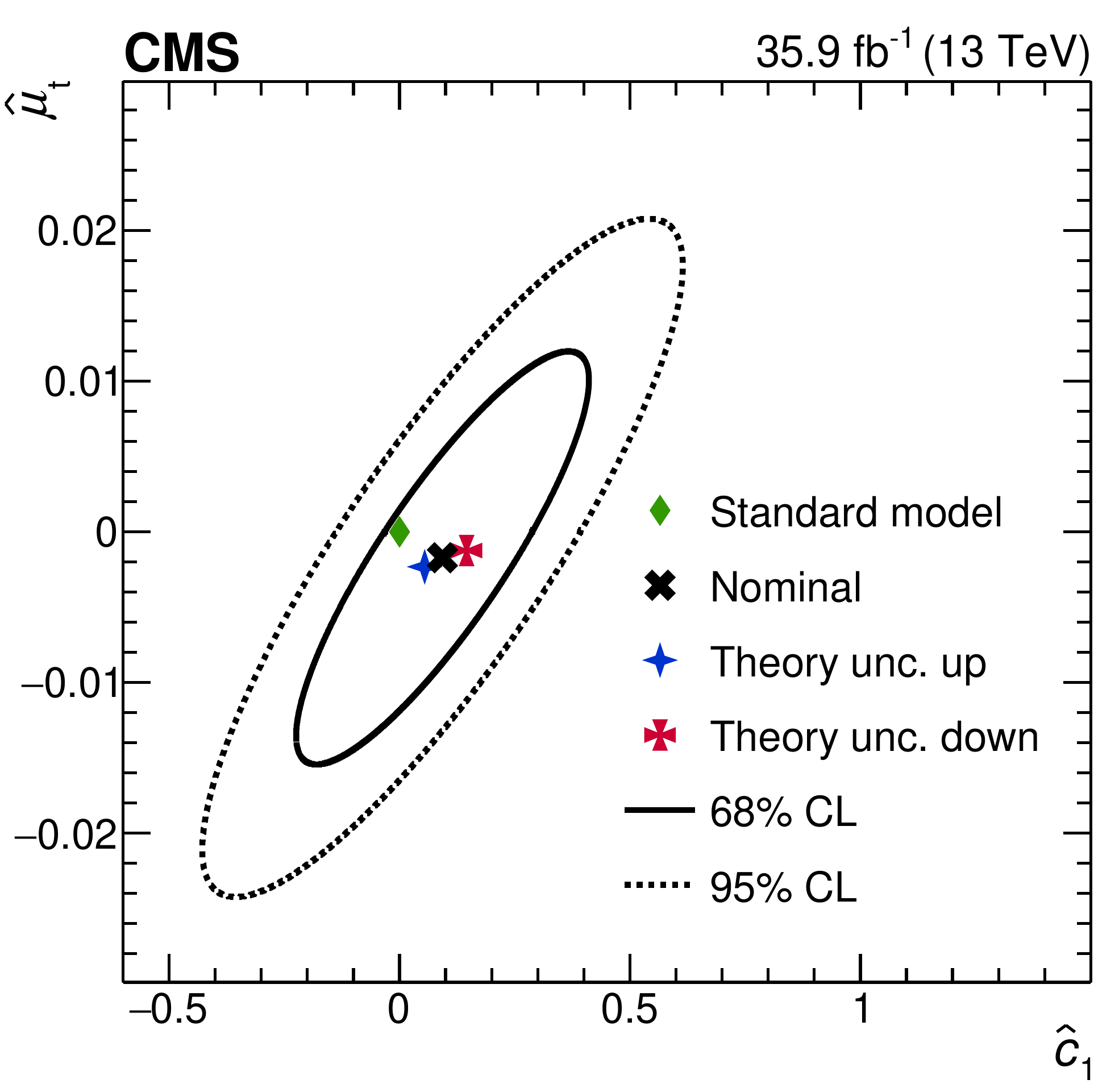}
\includegraphics[width=0.49\linewidth]{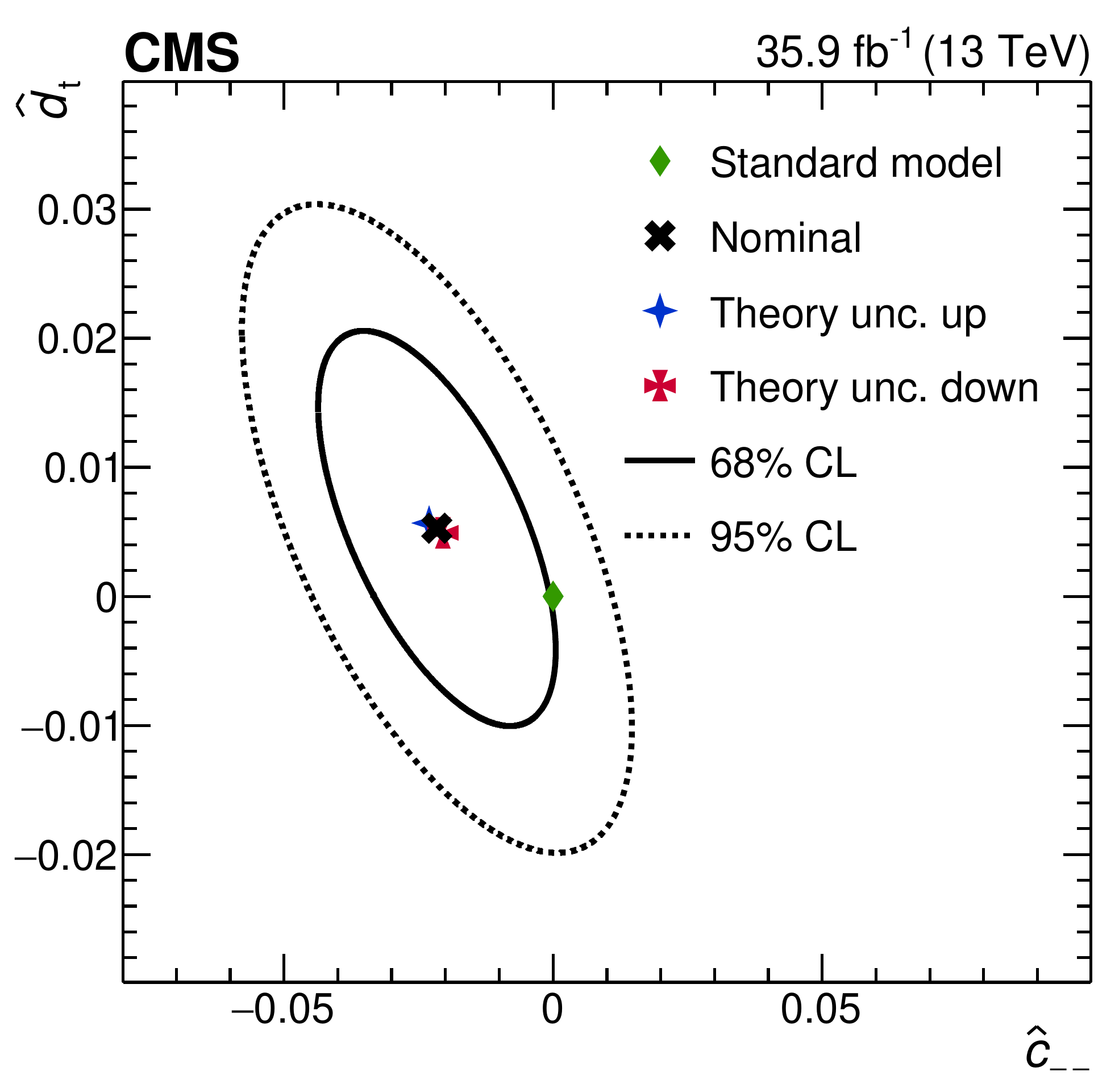}
\caption{The two-dimensional 68\% (solid curve) and 95\% (dotted curve) \CL limits on (upper left) $\mut$ vs. $\cVV$, (upper right) $\mut$ vs. $\cone$, and (lower) $\dt$ vs. $\cmm$.
The central value from the nominal fit is shown by the cross and the SM prediction by the diamond.
``Theory unc. up'' refers to the fit value when $\mu_\mathrm{R}$ and $\mu_\mathrm{F}$ are simultaneously increased by a factor of 2, and ``theory unc. down'' when they are decreased by the same factor.
}
\label{fig:wbern_2D_ut_cvv_c1}
\end{figure*}

For a direct comparison with the top quark CMDM results of Section~\ref{sec:CMDM}, we use the relationship $\ctgl = \mut/(2 \mt^2)$.
Taking the result for $\mut$ from Table~\ref{tab:fulllimits},
we find a central value of $\ctgl=-0.09\TeV^{-2}$, with \ctgfitCI\ at 95\% \CLnp.
The sensitivity to \ctgl\ (determined from the width of the confidence interval) is the same as that found in Section~\ref{sec:CMDM}, 
which suggests that the tree-level calculation of the interference terms,
using the linear approximation, is adequate for \ctgl.
The difference in central value is attributable to the difference in the SM predictions for the coefficients in the NLO calculations and the \MGvATNLO\ simulation. 
Since the SM prediction is of greater accuracy in the NLO calculations (which include EW corrections), we quote the \ctgl\ result of this section as the nominal result of the analysis.

In a similar way, $\dt$ is related to the imaginary part of the Wilson coefficient of the \otg\ operator \ctgi, and we find a constraint at 95\% \CL of \ctgifitCI, with a central value of $\ctgil=-0.07\TeV^{-2}$.
This represents a substantial improvement over
existing direct constraints on the top quark CEDM~\cite{Hioki:2013hva,Aguilar-Saavedra:2014iga},
 but it is still significantly weaker than the indirect constraint of $\abs{C^{I}_\text{tG}/\Lambda^{2}} < 0.007\TeV^{-2}$~\cite{AguilarSaavedra:2018nen} derived from the experimental limit on the neutron electric dipole moment~\cite{Baker:2006ts,Afach:2015sja}.

Analogous to the magnetic and electric dipole moments,
$\mut$ and $\dt$ can be expressed in
terms of the dimensionful parameters $C_{5}$ and $D_{5}$, which are
related to the former by a factor of 1/$\mt$~\cite{Sjolin:2003ah}. 
In this parametrization, we find constraints at 95\% \CL of
$(-1.6 < C_{5} < 0.5) \times 10^{-18} \, g_{S}\cm$ and
$(-2.3 < D_{5} < 1.4) \times 10^{-18} \, g_{S}\cm$.

\section{Summary}
\label{sec:summary}

Measurements of the
top quark polarization and \ttbar spin correlations have been presented, probing
all of the independent
coefficients of the top quark spin-dependent parts of the \ttbar
production density matrix for the first time in proton-proton collisions at $\sqrt{s}=13\TeV$.
Each coefficient was extracted from a normalized differential cross section,
unfolded to the parton level and extrapolated to the full phase space.
The measurements were made using a data sample of events containing 
two oppositely charged leptons (\ee, \emu, or \mumu)
and two or more jets, of which at least one was identified as coming from the hadronization of a bottom quark.
The data were recorded by the CMS experiment in 2016 and correspond to an integrated luminosity of
\lumivalue. 

The measured normalized differential cross sections and coefficients were
compared with standard model predictions 
from simulations with next-to-leading order (NLO) accuracy in quantum chromodynamics (QCD) and from NLO QCD calculations including electroweak corrections.
The measured distribution of $\dphi$, the absolute value of the difference in azimuthal angle between the two leptons in the laboratory frame, was additionally compared with a next-to-next-to-leading-order QCD prediction. 
All of the measurements were found to be consistent with the expectations of the standard model.
The distribution of $\cosphi$, equivalent to the dot product of the two lepton directions measured in their parent top quark and antiquark rest frames, is most sensitive to the presence of spin correlations, with a relative uncertainty below 5\%.

Statistical and systematic covariance matrices were provided for the set of all measured bins,
and were used in simultaneous fits to constrain
the contributions from ten dimension-six effective operators.
Two of these operators represent the anomalous chromomagnetic and chromoelectric dipole moments of the top quark, and constraints on their Wilson coefficients of \ctgfitCI\ 
and \ctgifitCI, respectively, were obtained at 95\% confidence level.
This constitutes a substantial improvement over previous direct constraints.

\begin{acknowledgments}
We congratulate our colleagues in the CERN accelerator departments for the excellent performance of the LHC and thank the technical and administrative staffs at CERN and at other CMS institutes for their contributions to the success of the CMS effort. In addition, we gratefully acknowledge the computing centers and personnel of the Worldwide LHC Computing Grid for delivering so effectively the computing infrastructure essential to our analyses. Finally, we acknowledge the enduring support for the construction and operation of the LHC and the CMS detector provided by the following funding agencies: BMBWF and FWF (Austria); FNRS and FWO (Belgium); CNPq, CAPES, FAPERJ, FAPERGS, and FAPESP (Brazil); MES (Bulgaria); CERN; CAS, MoST, and NSFC (China); COLCIENCIAS (Colombia); MSES and CSF (Croatia); RPF (Cyprus); SENESCYT (Ecuador); MoER, ERC IUT, PUT and ERDF (Estonia); Academy of Finland, MEC, and HIP (Finland); CEA and CNRS/IN2P3 (France); BMBF, DFG, and HGF (Germany); GSRT (Greece); NKFIA (Hungary); DAE and DST (India); IPM (Iran); SFI (Ireland); INFN (Italy); MSIP and NRF (Republic of Korea); MES (Latvia); LAS (Lithuania); MOE and UM (Malaysia); BUAP, CINVESTAV, CONACYT, LNS, SEP, and UASLP-FAI (Mexico); MOS (Montenegro); MBIE (New Zealand); PAEC (Pakistan); MSHE and NSC (Poland); FCT (Portugal); JINR (Dubna); MON, RosAtom, RAS, RFBR, and NRC KI (Russia); MESTD (Serbia); SEIDI, CPAN, PCTI, and FEDER (Spain); MOSTR (Sri Lanka); Swiss Funding Agencies (Switzerland); MST (Taipei); ThEPCenter, IPST, STAR, and NSTDA (Thailand); TUBITAK and TAEK (Turkey); NASU and SFFR (Ukraine); STFC (United Kingdom); DOE and NSF (USA).

\hyphenation{Rachada-pisek} Individuals have received support from the Marie-Curie program and the European Research Council and Horizon 2020 Grant, contract Nos.\ 675440, 752730, and 765710 (European Union); the Leventis Foundation; the A.P.\ Sloan Foundation; the Alexander von Humboldt Foundation; the Belgian Federal Science Policy Office; the Fonds pour la Formation \`a la Recherche dans l'Industrie et dans l'Agriculture (FRIA-Belgium); the Agentschap voor Innovatie door Wetenschap en Technologie (IWT-Belgium); the F.R.S.-FNRS and FWO (Belgium) under the ``Excellence of Science -- EOS" -- be.h project n.\ 30820817; the Beijing Municipal Science \& Technology Commission, No. Z181100004218003; the Ministry of Education, Youth and Sports (MEYS) of the Czech Republic; the Lend\"ulet (``Momentum") Programme and the J\'anos Bolyai Research Scholarship of the Hungarian Academy of Sciences, the New National Excellence Program \'UNKP, the NKFIA research grants 123842, 123959, 124845, 124850, 125105, 128713, 128786, and 129058 (Hungary); the Council of Science and Industrial Research, India; the HOMING PLUS program of the Foundation for Polish Science, cofinanced from European Union, Regional Development Fund, the Mobility Plus program of the Ministry of Science and Higher Education, the National Science Center (Poland), contracts Harmonia 2014/14/M/ST2/00428, Opus 2014/13/B/ST2/02543, 2014/15/B/ST2/03998, and 2015/19/B/ST2/02861, Sonata-bis 2012/07/E/ST2/01406; the National Priorities Research Program by Qatar National Research Fund; the Ministry of Science and Education, grant no. 3.2989.2017 (Russia); the Programa Estatal de Fomento de la Investigaci{\'o}n Cient{\'i}fica y T{\'e}cnica de Excelencia Mar\'{\i}a de Maeztu, grant MDM-2015-0509 and the Programa Severo Ochoa del Principado de Asturias; the Thalis and Aristeia programs cofinanced by EU-ESF and the Greek NSRF; the Rachadapisek Sompot Fund for Postdoctoral Fellowship, Chulalongkorn University and the Chulalongkorn Academic into Its 2nd Century Project Advancement Project (Thailand); the Welch Foundation, contract C-1845; and the Weston Havens Foundation (USA).

\end{acknowledgments}

\bibliography{auto_generated} 

\providecommand{\href}[2]{#2}\begingroup\raggedright\begin{thebibliography}{10}%
\makeatletter
\providecommand{\hrefCMSnoop }[0]{\@secondoftwo}%
\makeatother
\providecommand{\doi}{\texttt{doi:}\begingroup \urlstyle{tt}\Url}

\bibitem{PhysRevD.98.030001}
\hrefCMSnoop {}{{Particle Data Group}, ``Review of particle physics'',}
  \textit{ Phys. Rev. D} \textbf{ 98} (2018) 030001,
  \href{http://dx.doi.org/10.1103/PhysRevD.98.030001}{\doi{10.1103/PhysRevD.98.030001}}.

\bibitem{MahlonParke2010}
\hrefCMSnoop {}{G.~Mahlon and S.~J. Parke, ``{Spin correlation effects in top
  quark pair production at the LHC}'',} \textit{ Phys. Rev. D} \textbf{ 81}
  (2010) 074024,
  \href{http://dx.doi.org/10.1103/PhysRevD.81.074024}{\doi{10.1103/PhysRevD.81.074024}},
  \href{http://www.arXiv.org/abs/1001.3422}{\texttt{arXiv:1001.3422}}.

\bibitem{Bernreuther:2013aga}
\hrefCMSnoop {}{W.~Bernreuther and Z.-G. Si, ``{Top quark spin correlations and
  polarization at the LHC: Standard model predictions and effects of anomalous
  top chromo moments}'',} \textit{ Phys. Lett. B} \textbf{ 725} (2013) 115,
  \href{http://dx.doi.org/10.1016/j.physletb.2013.06.051}{\doi{10.1016/j.physletb.2013.06.051}},
  \href{http://www.arXiv.org/abs/1305.2066}{\texttt{arXiv:1305.2066}}.
[Erratum: \DOI{10.1016/j.physletb.2015.03.035}].

\bibitem{Bernreuther:2015yna}
\hrefCMSnoop {}{W.~Bernreuther, D.~Heisler, and Z.-G. Si, ``A set of top quark
  spin correlation and polarization observables for the {LHC}: Standard model
  predictions and new physics contributions'',} \textit{ JHEP} \textbf{ 12}
  (2015) 026,
  \href{http://dx.doi.org/10.1007/JHEP12(2015)026}{\doi{10.1007/JHEP12(2015)026}},
\href{http://www.arXiv.org/abs/1508.05271}{\texttt{arXiv:1508.05271}}.

\bibitem{bib:CMS-PAS-LUM-17-001}
\href {https://cds.cern.ch/record/2257069}{{CMS Collaboration}, ``{CMS
  luminosity measurements for the 2016 data taking period}'',} CMS Physics
  Analysis Summary CMS-PAS-LUM-17-001, 2017.

\bibitem{bib:ATLASspindensity8TeV}
\hrefCMSnoop {}{{ATLAS Collaboration}, ``{Measurements of top quark spin
  observables in \ttbar events using dilepton final states in $\sqrt{s} = 8$
  TeV pp collisions with the {ATLAS} detector}'',} \textit{ JHEP} \textbf{ 03}
  (2017) 113,
  \href{http://dx.doi.org/10.1007/JHEP03(2017)113}{\doi{10.1007/JHEP03(2017)113}},
\href{http://www.arXiv.org/abs/1612.07004}{\texttt{arXiv:1612.07004}}.

\bibitem{bib:Chatrchyan:2013wua}
\hrefCMSnoop {}{{CMS Collaboration}, ``{Measurements of \ttbar spin
  correlations and top-quark polarization using dilepton final states in pp
  collisions at $\sqrt{s} = 7$ TeV}'',} \textit{ Phys. Rev. Lett.} \textbf{
  112} (2014) 182001,
  \href{http://dx.doi.org/10.1103/PhysRevLett.112.182001}{\doi{10.1103/PhysRevLett.112.182001}},
\href{http://www.arXiv.org/abs/1311.3924}{\texttt{arXiv:1311.3924}}.

\bibitem{bib:PhysRevD.90.112016}
\hrefCMSnoop {}{{ATLAS Collaboration}, ``{Measurements of spin correlation in
  top-antitop quark events from proton-proton collisions at $\sqrt{s} = 7$ TeV
  using the {ATLAS} detector}'',} \textit{ Phys. Rev. D} \textbf{ 90} (2014)
  112016,
  \href{http://dx.doi.org/10.1103/PhysRevD.90.112016}{\doi{10.1103/PhysRevD.90.112016}},
\href{http://www.arXiv.org/abs/1407.4314}{\texttt{arXiv:1407.4314}}.

\bibitem{bib:PhysRevD.93.012002}
\hrefCMSnoop {}{{ATLAS Collaboration}, ``{Measurement of the correlation
  between the polar angles of leptons from top quark decays in the helicity
  basis at $\sqrt{s} = 7$ TeV using the {ATLAS} detector}'',} \textit{ Phys.
  Rev. D} \textbf{ 93} (2016) 012002,
  \href{http://dx.doi.org/10.1103/PhysRevD.93.012002}{\doi{10.1103/PhysRevD.93.012002}},
\href{http://www.arXiv.org/abs/1510.07478}{\texttt{arXiv:1510.07478}}.

\bibitem{bib:Khachatryan:2016xws}
\hrefCMSnoop {}{{CMS Collaboration}, ``{Measurements of \ttbar\ spin
  correlations and top quark polarization using dilepton final states in pp
  collisions at $\sqrt{s}$ = 8 TeV}'',} \textit{ Phys. Rev. D} \textbf{ 93}
  (2016) 052007,
  \href{http://dx.doi.org/10.1103/PhysRevD.93.052007}{\doi{10.1103/PhysRevD.93.052007}},
\href{http://www.arXiv.org/abs/1601.01107}{\texttt{arXiv:1601.01107}}.

\bibitem{Aaboud:2019hwz}
\hrefCMSnoop {}{{ATLAS Collaboration}, ``{Measurements of top-quark pair spin
  correlations in the e$\mu$ channel at $\sqrt{s} = 13$ TeV using pp collisions
  in the ATLAS detector}'',} (2019).
  \href{http://www.arXiv.org/abs/1903.07570}{\texttt{arXiv:1903.07570}}.
{Submitted to \textit{Eur. Phys. J. C}.}

\bibitem{Sirunyan:2018ucr}
\hrefCMSnoop {}{{CMS Collaboration}, ``{Measurements of
  $\mathrm{t\overline{t}}$ differential cross sections in proton-proton
  collisions at $\sqrt{s}=$ 13 TeV using events containing two leptons}'',}
  \textit{ JHEP} \textbf{ 02} (2019) 149,
  \href{http://dx.doi.org/10.1007/JHEP02(2019)149}{\doi{10.1007/JHEP02(2019)149}},
\href{http://www.arXiv.org/abs/1811.06625}{\texttt{arXiv:1811.06625}}.

\bibitem{bib:1212.4888}
\hrefCMSnoop {}{M.~Baumgart and B.~Tweedie, ``A new twist on top quark spin
  correlations'',} \textit{ JHEP} \textbf{ 03} (2013) 117,
  \href{http://dx.doi.org/10.1007/JHEP03(2013)117}{\doi{10.1007/JHEP03(2013)117}},
  \href{http://www.arXiv.org/abs/1212.4888}{\texttt{arXiv:1212.4888}}.

\bibitem{Fabbrichesi:2014wva}
\hrefCMSnoop {}{M.~Fabbrichesi, M.~Pinamonti, and A.~Tonero, ``{Limits on
  anomalous top quark gauge couplings from Tevatron and LHC data}'',} \textit{
  Eur. Phys. J. C} \textbf{ 74} (2014) 3193,
  \href{http://dx.doi.org/10.1140/epjc/s10052-014-3193-8}{\doi{10.1140/epjc/s10052-014-3193-8}},
\href{http://www.arXiv.org/abs/1406.5393}{\texttt{arXiv:1406.5393}}.

\bibitem{Cao:2015doa}
\hrefCMSnoop {}{Q.-H. Cao, B.~Yan, J.-H. Yu, and C.~Zhang, ``A general analysis
  of {Wtb} anomalous couplings'',} \textit{ Chin. Phys. C} \textbf{ 41} (2017)
  063101,
  \href{http://dx.doi.org/10.1088/1674-1137/41/6/063101}{\doi{10.1088/1674-1137/41/6/063101}},
\href{http://www.arXiv.org/abs/1504.03785}{\texttt{arXiv:1504.03785}}.

\bibitem{Degrande2011}
C.~Degrande\hrefCMSnoop {}{ {et~al.}, ``Non-resonant new physics in top pair
  production at hadron colliders'',} \textit{ JHEP} \textbf{ 03} (2011) 125,
  \href{http://dx.doi.org/10.1007/JHEP03(2011)125}{\doi{10.1007/JHEP03(2011)125}},
\href{http://www.arXiv.org/abs/1010.6304}{\texttt{arXiv:1010.6304}}.

\bibitem{Brandenburg2002235}
\hrefCMSnoop {}{A.~Brandenburg, Z.-G. Si, and P.~Uwer, ``{QCD-corrected spin
  analysing power of jets in decays of polarized top quarks}'',} \textit{ Phys.
  Lett. B} \textbf{ 539} (2002) 235,
  \href{http://dx.doi.org/10.1016/S0370-2693(02)02098-1}{\doi{10.1016/S0370-2693(02)02098-1}},
  \href{http://www.arXiv.org/abs/hep-ph/0205023}{\texttt{arXiv:hep-ph/0205023}}.

\bibitem{Khachatryan:2016bia}
\hrefCMSnoop {}{{CMS Collaboration}, ``{The CMS trigger system}'',} \textit{
  JINST} \textbf{ 12} (2017) P01020,
  \href{http://dx.doi.org/10.1088/1748-0221/12/01/P01020}{\doi{10.1088/1748-0221/12/01/P01020}},
\href{http://www.arXiv.org/abs/1609.02366}{\texttt{arXiv:1609.02366}}.

\bibitem{bib:Chatrchyan:2008zzk}
\hrefCMSnoop {}{{CMS Collaboration}, ``The {CMS} experiment at the {CERN}
  {LHC}'',} \textit{ JINST} \textbf{ 3} (2008) S08004,
\href{http://dx.doi.org/10.1088/1748-0221/3/08/S08004}{\doi{10.1088/1748-0221/3/08/S08004}}.

\bibitem{Frixione:2007nw}
\hrefCMSnoop {}{S.~Frixione, P.~Nason, and G.~Ridolfi, ``{A positive-weight
  next-to-leading-order Monte Carlo for heavy flavour hadroproduction}'',}
  \textit{ JHEP} \textbf{ 09} (2007) 126,
  \href{http://dx.doi.org/10.1088/1126-6708/2007/09/126}{\doi{10.1088/1126-6708/2007/09/126}},
\href{http://www.arXiv.org/abs/0707.3088}{\texttt{arXiv:0707.3088}}.

\bibitem{bib:powheg0}
\hrefCMSnoop {}{P.~Nason, ``A new method for combining {NLO} {QCD} with shower
  {Monte Carlo} algorithms'',} \textit{ JHEP} \textbf{ 11} (2004) 040,
  \href{http://dx.doi.org/10.1088/1126-6708/2004/11/040}{\doi{10.1088/1126-6708/2004/11/040}},
  \href{http://www.arXiv.org/abs/hep-ph/0409146}{\texttt{arXiv:hep-ph/0409146}}.

\bibitem{bib:powheg}
\hrefCMSnoop {}{S.~Frixione, P.~Nason, and C.~Oleari, ``{Matching NLO QCD
  computations with parton shower simulations: the POWHEG method}'',} \textit{
  JHEP} \textbf{ 11} (2007) 070,
  \href{http://dx.doi.org/10.1088/1126-6708/2007/11/070}{\doi{10.1088/1126-6708/2007/11/070}},
  \href{http://www.arXiv.org/abs/0709.2092}{\texttt{arXiv:0709.2092}}.

\bibitem{bib:powheg2}
\hrefCMSnoop {}{S.~Alioli, P.~Nason, C.~Oleari, and E.~Re, ``{A general
  framework for implementing NLO calculations in shower Monte Carlo programs:
  the POWHEG BOX}'',} \textit{ JHEP} \textbf{ 06} (2010) 043,
  \href{http://dx.doi.org/10.1007/JHEP06(2010)043}{\doi{10.1007/JHEP06(2010)043}},
  \href{http://www.arXiv.org/abs/1002.2581}{\texttt{arXiv:1002.2581}}.

\bibitem{bib:CMS:2016kle}
\href {https://cds.cern.ch/record/2235192}{{CMS Collaboration},
  ``{Investigations of the impact of the parton shower tuning in \PYTHIA~8 in
  the modelling of $\mathrm{t\overline{t}}$ at $\sqrt{s}=8$ and 13 TeV}'',} CMS
  Physics Analysis Summary CMS-PAS-TOP-16-021, 2016.

\bibitem{Sjostrand:2007gs}
T.~Sj{\"o}strand\hrefCMSnoop {}{ {et~al.}, ``{An introduction to PYTHIA
  8.2}'',} \textit{ Comput. Phys. Commun.} \textbf{ 191} (2015) 159,
  \href{http://dx.doi.org/10.1016/j.cpc.2015.01.024}{\doi{10.1016/j.cpc.2015.01.024}},
  \href{http://www.arXiv.org/abs/1410.3012}{\texttt{arXiv:1410.3012}}.

\bibitem{bib:CUETP8tune}
\hrefCMSnoop {}{{CMS Collaboration}, ``{Event generator tunes obtained from
  underlying event and multiparton scattering measurements}'',} \textit{ Eur.
  Phys. J. C} \textbf{ 76} (2016) 155,
  \href{http://dx.doi.org/10.1140/epjc/s10052-016-3988-x}{\doi{10.1140/epjc/s10052-016-3988-x}},
\href{http://www.arXiv.org/abs/1512.00815}{\texttt{arXiv:1512.00815}}.

\bibitem{Skands:2014pea}
\hrefCMSnoop {}{P.~Skands, S.~Carrazza, and J.~Rojo, ``{Tuning PYTHIA 8.1: the
  Monash 2013 tune}'',} \textit{ Eur. Phys. J. C} \textbf{ 74} (2014) 3024,
  \href{http://dx.doi.org/10.1140/epjc/s10052-014-3024-y}{\doi{10.1140/epjc/s10052-014-3024-y}},
  \href{http://www.arXiv.org/abs/1404.5630}{\texttt{arXiv:1404.5630}}.

\bibitem{Alwall:2014hca}
J.~Alwall\hrefCMSnoop {}{ {et~al.}, ``{The automated computation of tree-level
  and next-to-leading order differential cross sections, and their matching to
  parton shower simulations}'',} \textit{ JHEP} \textbf{ 07} (2014) 079,
  \href{http://dx.doi.org/10.1007/JHEP07(2014)079}{\doi{10.1007/JHEP07(2014)079}},
  \href{http://www.arXiv.org/abs/1405.0301}{\texttt{arXiv:1405.0301}}.

\bibitem{bib:madspin}
\hrefCMSnoop {}{P.~Artoisenet, R.~Frederix, O.~Mattelaer, and R.~Rietkerk,
  ``{Automatic spin-entangled decays of heavy resonances in Monte Carlo
  simulations}'',} \textit{ JHEP} \textbf{ 03} (2013) 015,
  \href{http://dx.doi.org/10.1007/JHEP03(2013)015}{\doi{10.1007/JHEP03(2013)015}},
  \href{http://www.arXiv.org/abs/1212.3460}{\texttt{arXiv:1212.3460}}.

\bibitem{Frederix:2012ps}
\hrefCMSnoop {}{R.~Frederix and S.~Frixione, ``{Merging meets matching in
  MC@NLO}'',} \textit{ JHEP} \textbf{ 12} (2012) 061,
  \href{http://dx.doi.org/10.1007/JHEP12(2012)061}{\doi{10.1007/JHEP12(2012)061}},
\href{http://www.arXiv.org/abs/1209.6215}{\texttt{arXiv:1209.6215}}.

\bibitem{Alwall:2007fs}
J.~Alwall\hrefCMSnoop {}{ {et~al.}, ``{Comparative study of various algorithms
  for the merging of parton showers and matrix elements in hadronic
  collisions}'',} \textit{ Eur. Phys. J. C} \textbf{ 53} (2008) 473,
  \href{http://dx.doi.org/10.1140/epjc/s10052-007-0490-5}{\doi{10.1140/epjc/s10052-007-0490-5}},
\href{http://www.arXiv.org/abs/0706.2569}{\texttt{arXiv:0706.2569}}.

\bibitem{bib:powheg1}
\hrefCMSnoop {}{S.~Alioli, P.~Nason, C.~Oleari, and E.~Re, ``{NLO single-top
  production matched with shower in POWHEG: $s$- and $t$-channel
  contributions}'',} \textit{ JHEP} \textbf{ 09} (2009) 111,
  \href{http://dx.doi.org/10.1088/1126-6708/2009/09/111}{\doi{10.1088/1126-6708/2009/09/111}},
  \href{http://www.arXiv.org/abs/0907.4076}{\texttt{arXiv:0907.4076}}.

\bibitem{bib:powheg3}
\hrefCMSnoop {}{E.~Re, ``{Single-top Wt-channel production matched with parton
  showers using the POWHEG method}'',} \textit{ Eur. Phys. J. C} \textbf{ 71}
  (2011) 1547,
  \href{http://dx.doi.org/10.1140/epjc/s10052-011-1547-z}{\doi{10.1140/epjc/s10052-011-1547-z}},
  \href{http://www.arXiv.org/abs/1009.2450}{\texttt{arXiv:1009.2450}}.

\bibitem{bib:NNPDF}
\hrefCMSnoop {}{{NNPDF} Collaboration, ``{Unbiased global determination of
  parton distributions and their uncertainties at NNLO and LO}'',} \textit{
  Nucl. Phys. B} \textbf{ 855} (2012) 153,
  \href{http://dx.doi.org/10.1016/j.nuclphysb.2011.09.024}{\doi{10.1016/j.nuclphysb.2011.09.024}},
  \href{http://www.arXiv.org/abs/1107.2652}{\texttt{arXiv:1107.2652}}.

\bibitem{Ball2015}
\hrefCMSnoop {}{{NNPDF} Collaboration, ``{Parton distributions for the LHC Run
  II}'',} \textit{ JHEP} \textbf{ 04} (2015) 040,
  \href{http://dx.doi.org/10.1007/JHEP04(2015)040}{\doi{10.1007/JHEP04(2015)040}},
  \href{http://www.arXiv.org/abs/1410.8849}{\texttt{arXiv:1410.8849}}.

\bibitem{Li:2012wna}
\hrefCMSnoop {}{Y.~Li and F.~Petriello, ``{Combining QCD and electroweak
  corrections to dilepton production in FEWZ}'',} \textit{ Phys. Rev. D}
  \textbf{ 86} (2012) 094034,
  \href{http://dx.doi.org/10.1103/PhysRevD.86.094034}{\doi{10.1103/PhysRevD.86.094034}},
\href{http://www.arXiv.org/abs/1208.5967}{\texttt{arXiv:1208.5967}}.

\bibitem{bib:twchan}
\hrefCMSnoop {}{N.~Kidonakis, ``{Two-loop soft anomalous dimensions for single
  top quark associated production with $\mathrm{W^-}$ or $\mathrm{H^-}$}'',}
  \textit{ Phys. Rev. D} \textbf{ 82} (2010) 054018,
  \href{http://dx.doi.org/10.1103/PhysRevD.82.054018}{\doi{10.1103/PhysRevD.82.054018}},
  \href{http://www.arXiv.org/abs/hep-ph/1005.4451}{\texttt{arXiv:hep-ph/1005.4451}}.

\bibitem{bib:mcfm:diboson}
\hrefCMSnoop {}{J.~M. Campbell, R.~K. Ellis, and C.~Williams, ``{Vector boson
  pair production at the LHC}'',} \textit{ JHEP} \textbf{ 07} (2011) 018,
  \href{http://dx.doi.org/10.1007/JHEP07(2011)018}{\doi{10.1007/JHEP07(2011)018}},
  \href{http://www.arXiv.org/abs/1105.0020}{\texttt{arXiv:1105.0020}}.

\bibitem{bib:Maltoni:2015ena}
\hrefCMSnoop {}{F.~Maltoni, D.~Pagani, and I.~Tsinikos, ``{Associated
  production of a top-quark pair with vector bosons at NLO in QCD: impact on $
  \mathrm{t}\overline{\mathrm{t}}\mathrm{H} $ searches at the LHC}'',} \textit{
  JHEP} \textbf{ 02} (2016) 113,
  \href{http://dx.doi.org/10.1007/JHEP02(2016)113}{\doi{10.1007/JHEP02(2016)113}},
\href{http://www.arXiv.org/abs/1507.05640}{\texttt{arXiv:1507.05640}}.

\bibitem{Czakon:2011xx}
\hrefCMSnoop {}{M.~Czakon and A.~Mitov, ``{TOP++: a program for the calculation
  of the top-pair cross-section at hadron colliders}'',} \textit{ Comput. Phys.
  Commun.} \textbf{ 185} (2014) 2930,
  \href{http://dx.doi.org/10.1016/j.cpc.2014.06.021}{\doi{10.1016/j.cpc.2014.06.021}},
  \href{http://www.arXiv.org/abs/1112.5675}{\texttt{arXiv:1112.5675}}.

\bibitem{bib:geant}
\hrefCMSnoop {}{{GEANT4} Collaboration, ``{\GEANTfour}---a simulation
  toolkit'',} \textit{ Nucl. Instrum. Meth. A} \textbf{ 506} (2003) 250,
  \href{http://dx.doi.org/10.1016/S0168-9002(03)01368-8}{\doi{10.1016/S0168-9002(03)01368-8}}.

\bibitem{bib:Sirunyan:2017ulk}
\hrefCMSnoop {}{{CMS Collaboration}, ``{Particle-flow reconstruction and global
  event description with the CMS detector}'',} \textit{ JINST} \textbf{ 12}
  (2017) P10003,
  \href{http://dx.doi.org/10.1088/1748-0221/12/10/P10003}{\doi{10.1088/1748-0221/12/10/P10003}},
\href{http://www.arXiv.org/abs/1706.04965}{\texttt{arXiv:1706.04965}}.

\bibitem{1748-0221-10-06-P06005}
\hrefCMSnoop {}{{CMS Collaboration}, ``{Performance of electron reconstruction
  and selection with the CMS detector in proton-proton collisions at $\sqrt{s}$
  = 8 \TeV}'',} \textit{ JINST} \textbf{ 10} (2015) P06005,
  \href{http://dx.doi.org/10.1088/1748-0221/10/06/P06005}{\doi{10.1088/1748-0221/10/06/P06005}},
\href{http://www.arXiv.org/abs/1502.02701}{\texttt{arXiv:1502.02701}}.

\bibitem{Sirunyan:2018fpa}
\hrefCMSnoop {}{{CMS Collaboration}, ``{Performance of the CMS muon detector
  and muon reconstruction with proton-proton collisions at $\sqrt{s}=$ 13
  TeV}'',} \textit{ JINST} \textbf{ 13} (2018) P06015,
  \href{http://dx.doi.org/10.1088/1748-0221/13/06/P06015}{\doi{10.1088/1748-0221/13/06/P06015}},
\href{http://www.arXiv.org/abs/1804.04528}{\texttt{arXiv:1804.04528}}.

\bibitem{bib:antikt}
\hrefCMSnoop {}{M.~Cacciari, G.~P. Salam, and G.~Soyez, ``{The anti-\kt jet
  clustering algorithm}'',} \textit{ JHEP} \textbf{ 04} (2008) 063,
  \href{http://dx.doi.org/10.1088/1126-6708/2008/04/063}{\doi{10.1088/1126-6708/2008/04/063}},
  \href{http://www.arXiv.org/abs/0802.1189}{\texttt{arXiv:0802.1189}}.

\bibitem{bib:Cacciari:2011ma}
\hrefCMSnoop {}{M.~Cacciari, G.~P. Salam, and G.~Soyez, ``{FastJet user
  manual}'',} \textit{ Eur. Phys. J. C} \textbf{ 72} (2012) 1896,
  \href{http://dx.doi.org/10.1140/epjc/s10052-012-1896-2}{\doi{10.1140/epjc/s10052-012-1896-2}},
\href{http://www.arXiv.org/abs/1111.6097}{\texttt{arXiv:1111.6097}}.

\bibitem{Chatrchyan:2009hy}
\hrefCMSnoop {}{{CMS Collaboration}, ``Identification and filtering of
  uncharacteristic noise in the {CMS} hadron calorimeter'',} \textit{ JINST}
  \textbf{ 5} (2010) T03014,
  \href{http://dx.doi.org/10.1088/1748-0221/5/03/T03014}{\doi{10.1088/1748-0221/5/03/T03014}},
\href{http://www.arXiv.org/abs/0911.4881}{\texttt{arXiv:0911.4881}}.

\bibitem{Sirunyan:2017ezt}
\hrefCMSnoop {}{{CMS Collaboration}, ``{Identification of heavy-flavour jets
  with the CMS detector in pp collisions at 13\TeV}'',} \textit{ JINST}
  \textbf{ 13} (2018) P05011,
  \href{http://dx.doi.org/10.1088/1748-0221/13/05/P05011}{\doi{10.1088/1748-0221/13/05/P05011}},
  \href{http://www.arXiv.org/abs/1712.07158}{\texttt{arXiv:1712.07158}}.

\bibitem{bib:TOP-12-028_paper}
\hrefCMSnoop {}{{CMS Collaboration}, ``{Measurement of the differential cross
  section for top quark pair production in pp collisions at $\sqrt{s} =
  8\,\text {TeV} $}'',} \textit{ Eur. Phys. J. C} \textbf{ 75} (2015) 542,
  \href{http://dx.doi.org/10.1140/epjc/s10052-015-3709-x}{\doi{10.1140/epjc/s10052-015-3709-x}},
  \href{http://www.arXiv.org/abs/1505.04480}{\texttt{arXiv:1505.04480}}.

\bibitem{bib:TOP-15-003_paper}
\hrefCMSnoop {}{{CMS Collaboration}, ``{Measurement of the top quark pair
  production cross section in proton-proton collisions at $\sqrt{s}$ = 13 TeV
  with the CMS detector}'',} \textit{ Phys. Rev. Lett.} \textbf{ 116} (2016)
  052002,
  \href{http://dx.doi.org/10.1103/PhysRevLett.116.052002}{\doi{10.1103/PhysRevLett.116.052002}},
  \href{http://www.arXiv.org/abs/1510.05302}{\texttt{arXiv:1510.05302}}.

\bibitem{bib:TUnfold}
\hrefCMSnoop {}{S.~Schmitt, ``{TUnfold, an algorithm for correcting migration
  effects in high energy physics}'',} \textit{ JINST} \textbf{ 7} (2012)
  T10003,
  \href{http://dx.doi.org/10.1088/1748-0221/7/10/T10003}{\doi{10.1088/1748-0221/7/10/T10003}},
\href{http://www.arXiv.org/abs/1205.6201}{\texttt{arXiv:1205.6201}}.

\bibitem{TOP-17-001}
\hrefCMSnoop {}{{CMS Collaboration}, ``{Measurement of the
  $\mathrm{t\overline{t}}$ production cross section, the top quark mass, and
  the strong coupling constant using dilepton events in pp collisions at
  $\sqrt{s}=$ 13 TeV}'',} \textit{ Eur. Phys. J. C} \textbf{ 79} (2019) 368,
  \href{http://dx.doi.org/10.1140/epjc/s10052-019-6863-8}{\doi{10.1140/epjc/s10052-019-6863-8}},
\href{http://www.arXiv.org/abs/1812.10505}{\texttt{arXiv:1812.10505}}.

\bibitem{bib:tp}
\hrefCMSnoop {}{{CMS Collaboration}, ``{Measurement of the Drell--Yan cross
  sections in pp collisions at $\sqrt{s} = 7$ TeV with the CMS experiment}'',}
  \textit{ JHEP} \textbf{ 10} (2011) 007,
  \href{http://dx.doi.org/10.1007/JHEP10(2011)007}{\doi{10.1007/JHEP10(2011)007}},
  \href{http://www.arXiv.org/abs/1108.0566}{\texttt{arXiv:1108.0566}}.

\bibitem{Aaboud:2016mmw}
\hrefCMSnoop {}{{ATLAS Collaboration}, ``{Measurement of the inelastic
  proton-proton cross section at $\sqrt{s} = 13$\TeV with the ATLAS detector at
  the LHC}'',} \textit{ Phys. Rev. Lett.} \textbf{ 117} (2016) 182002,
  \href{http://dx.doi.org/10.1103/PhysRevLett.117.182002}{\doi{10.1103/PhysRevLett.117.182002}},
\href{http://www.arXiv.org/abs/1606.02625}{\texttt{arXiv:1606.02625}}.

\bibitem{Khachatryan:2016kdb}
\hrefCMSnoop {}{{CMS Collaboration}, ``{Jet energy scale and resolution in the
  CMS experiment in pp collisions at 8\TeV}'',} \textit{ JINST} \textbf{ 12}
  (2017) P02014,
  \href{http://dx.doi.org/10.1088/1748-0221/12/02/P02014}{\doi{10.1088/1748-0221/12/02/P02014}},
\href{http://www.arXiv.org/abs/1607.03663}{\texttt{arXiv:1607.03663}}.

\bibitem{Argyropoulos:2014zoa}
\hrefCMSnoop {}{S.~Argyropoulos and T.~Sj{\"o}strand, ``{Effects of color
  reconnection on \ttbar\ final states at the LHC}'',} \textit{ JHEP} \textbf{
  11} (2014) 043,
  \href{http://dx.doi.org/10.1007/JHEP11(2014)043}{\doi{10.1007/JHEP11(2014)043}},
\href{http://www.arXiv.org/abs/1407.6653}{\texttt{arXiv:1407.6653}}.

\bibitem{Christiansen:2015yqa}
\hrefCMSnoop {}{J.~R. Christiansen and P.~Z. Skands, ``{String formation beyond
  leading colour}'',} \textit{ JHEP} \textbf{ 08} (2015) 003,
  \href{http://dx.doi.org/10.1007/JHEP08(2015)003}{\doi{10.1007/JHEP08(2015)003}},
\href{http://www.arXiv.org/abs/1505.01681}{\texttt{arXiv:1505.01681}}.

\bibitem{Bowler:1981sb}
\hrefCMSnoop {}{M.~G. Bowler, ``{$\text{e}^{+}\text{e}^{-}$ production of heavy
  quarks in the string model}'',} \textit{ Z. Phys. C} \textbf{ 11} (1981) 169,
\href{http://dx.doi.org/10.1007/BF01574001}{\doi{10.1007/BF01574001}}.

\bibitem{PhysRevD.27.105}
\hrefCMSnoop {}{C.~Peterson, D.~Schlatter, I.~Schmitt, and P.~M. Zerwas,
  ``Scaling violations in inclusive $\text{e}^{+}\text{e}^{\ensuremath{-}}$
  annihilation spectra'',} \textit{ Phys. Rev. D} \textbf{ 27} (1983) 105,
  \href{http://dx.doi.org/10.1103/PhysRevD.27.105}{\doi{10.1103/PhysRevD.27.105}}.

\bibitem{Butterworth:2015oua}
\hrefCMSnoop {}{J.~Butterworth {et~al.}, ``{PDF4LHC recommendations for LHC Run
  II}'',} \textit{ J. Phys. G} \textbf{ 43} (2016) 023001,
  \href{http://dx.doi.org/10.1088/0954-3899/43/2/023001}{\doi{10.1088/0954-3899/43/2/023001}},
\href{http://www.arXiv.org/abs/1510.03865}{\texttt{arXiv:1510.03865}}.

\bibitem{bib:TOP-11-013_paper}
\hrefCMSnoop {}{{CMS Collaboration}, ``{Measurement of differential top-quark
  pair production cross sections in pp collisions at \texorpdfstring{$\sqrt{s}
  = 7\TeV$}{sqrt(s) = 7 TeV}}'',} \textit{ Eur. Phys. J. C} \textbf{ 73} (2013)
  2339,
  \href{http://dx.doi.org/10.1140/epjc/s10052-013-2339-4}{\doi{10.1140/epjc/s10052-013-2339-4}},
  \href{http://www.arXiv.org/abs/1211.2220}{\texttt{arXiv:1211.2220}}.

\bibitem{Khachatryan:2015fwh}
\hrefCMSnoop {}{{CMS Collaboration}, ``{Measurement of the
  $\mathrm{t}\overline{{\mathrm{t}}}$ production cross section in the all-jets
  final state in pp collisions at $\sqrt{s}=8$ $\,\text {TeV}$}'',} \textit{
  Eur. Phys. J. C} \textbf{ 76} (2016) 128,
  \href{http://dx.doi.org/10.1140/epjc/s10052-016-3956-5}{\doi{10.1140/epjc/s10052-016-3956-5}},
\href{http://www.arXiv.org/abs/1509.06076}{\texttt{arXiv:1509.06076}}.

\bibitem{bib:TOP-16-008}
\hrefCMSnoop {}{{CMS Collaboration}, ``{Measurement of differential cross
  sections for top quark pair production using the lepton+jets final state in
  proton-proton collisions at 13 TeV}'',} \textit{ Phys. Rev. D} \textbf{ 95}
  (2017) 092001,
  \href{http://dx.doi.org/10.1103/PhysRevD.95.092001}{\doi{10.1103/PhysRevD.95.092001}},
\href{http://www.arXiv.org/abs/1610.04191}{\texttt{arXiv:1610.04191}}.

\bibitem{bib:difftop}
\hrefCMSnoop {}{M.~Guzzi, K.~Lipka, and S.-O. Moch, ``Top-quark pair production
  at hadron colliders: differential cross section and phenomenological
  applications with {DiffTop}'',} \textit{ JHEP} \textbf{ 01} (2015) 082,
  \href{http://dx.doi.org/10.1007/JHEP01(2015)082}{\doi{10.1007/JHEP01(2015)082}},
  \href{http://www.arXiv.org/abs/1406.0386}{\texttt{arXiv:1406.0386}}.

\bibitem{bib:kidonakis_13TeV}
\hrefCMSnoop {}{N.~Kidonakis, ``{NNNLO} soft-gluon corrections for the
  top-quark \pt and rapidity distributions'',} \textit{ Phys. Rev. D} \textbf{
  91} (2015) 031501(R),
  \href{http://dx.doi.org/10.1103/PhysRevD.91.031501}{\doi{10.1103/PhysRevD.91.031501}},
  \href{http://www.arXiv.org/abs/1411.2633}{\texttt{arXiv:1411.2633}}.

\bibitem{Czakon:2017wor}
M.~Czakon\hrefCMSnoop {}{ {et~al.}, ``Top-pair production at the {LHC} through
  {NNLO} {QCD} and {NLO} {EW}'',} \textit{ JHEP} \textbf{ 10} (2017) 186,
  \href{http://dx.doi.org/10.1007/JHEP10(2017)186}{\doi{10.1007/JHEP10(2017)186}},
  \href{http://www.arXiv.org/abs/1705.04105}{\texttt{arXiv:1705.04105}}.

\bibitem{Czakon:2018nun}
M.~Czakon\hrefCMSnoop {}{ {et~al.}, ``Resummation for (boosted) top-quark pair
  production at {NNLO+NNLL'} in {QCD}'',} \textit{ JHEP} \textbf{ 05} (2018)
  149,
  \href{http://dx.doi.org/10.1007/JHEP05(2018)149}{\doi{10.1007/JHEP05(2018)149}},
  \href{http://www.arXiv.org/abs/1803.07623}{\texttt{arXiv:1803.07623}}.

\bibitem{Bernreuther2010}
\hrefCMSnoop {}{W.~Bernreuther and Z.-G. Si, ``{Distributions and correlations
  for top quark pair production and decay at the Tevatron and LHC}'',} \textit{
  Nucl. Phys. B} \textbf{ 837} (2010) 90,
  \href{http://dx.doi.org/10.1016/j.nuclphysb.2010.05.001}{\doi{10.1016/j.nuclphysb.2010.05.001}},
  \href{http://www.arXiv.org/abs/1003.3926}{\texttt{arXiv:1003.3926}}.

\bibitem{Behring:2019iiv}
A.~Behring\hrefCMSnoop {}{ {et~al.}, ``Higher order corrections to spin
  correlations in top quark pair production at the {LHC}'',} \textit{ Phys.
  Rev. Lett.} \textbf{ 123} (2019) 082001,
  \href{http://dx.doi.org/10.1103/PhysRevLett.123.082001}{\doi{10.1103/PhysRevLett.123.082001}},
\href{http://www.arXiv.org/abs/1901.05407}{\texttt{arXiv:1901.05407}}.

\bibitem{10.2307/2958830}
\hrefCMSnoop {}{B.~Efron, ``Bootstrap methods: Another look at the
  jackknife'',} \textit{ Ann. Stat.} \textbf{ 7} (1979) 1,
  \href{http://dx.doi.org/10.1214/aos/1176344552}{\doi{10.1214/aos/1176344552}}.

\bibitem{bib:zhang}
\hrefCMSnoop {}{D.~Buarque~Franzosi and C.~Zhang, ``{Probing the top-quark
  chromomagnetic dipole moment at next-to-leading order in QCD}'',} \textit{
  Phys. Rev. D} \textbf{ 91} (2015) 114010,
  \href{http://dx.doi.org/10.1103/PhysRevD.91.114010}{\doi{10.1103/PhysRevD.91.114010}},
  \href{http://www.arXiv.org/abs/1503.08841}{\texttt{arXiv:1503.08841}}.

\bibitem{Martinez:2007qf}
\hrefCMSnoop {}{R.~Martinez, M.~A. Perez, and N.~Poveda, ``{Chromomagnetic
  dipole moment of the top quark revisited}'',} \textit{ Eur. Phys. J. C}
  \textbf{ 53} (2008) 221,
  \href{http://dx.doi.org/10.1140/epjc/s10052-007-0457-6}{\doi{10.1140/epjc/s10052-007-0457-6}},
  \href{http://www.arXiv.org/abs/hep-ph/0701098}{\texttt{arXiv:hep-ph/0701098}}.

\bibitem{Buchmuller:1985jz}
\hrefCMSnoop {}{W.~Buchm{\"u}ller and D.~Wyler, ``{Effective Lagrangian
  analysis of new interactions and flavor conservation}'',} \textit{ Nucl.
  Phys. B} \textbf{ 268} (1986) 621,
\href{http://dx.doi.org/10.1016/0550-3213(86)90262-2}{\doi{10.1016/0550-3213(86)90262-2}}.

\bibitem{Grzadkowski:2010es}
\hrefCMSnoop {}{B.~Grzadkowski, M.~Iskrzynski, M.~Misiak, and J.~Rosiek,
  ``{Dimension-six terms in the standard model Lagrangian}'',} \textit{ JHEP}
  \textbf{ 10} (2010) 085,
  \href{http://dx.doi.org/10.1007/JHEP10(2010)085}{\doi{10.1007/JHEP10(2010)085}},
\href{http://www.arXiv.org/abs/1008.4884}{\texttt{arXiv:1008.4884}}.

\bibitem{Buckley:2010ar}
A.~Buckley\hrefCMSnoop {}{ {et~al.}, ``{Rivet user manual}'',} \textit{ Comput.
  Phys. Commun.} \textbf{ 184} (2013) 2803,
  \href{http://dx.doi.org/10.1016/j.cpc.2013.05.021}{\doi{10.1016/j.cpc.2013.05.021}},
\href{http://www.arXiv.org/abs/1003.0694}{\texttt{arXiv:1003.0694}}.

\bibitem{Hioki:2013hva}
\hrefCMSnoop {}{Z.~Hioki and K.~Ohkuma, ``{Latest constraint on nonstandard
  top-gluon couplings at hadron colliders and its future prospect}'',} \textit{
  Phys. Rev. D} \textbf{ 88} (2013) 017503,
  \href{http://dx.doi.org/10.1103/PhysRevD.88.017503}{\doi{10.1103/PhysRevD.88.017503}},
\href{http://www.arXiv.org/abs/1306.5387}{\texttt{arXiv:1306.5387}}.

\bibitem{Aguilar-Saavedra:2014iga}
\hrefCMSnoop {}{J.~A. Aguilar-Saavedra, B.~Fuks, and M.~L. Mangano, ``{Pinning
  down top dipole moments with ultra-boosted tops}'',} \textit{ Phys. Rev. D}
  \textbf{ 91} (2015) 094021,
  \href{http://dx.doi.org/10.1103/PhysRevD.91.094021}{\doi{10.1103/PhysRevD.91.094021}},
\href{http://www.arXiv.org/abs/1412.6654}{\texttt{arXiv:1412.6654}}.

\bibitem{AguilarSaavedra:2018nen}
\hrefCMSnoop {}{D.~Barducci {et~al.}, ``{Interpreting top-quark LHC
  measurements in the standard-model effective field theory}'',} (2018).
\href{http://www.arXiv.org/abs/1802.07237}{\texttt{arXiv:1802.07237}}.

\bibitem{Baker:2006ts}
C.~A. Baker\hrefCMSnoop {}{ {et~al.}, ``An improved experimental limit on the
  electric dipole moment of the neutron'',} \textit{ Phys. Rev. Lett.} \textbf{
  97} (2006) 131801,
  \href{http://dx.doi.org/10.1103/PhysRevLett.97.131801}{\doi{10.1103/PhysRevLett.97.131801}},
\href{http://www.arXiv.org/abs/hep-ex/0602020}{\texttt{arXiv:hep-ex/0602020}}.

\bibitem{Afach:2015sja}
\hrefCMSnoop {}{J.~M. Pendlebury {et~al.}, ``Revised experimental upper limit
  on the electric dipole moment of the neutron'',} \textit{ Phys. Rev. D}
  \textbf{ 92} (2015) 092003,
  \href{http://dx.doi.org/10.1103/PhysRevD.92.092003}{\doi{10.1103/PhysRevD.92.092003}},
\href{http://www.arXiv.org/abs/1509.04411}{\texttt{arXiv:1509.04411}}.

\bibitem{Sjolin:2003ah}
\hrefCMSnoop {}{J.~Sj{\"o}lin, ``{LHC experimental sensitivity to CP violating
  g\ttbar couplings}'',} \textit{ J. Phys. G} \textbf{ 29} (2003) 543,
\href{http://dx.doi.org/10.1088/0954-3899/29/3/308}{\doi{10.1088/0954-3899/29/3/308}}.

\end{thebibliography}\endgroup

\cleardoublepage \appendix\section{The CMS Collaboration \label{app:collab}}\begin{sloppypar}\hyphenpenalty=5000\widowpenalty=500\clubpenalty=5000\vskip\cmsinstskip
\textbf{Yerevan Physics Institute, Yerevan, Armenia}\\*[0pt]
A.M.~Sirunyan, A.~Tumasyan
\vskip\cmsinstskip
\textbf{Institut für Hochenergiephysik, Wien, Austria}\\*[0pt]
W.~Adam, F.~Ambrogi, E.~Asilar, T.~Bergauer, J.~Brandstetter, M.~Dragicevic, J.~Erö, A.~Escalante~Del~Valle, M.~Flechl, R.~Frühwirth\cmsAuthorMark{1}, V.M.~Ghete, J.~Hrubec, M.~Jeitler\cmsAuthorMark{1}, N.~Krammer, I.~Krätschmer, D.~Liko, T.~Madlener, I.~Mikulec, N.~Rad, H.~Rohringer, J.~Schieck\cmsAuthorMark{1}, R.~Schöfbeck, M.~Spanring, D.~Spitzbart, W.~Waltenberger, J.~Wittmann, C.-E.~Wulz\cmsAuthorMark{1}, M.~Zarucki
% ...
\vskip\cmsinstskip
% ...
1:  Also at Vienna University of Technology, Vienna, Austria\\
% ...
\end{sloppypar}
\end{document}